\documentclass{aa}  
\usepackage[colorlinks=true,linkcolor=blue,citecolor=blue]{hyperref} 
\usepackage{graphicx}
\graphicspath{{images/}}

\usepackage{txfonts}
\usepackage{longtable}
\usepackage{amssymb} 
\usepackage{multirow}
\usepackage{blindtext}
\usepackage{verbatim}
\usepackage{adjustbox}
\usepackage{subfig}
\usepackage{threeparttable}
\usepackage{setspace}
\usepackage[normalem]{ulem}
\usepackage{color}
\newcommand{\redcomment}[1]{{\color{red}#1}}

\newcommand{\vsys}{$v_{\rm sys}$}
\newcommand{\kp}{$K_{\rm p}$}
\def \Hei{\ion{He}{I}\,\,} 

\usepackage{soul} 

\usepackage{acronym}
\acrodef{LC}[LC]{light curve}
\acrodef{GP}[GP]{Gaussian process}
\acrodef{MCMC}[MCMC]{Markov Chain Monte Carlo}
\acrodef{HJ}[HJ]{Hot Jupiter}
\acrodef{LD}[LD]{limb darkening}
\acrodef{PSD}[PSD]{power spectral density}
\acrodef{CCF}[CCF]{cross-correlation function}
\acrodef{RV}[RV]{radial velocity}
\acrodef{RML}[RML]{Rossiter–McLaughlin}
\acrodef{CLV}[CLV]{center-to-limb variation}

\begin{document} 

\newcommand{\teff}{T$_{\rm eff}$}
\newcommand{\logg}{$\log${(g)}}
\newcommand{\hatssb}{HAT-P-67~b}
\newcommand{\hatss}{HAT-P-67}
\newcommand{\tess}{TESS}

\title{The GAPS Programme at TNG. TBD: Characterization of the low-density gas giant HAT-P-67 b with GIARPS
\thanks{Based on observations made with the Italian \textit{Telescopio Nazionale Galileo} (TNG) operated on the island of La Palma by the \textit{Fundación Galileo Galilei} (FGG) of the \textit{Istituto Nazionale di Astrofisica} (INAF) at the Spanish \textit{Observatorio del Roque de los Muchachos} of the \textit{Instituto de Astrofisica de Canarias}. }}
\author{D. Sicilia\inst{1}, G. Scandariato\inst{1}, G. Guilluy\inst{2}, M. Esposito\inst{3}, F. Borsa\inst{4}, M. Stangret\inst{5}, C. Di Maio\inst{6}, A. F. Lanza\inst{1},  A. S. Bonomo\inst{2}, S. Desidera\inst{5}, L. Fossati\inst{7}, D. Nardiello\inst{8,5}, A. Sozzetti\inst{2}, 
L. Malavolta\inst{8}, V. Nascimbeni\inst{5}, M. Rainer\inst{4}, M. C. D'Arpa\inst{6,9}, L. Mancini\inst{10,2,11}, V. Singh\inst{1}, T. Zingales\inst{8}, L. Affer\inst{6}, A. Bignamini\inst{12}, R. Claudi\inst{5}, S. Colombo\inst{6}, R. Cosentino\inst{13}, A. Ghedina\inst{13}, G. Micela\inst{6}, E. Molinari\inst{4}, M. Molinaro\inst{12}, I. Pagano\inst{1}, G. Piotto\inst{8}}
\institute{
\label{inst:1}INAF - Osservatorio Astrofisico di Catania, Via S. Sofia 78, 95123 Catania, Italy 
\and 
\label{inst:2}INAF – Osservatorio Astrofisico di Torino, Via Osservatorio 20, 10025, Pino Torinese, Italy
\and 
\label{inst:3}Thüringer Landessternwarte Tautenburg, Sternwarte 5, 07778 Tautenburg, Germany
\and
\label{inst:4}INAF – Osservatorio Astronomico di Brera, Via E. Bianchi 46, 23807 Merate, Italy
\and
\label{inst:5}INAF, Osservatorio Astronomico di Padova, Vicolo dell'Osservatorio 5, Padova I-35122, Italy
\and
\label{inst:6}INAF – Osservatorio Astronomico di Palermo, Piazza del Parlamento, 1, I-90134 Palermo, Italy
\and
\label{inst:7}Space Research Institute, Austrian Academy of Sciences, Schmiedl- strasse 6, 8042 Graz, Austria
\and
\label{inst:8}Dipartimento di Fisica e Astronomia "Galileo Galilei" -- Universit\`a degli Studi di Padova, Vicolo dell'Osservatorio 3, I-35122 Padova, Italy 
\and
\label{inst:9}University of Palermo, Department of Physics and Chemistry “Emilio Segrè, Via Archirafi 36, Palermo, Italy
\and
\label{inst:10}Department of Physics, University of Rome "Tor Vergata", Via della Ricerca Scientifica 1, I-00133, Roma, Italy
\and
\label{inst:11}Max Planck Institute for Astronomy, Königstuhl 17, D-69117, Heidelberg, Germany
\and
\label{inst:12}INAF – Osservatorio Astronomico di Trieste, via Tiepolo 11, 34143 Trieste, Italy
\and
\label{inst:13}Fundación Galileo Galilei-INAF, Rambla José Ana Fernandez Pérez 7, 38712 Breña Baja, TF, Spain
}

\date{}
 
\abstract{
\textit{Context}. 
\hatssb\ is one of the lowest-density gas giants known to date, making it an excellent target for atmospheric characterization through the transmission spectroscopy technique. 
\\
\textit{Aims}. In the framework of the GAPS large programme, we collected four transit events of \hatssb \, with the aim of studying the exoplanet atmosphere and deriving the orbital projected obliquity. \\
\textit{Methods}. We exploited the high-precision GIARPS (GIANO-B + HARPS-N) observing mode of the Telescopio Nazionale Galileo (TNG), along with additional archival TESS photometry, to explore the activity level of the host star. We performed transmission spectroscopy, both in the visible (VIS) and in the near-infrared (nIR) wavelength range, and analysed the \ac{RML} effect both fitting the radial velocities and the Doppler shadow.
Based on the TESS photometry, we redetermined the transit parameters of \hatssb. \\
\textit{Results}. By modelling the \ac{RML} effect, we derived a sky-projected obliquity 
of ($2.2\pm0.4$)$^{\circ}$
indicating an aligned planetary orbit. The chromospheric activity index $\log\,R^{\prime}_{\rm HK}$, the CCF profile, and the variability in the transmission spectrum of the H$\rm\alpha$ line suggest that the host star shows signatures of stellar activity and/or pulsations. We found no evidence of atomic or molecular species in the optical transmission spectra, with the exception of pseudo-signals corresponding to \ion{Cr}{I}, \ion{Fe}{I}, H$\rm\alpha$, \ion{Na}{I}, and \ion{Ti}{I}. In the nIR range, we found an absorption signal of the \Hei triplet of 5.56$^{+ 0.29 }_{ -0.30 }$\% (19.0$\sigma$), corresponding to an effective planetary radius of $\sim$ 3 $R\rm_p$ (where $R\rm_p \sim$ 2 $R\rm_J$) which extends beyond the planet's Roche Lobe radius.
\\
\textit{Conclusions}. 
Owing to the stellar 
variability, together with the high uncertainty of the model, we could not confirm the
planetary origin of the signals found in the optical 
transmission spectrum. On the other hand, we confirmed previous detections of the infrared \Hei triplet, providing a 19.0$\sigma$ detection. Our finding indicates that the planet's atmosphere is evaporating.
}


\titlerunning{Characterization of HAT-P-67 b with GIARPS}
\authorrunning{D. Sicilia et al.}
\maketitle

\begin{table*}
\centering
\begin{threeparttable}
\caption{Summary of TESS and GIARPS data used in this work.}\label{tab:observing_logs}
    \centering
     \begin{tabular}{c c c c c c}
    
         \multicolumn{6}{c}{\textit{TESS}}\\
      \hline
    \hline
    Sector & Date start [UTC] & Date end [UTC] & $N_{\rm transits}$ & $t_{\rm exp}$[s] & Simultaneous spectroscopy\\
      \hline
     24 & 2020-04-16 06:55:19 & 2020-05-12 18:41:18 & 6 & 120 & No \\
     26 & 2020-06-09 18:15:17 & 2020-07-04 15:11:16 & 6 & 120 & Yes (N2)\\
     51 & 2022-04-23 10:34:51 & 2022-05-18 00:46:50 & 3 & 120 & No\\
     52 & 2022-05-19 03:04:50 & 2022-06-12 13:46:49 & 5 & 120 & No\\
     53 & 2022-06-13 11:44:48 & 2022-07-08 11:26:47 & 4 & 120 & No\\
     \hline
    \\
       \end{tabular}
\end{threeparttable}
\end{table*}
       
\begin{table*}
\centering
\begin{threeparttable}
    \begin{tabular}{c |cc|cc| c c c}
    \multicolumn{8}{c}{\textit{GIARPS}}\\
    \hline
    \hline
    Night & $N_{\rm obs}$\tnote{\textit{a}} & $t_{\rm exp}$[s] &  $N_{\rm obs}$\tnote{\textit{a}} & $t_{\rm exp}$[s] & Airmass\tnote{\textit{b}} & S/N (min-max)\tnote{\textit{c}} & Seeing\\
    & \multicolumn{2}{c|}{\textit{HARPS-N}} & \multicolumn{2}{c|}{\textit{GIANO-B}} & & & \\
      \hline
    N1: 2020-05-26 & 45 (42) & 600 & 80 (71) & 300 & 1.69-1.04-1.34 &  42.6-67.1 & 1.0" \\
    N2: 2020-06-24 & 31 (23) & 600 & 50 (39) & 300 & 1.18-1.04-1.27 & 29.8-52.0 & 1-3" very variable\\
    N3: 2021-06-10 & 45 (40) & 600 & - & - & 1.47-1.04-1.56 & 22.2-47.0 & 0.9"\\
    N4: 2023-07-10 & 42 (37) & 600 & 68 (59) & 300 & 1.14-1.04-2.08 & 23.0-36.3 & 0.7"\\
       \hline
    \end{tabular}
   \rule{0pt}{3ex}  
\begin{tablenotes}\footnotesize
\item[\textit{a}] In parenthesis: the number of analysed spectra that are considered in-transit (between the first and fourth contact).
\item[\textit{b}] The values are extracted from the FITS header of the HARPS-N spectra, and indicate the airmass at the beginning, during its minimum, and at the end of the transit.
\item[\textit{c}] The signal-to-noise ratio (S/N) indicated is extracted from the FITS header of the HARPS-N spectra on the 53rd order that contains the sodium feature. The values are more or less the same as those of the GIANO-B spectra extracted in the region of the \Hei triplet.
\end{tablenotes}
\end{threeparttable}
\end{table*}

\section{Introduction}
Over the last few decades, the field of exoplanet research has grown rapidly, revealing that extrasolar systems are very common and extremely diverse in masses, radii, temperatures, and orbital parameters. 
Thanks to increasingly efficient ground- and space-based surveys, today we are able to explore the exoplanet compositions and atmospheres in an ever larger sample.

Puffy planets (i.e., planets with large radii and very low densities) constitute some of the most favorable targets for characterization through the transmission spectroscopy technique (e.g., \citealt{Sedaghati_2016, Allart_2020, Colon_2020, Czesla_2022}). Due to their large radius and low surface gravity, they are expected to present a high atmospheric pressure scale height ($H$), which is the characteristic length scale of the atmosphere. The amplitude of an absorption signal in transmission spectroscopy, i.e. the change in measured transit depth, 
is proportional to $H$ ($2HR\rm_p/$$R_\star^2$, \citealt{Brown}). Transmission spectral signals are typically on the order of 1 to $\sim 5 \, H $ in size, thus if the transit depth can be measured to about 1 $H$ in precision with sufficient spectral resolution, detectable spectral features would begin to appear. Since puffy planets are characterized by atmospheres with higher $H$ values, they are expected to present stronger absorption signals (assuming the host star's brightness and the signal-to-noise ratio high enough).
Some of them that have been explored so far, such as KELT-11 b \citep{Pepper_2017}, WASP-17 b \citep{Anderson_2010}, and WASP-127 b \citep{Lam_2017}, show even $H \gtrsim 1900$ km (compared to $\sim$10 km on Earth, or $\sim$27 km on Jupiter). In these cases, the signal in transmission for 1 $H$ is of the order of $10^{-2}-1 \%$, (compared to $\sim2\cdot10^{-5} \%$ on Earth, and  $\sim10^{-3} \%$ on Jupiter).

\hatssb\ \citep{Zhou_2017} is a gas giant puffy planet transiting a rapidly rotating ($v\sin{i}$ = 35.8 $\pm$ 1.1 km$\,$s$^{-1}$) F-subgiant star every $\sim$ 4.8 days, at the orbital distance of 0.065 au. It is one of the largest ($R\rm_p \sim$ 2 $R\rm_J$) and lowest density ($\rho \sim$  0.05 
g cm$^{-3}$) planets found to date. Due to the rapid rotation of the host star, the \ac{RV} technique does not allow to precisely determine the semi-amplitude of the planet $K\rm_p$ and, consequently, the mass of the planet, for which there is only an upper limit ($M\rm_p$ < 0.59 $M\rm_J$). A lower limit ($M\rm_p$ > 0.056 $M\rm_J$) was also be applied by \citet{Zhou_2017} when assuming the planet is not undergoing Roche lobe overflow. The host star belongs to a binary stellar system; however, the M dwarf companion (HAT-P-67B, \citealt{Mugrauer2019}), 
has a projected separation of about 3400 au, thus not being a source of contamination for observations.

The low density and high irradiation of \hatssb \, (it receives approximately two times
the incident flux of a zero-age main-sequence star) also results in a bloated atmosphere with a large $H$ of $\sim$ 3500 km (assuming an H$_2$/He mixture of near-solar composition atmosphere). This makes the planet another good candidate for transmission spectroscopy studies. 

Recently, the atmosphere of \hatssb \, was explored by \citet{Bello-Arufe2023}, through the analysis of one full transit acquired with CARMENES \citep{CARMENES}. The authors reported the detection of \ion{Ca}{II} and \ion{Na}{I} (with a statistical significance of 13.2$\sigma$ and 4.6$\sigma$ respectively), which they ascribe to the planetary signal. Besides, they found strong variability in the H$\rm\alpha$ line and in the He triplet, suggesting the possible presence of an extended planetary outflow. This extended atmosphere was confirmed by \citet{Gully2023} who reported an absorption depth up to 10\% in the stellar \Hei triplet, thanks to a series of observations taken with the Habitable Zone Planet Finder Spectrograph (HPF, \citealt{Mahadevan2012}). Besides, \citet{Gully2023} derived an increase of the stellar radius (2.65 $\pm$ 0.12 $R\rm_J$), to match the updated Gaia DR3 distance (8.7 \% farther than previously estimated by \citealt{Zhou_2017}). However, this update does not entail any major changes to the rest of the parameters derived by \citet{Zhou_2017}.

\hatssb \, is one of the targets of the atmospheric sample of the Global Architecture of Planetary Systems (GAPS) \citep{Guilluy2022}. Thanks to the availability of the HARPS-N spectrograph \citep{Cosentino_2012}, mounted at the Italian Telescopio Nazionale Galileo (TNG) in La Palma,
we have collected and analysed 
four transits of \hatssb. The GIARPS observing mode \citep{Claudi_2018}, allowed us to gather simultaneous observations both in the visible (VIS) with HARPS-N (0.39 - 0.68 $\mu$m, $R \simeq$ 115,000), and in the near infrared
(nIR) with GIANO-B (0.95 - 2.45 $\mu$m, $R \simeq$ 50,000).
Thereby providing capability to 
investigate the system architecture and the exoplanet atmosphere in a wider wavelength range.

A description of the observations used in this work is presented in Sect. \ref{sec:observations}. Taking advantage of multiple Transiting Exoplanet Survey Satellite (TESS, \citealt{Ricker2014}) light curves, and high precision HARPS-N RV time series, we redetermined the 
orbital parameters 
and 
analysed the \ac{RML} effect, 
in Sect. \ref{sec:orbit}.
We also tried to explore the activity level of the host star using different indicators in Sect. \ref{sec:activity}.
We presented the characterization of the planetary atmosphere through the analysis of the transmission spectrum both in the VIS and in the nIR in Sect. \ref{sec:atmosphere}. 
Finally, we discussed our findings in Sect. \ref{sec:summary}.

\section{Observations}\label{sec:observations}
For the analysis of the HAT-P-67 system we used both high-resolution VIS and nIR spectroscopy with 
the GIARPS (GIANO-B + HARPS-N) observing mode of the TNG, 
as well as photometry from TESS.
A summary of the observations and stellar parameters adopted in this work for analysis is reported in Tables \ref{tab:observing_logs} and \ref{tab:parameters}, respectively.

\begin{table}[]
\centering
\caption{Adopted stellar parameters. All parameters listed here are taken from \citet{Zhou_2017}, except the stellar radius  which is taken from \citet{Gully2023}.}\label{tab:parameters}
    \centering
    \begin{spacing}{1.4}
    \begin{tabular}{l c}
    \hline
    \hline
     Symbol & Value \\
      \hline 
         $T_{\rm eff}$ [K] & 6406$^{+65}_{-61}$  \\
        $\rm [Fe/H]$ & -0.080 $\pm$ 0.050 \\
          $v\sin{i}$ [km$\,$s$^{-1}$] & 35.8 $\pm$ 1.1 \\
          $M_\star [M_\odot]$ & $1.642^{+0.155}_{-0.072}$ \\
         $R_\star [R_\odot]$ & 2.65 $\pm$ 0.12 \\
         $\log g_\star[\log_{\rm10}$ (cm$\,$s$^{-2}$)$]$ &$3.854^{+0.014}_{-0.023}$\\
        \hline
 
    \end{tabular}
    \end{spacing}
\end{table}

\begin{figure}
    \centering
    \includegraphics[width=\linewidth]{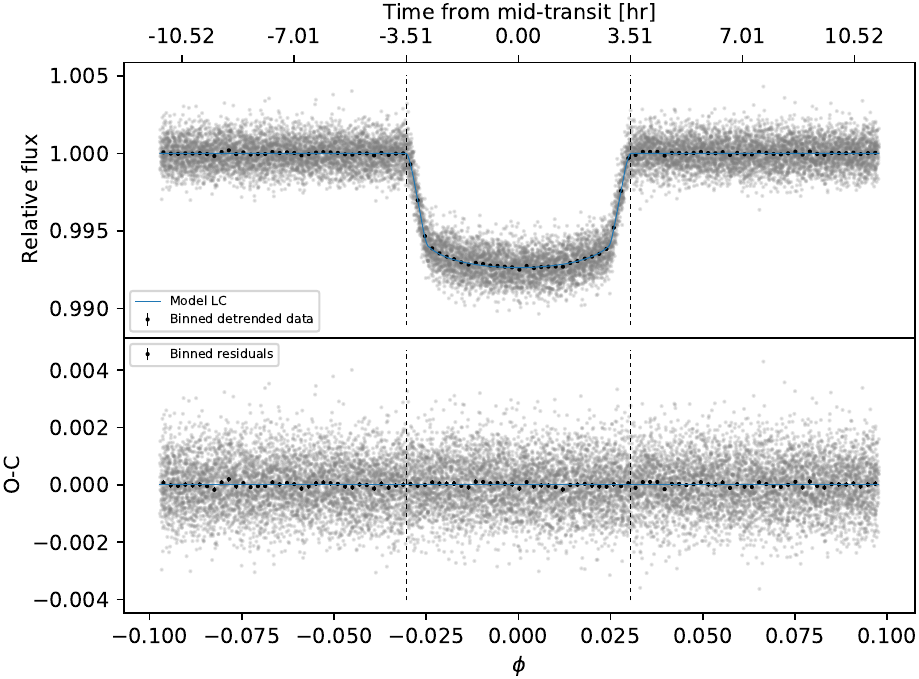}\\
    \includegraphics[width=\linewidth]{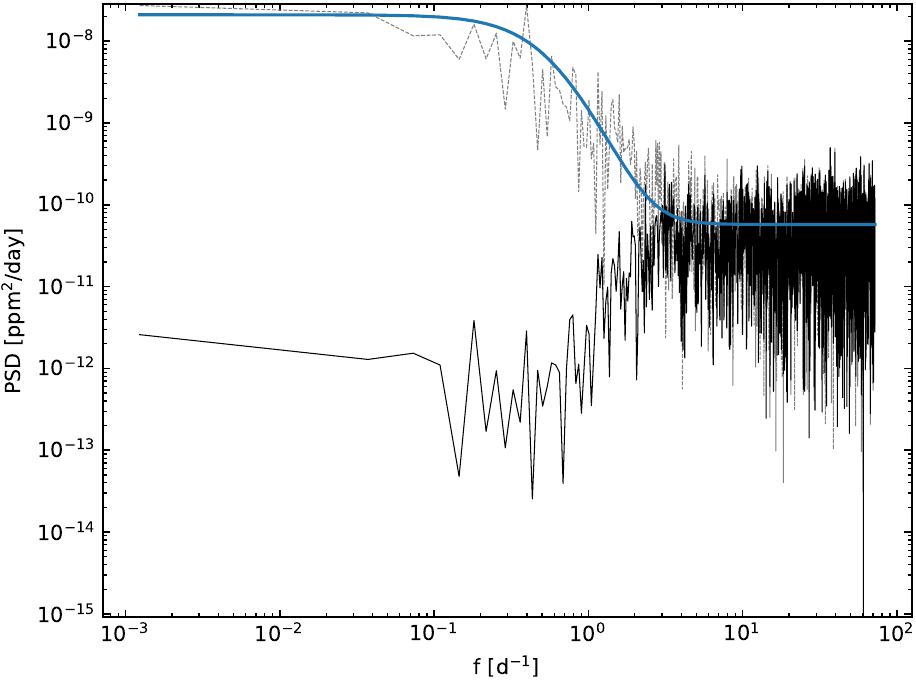}
    \caption{Results of the fit of the transit \acp{LC}. \textit{Top - } Phase-folding of the detrended data (\textit{top pane}l) and the corresponding O-C diagram (\textit{bottom panel}). For clarity, in each panel, we show the binned data with black dots. \textit{Bottom - }The gray dashed line shows the \ac{PSD} of the data after removing the best-fit transit model, while the blue solid line represents the combined \ac{PSD} of the best-fit \ac{GP} (the shoulder at f$\lesssim$1~d$^{-1}$) and the white noise in the data (the plateau at f$\gtrsim$10~d$^{-1}$). The black line is the \ac{PSD} of the residuals of the best fit, showing that the power excess at low frequencies has been effectively removed by the \ac{GP} in the model.}\label{fig:transitFit}
\end{figure}

\begin{table*}
\begin{center}
\caption{Model parameters for the fit of the \tess\ data.}\label{tab:orbit}
\begin{tabular}{llllll}
\hline\hline
Free parameters & Symbol & Units & C.I.\tablefootmark{a} & Prior\\
\hline
GP amplitude & $\log h$ & --- &  -7.49(7) & $U$(-8,-5) \\
GP timescale & $\log\frac{\lambda}{\rm 1~day}$ & --- &  -0.76(9) & $U$(-4,2) \\
Time of transit & $T_0$ & BJD$_{\rm TDB}$ &  2459338.0790(1) & $U$(2459338.078,2459338.080)\\
Orbital frequency & $\nu_{\rm orb}$ & d$^{-1}$ & 0.20789543(7) & $U$(0.207895,0.207897) \\
Stellar density & $\rho_\star$ & $\rho_\sun$  & 0.083(2) & $U$(0.07,0.09)\\
Radii ratio\tablefootmark & $R_{\rm p}$/$R_\star$ & ---  & 0.0821(2) & $U$(0.075,0.090) \\
Impact parameter & $b$ & --- & 0.44(2) & $U$(0.3,0.5) \\
First LD coef.\tablefootmark & $q_{\rm 1}$ & ---  & 0.16(2) & $U$(0,1) \\
Second LD coef.\tablefootmark & $q_{\rm 2}$ & ---  & 0.33(8) & $U$(0,1) \\
\hline
\hline
Derived parameters & Symbol & Units  & C.I. & \\
\hline
Planetary radius & $R\rm _p$ & $R\rm _J$ & 2.1(1) & including the stellar radius uncertainty\\
Orbital period & $P_{\rm orb}$ & d & 4.810110(2) & \\
Total transit duration & $T_{\rm 14}$ & hr & 7.01(1) & \\
Full transit duration & $T_{\rm 23}$ & hr & 5.70(2) & \\
Scaled semi-major axis & $a/R_\star$ & --- & 5.22(4) & \\
Orbital inclination & $i$ & deg & 85.1(2) & \\
\hline
\end{tabular}
\tablefoot{
        \tablefoottext{a}{The 68\% confidence interval. Uncertainties expressed in parentheses refer to the last digit.}
}
\end{center}
\end{table*}

\subsection{Spectroscopy}\label{sec:spectroscopy}
We observed four transits of \hatssb
, three of which in the GIARPS mode, in the framework of the 
GAPS large/long-term program \citep{GAPS_Covino}. 

The exposure time ($t\rm_{exp}$) of the observations of \hatssb \, was fixed to 600 s for HARPS-N, and 300 s for GIANO-B, yielding runs of 46, 31, 46 and 42 VIS spectra, and 80, 50 and 68 nIR spectra during each transit. However, 
the very long duration of the transit ($\sim$ 7 hours) results in quite a low number of out-of-transit spectra with respect to the total acquired spectra (only 23 out of 165 in the VIS, and 29 of 198 in the nIR). To determine whether a spectrum was (fully)in-transit or out-of-transit, we considered the half of $t\rm_{exp}$ in addition to the BJD (Barycentric Julian Date) of the observation. Sky spectra were retrieved simultaneously with the science observations thanks to a dedicated fiber, named fiber B, pointing at a fixed position at around 10 arcsec from the target star, ensuring the same atmospheric conditions in both spectra.

In the first run (N1), there are hints of stellar activity which we discussed in Section \ref{sec:activity}; the second run (N2) covers only the second half of the transit and is characterized by a strong variability in seeing (1 - 3 arcsec), but it is the run with the highest number of out-of-transit spectra; during the third run (N3), GIANO-B was offline so we collected only HARPS-N observations which present some spectra with a lower signal-to-noise-ratio (S/N), probably due to thin clouds and the presence of some calima
; the fourth run (N4) does not cover the egress and the post-transit phase and is characterized by a lower S/N compared to the other nights (see Fig. \ref{fig:airmass_snr}, right panel). We decided to remove the first exposure of N1 since the corresponding RV measurement deviates from the expected pattern, creating a distortion of the fit.
For transmission spectrum analysis only, we also discarded the exposure with the lowest S/N of N3 (< 20, at the center of the 53rd order, with the sodium doublet). 

\subsection{Photometry}\label{sec:photometry}

The time span covered by our spectroscopic survey almost completely overlaps with the \tess\ \citep{Ricker2014} observations of \hatss. This allowed us to determine the precise ephemeris of \hatssb\ and closest to the epochs of our GIARPS data, thus best suited for the extraction of the transmission signal of \hatssb.
\tess\ observed the \hatss\ system with a 2 min cadence in sector 24 (S24, from 2020 April 16 to 2020 May 12, six transits observed), sector 26 (S26, from 2020 June 9 to 2020 July 4, six transits), sector 51 (S51, from 2022 April 23 to 2022 May 18, three transits), sector 52 (S52, from 2022 May 19 to 2022 June 12, five transits), and sector 53 (S53, from 2022 June 13 to 2022 July 8, four transits). Of the analysed \tess\ sectors, only S26 covers one of the simultaneously observed transits with GIARPS which corresponds to the half-transit retrieved during the second night.

Using the package \texttt{lightkurve} \citep{Lightkurve2018}, we retrieved the Pre-search Data Conditioning Single Aperture Photometry (PDCSAP), which is corrected for instrumental systematics and for contamination from some nearby stars \citep{Smith2012,Stumpe2012,Stumpe2014}. We took into consideration only the photometry with good quality flag.

\section{Orbital parameters}\label{sec:orbit}
We redetermined the \hatssb \, transit parameters. Given the difficulty in the interpretation of the long-term stellar variability (Sect.~\ref{sec:activity}), we trimmed segments of the \acp{LC} centered on the transit events and as wide as three times the expected transit duration. Each photometric segment was normalised to the median out-of-transit flux. A total of 22 transits were extracted from the \tess\ \acp{LC}.

We adopted the same Bayesian approach described in \citet{Scandariato2022}: we maximized the likelihood of a model that includes the 
combined transit fit along with a \ac{GP}
to detrend against long-term stellar/instrumental systematics. The presence of long-term trends is testified by the \ac{PSD} of the data, that  monotonically increases with decreasing frequency. We thus used a Mat\'ern~3/2 kernel for the \ac{GP} (see Fig.~\ref{fig:transitFit}), being it characterized by a \ac{PSD} similar to the data. This approach is less time-consuming than fitting for each transit a polynomial trend. Moreover, it greatly reduces the dimensionality of the model: 2 free parameters of the \ac{GP} model against $22\times(deg+1)$ parameters for the polynomial detrending, where $deg$ is the polynomial degree used for the detrending.

The transit profile was computed using the quadratic \ac{LD} law of \citet{Mandel2002} with the reparametrization of the \ac{LD} coefficients of \citet{Kipping2013}. We also included in the model a re-normalisation factor and a jitter term to fit the white noise not included in the nominal photometric uncertainties.  We sampled the parameter space in a \ac{MCMC} framework using the Python \texttt{emcee} package version 3.1.3 \citep{Foreman2013}. We used 44 walkers, corresponding to four times the number of free parameters. We ran the chains for 50\,000 samplings, enough to ensure convergence following the criterion described in \citet{Goodman&Weare2010}. We used flat priors (listed in Table \ref{tab:orbit}) for all the fitting parameters. We ran the code in the HOTCAT computing infrastructure \citep{Bertocco2020,Taffoni2020}.
The results of the \ac{MCMC} fit are listed in Table~\ref{tab:orbit}. 
The best-fitting model is over-plotted to the phase-folded data in Fig.~\ref{fig:transitFit}.
\begin{figure}
    \centering
    \includegraphics[width=\linewidth]{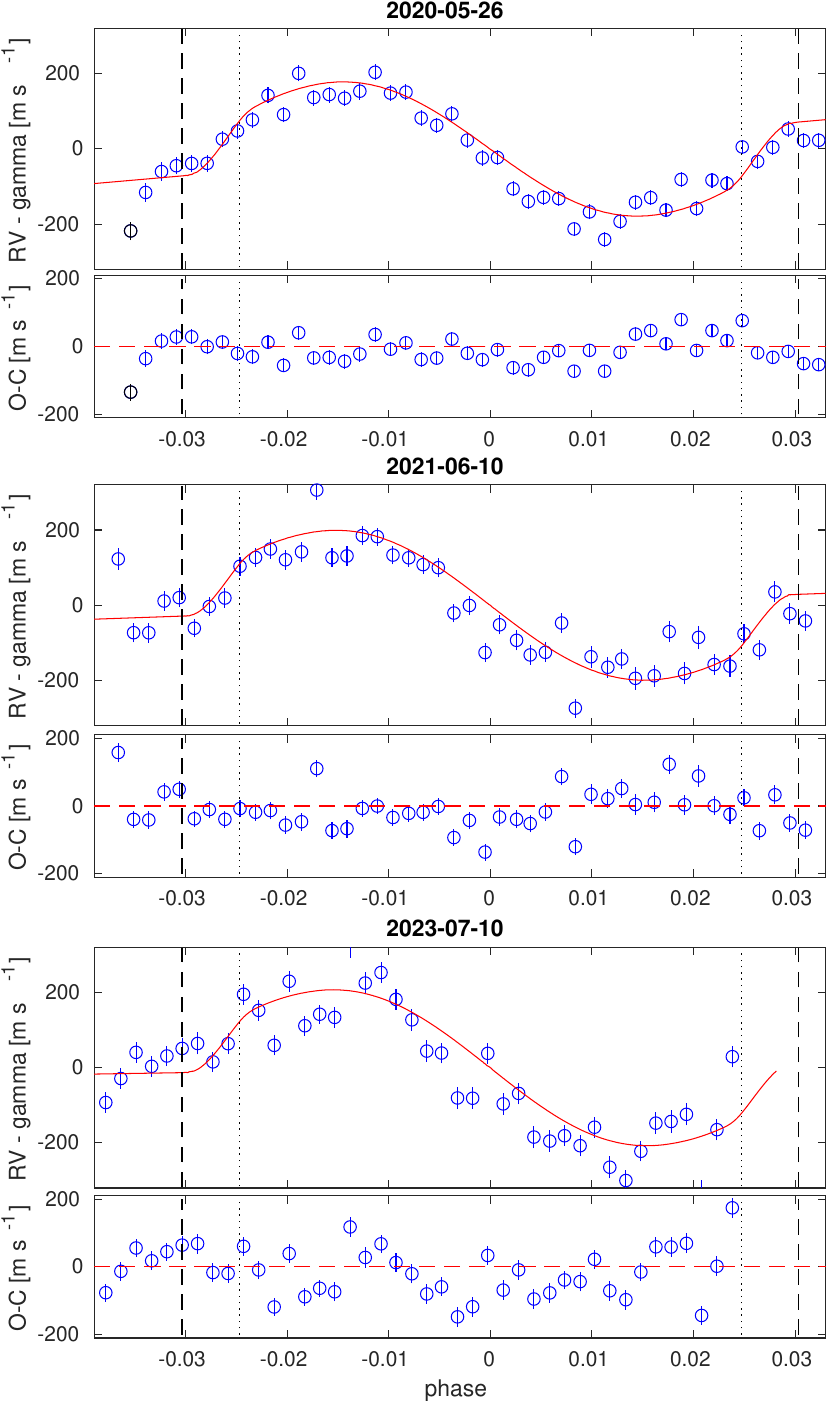}
     \caption{The Rossiter-McLaughlin effect analysis. \textit{Top - } RV time series taken during the transit on 2020-05-26 (N1). The best-fit model is superimposed and the corresponding residuals are shown in the lower panel. The dashed black lines indicate the points of first and fourth contact of the transit, while the dotted black lines represent the points of second and third contact.
     \textit{Middle - } The same for the transit 
     on 2021-06-10 (N3). \textit{Bottom - } The same for the transit on 2023-07-10 (N4).}
     \label{fig:RMeffect}
\end{figure}

Then, we used the \ac{RV} time series 
to detect the \ac{RML} effect (see e.g., \citealt{2022Albrecht}) and measure the sky-projected angle 
between the
stellar spin axis and the planet orbital axis. The data were reduced using the 3.7 version of the HARPS-N Data Reduction Software (DRS) through the Yabi web application, hosted at the Italian center for Astronomical Archive (IA2)\footnote{\url{https://www.ia2.inaf.it}}. The \ac{RV} measurements were obtained using a G2 mask template and a \ac{CCF} width of 250 km$\,$s$^{-1}$, with a step of 0.25 km$\,$s$^{-1}$. The lists of the \acp{RV} are presented in Tables \ref{tab:rvs-rhk}-\ref{tab:rvs-rhk4}, together with the stellar activity index $\log\,R^{\prime}_{\rm HK}$ (see Sect. \ref{sec:activity}).
We did not consider the observations taken on N2 as they do not cover
a full transit and were affected by highly variable seeing (see Table \ref{tab:observing_logs} and 
Fig. \ref{fig:airmass_snr}). The \ac{RML} effect modelling and the \acs{RV} fitting were carried out using a code
developed by us within the MATLAB software environment. A detailed explanation of the working principles of the code can be found in \citet{Esposito2017}.

For the fit of the \ac{RML} effect, most of the relevant parameters were adopted from the literature
($M_\star$, $R_\star$, $v \sin{i}$, see Table \ref{tab:parameters}) and from our analysis of the TESS LCs ($T_0$, $P\rm_{orb}$, $i$, $R\rm _p$, see Table \ref{tab:orbit}). The only parameters which were left free to vary were the barycentric \ac{RV} 
(\vsys)
and the projected
spin-orbit angle ($\lambda$). 
Given the large uncertainty on the planet's mass and considering that 
significant RV variations induced by stellar activity are expected
on a time scale of a few hours, we also added as a free parameter an RV linear trend (LT).

We fitted separately the three transit RV time series taken on 2020-05-26, 2021-06-10 and 2023-07-10.
For the first (second, third) transit the best-fit results are: 
$\lambda$ = 6.4 $\pm$ 7.5 deg (8 $\pm$ 11 deg, 10 $\pm$ 12 deg ), 
\vsys = -1968 $\pm$ 22 m s$^{-1}$
(-2225 $\pm$ 33 m s$^{-1}$, -2250 $\pm$ 45 m s$^{-1}$) and LT = 0.0057 $\pm$ 0.0015 m s$^{-2}$ (0.0021 $\pm$ 0.0020 m s$^{-2}$, 0.0010 $\pm$ 0.0020 m s$^{-2}$). The measure of \vsys \, found for N1 is statistically far from the values found for the other nights; this offset is likely due to a higher stellar variability (see Sect. \ref{sec:activity}).
The three RV time series with the best-fit models superimposed are shown in Fig.~\ref{fig:RMeffect}. 

In order to improve the precision on $\lambda$, we then performed a fit to the Doppler shadow, in the same way as in \citet{borsa21}. Indeed, for relatively fast rotators, such as \hatss, the tomography method can better constrain $\lambda$ than \acp{RV}. 
The Doppler shadow model was taken from EXOFASTv2 \citep{exofastv2}, and fitted to the data in a Bayesian framework by employing a Differential Evolution Markov chain Monte Carlo (DE-MCMC)
technique \citep{TerBraak2006, Eastman2013}, running ten DE-MCMC chains of 50,000 steps and discarding the burn-in, until convergence was reached. 
We fixed $T_0$ and $P_{\rm orb}$ as in Table \ref{tab:orbit}, as well as the quadratic limb darkening parameters ($\mu_1=0.43$ and $\mu_2=0.25$, taken from ExoCTK\footnote{\url{https://exoctk.stsci.edu/limb_darkening}}) and \vsys \, = -2000 $\mathrm{m\,s}^{-1}$. We note that the \vsys \, value cannot be well constrained by the Doppler tomography fit for fast rotators, on the contrary of what happens for \acp{RV}, and that changing this value within the differences found by the \acp{RV} analysis of the different transits does not affect the results.
We left $i$, $a/R_\star$, $R_{\rm p}$/$R_\star$ as free parameters with values and error-bars in Table \ref{tab:orbit} as priors. 
The $v \sin{i}$ (which includes macroturbolence) and $\lambda$ parameters were left free with uniform priors. 
The medians and the 15.86\% and 84.14\% quantiles of the posterior distributions were taken as the best values and $1\sigma$ uncertainties. 
We fitted independently all the four transits, finding $v \sin{i}$ = $39.2\pm0.5$, $36.7\pm0.9$, $38.3\pm0.7$, $39.5\pm0.8$ $\mathrm{km\,s}^{-1}$ and $\lambda$ = 1.0 $\pm$ 0.6, 5.5 $\pm$ 1.1, 1.2 $\pm$ 0.8, 3.2 $\pm$ 0.9 degrees for night 1 to 4, respectively. When fitting the four transits together, we find $v \sin{i}$ = 39.3 $\pm$ 0.4 $\mathrm{km\,s}^{-1}$ and $\lambda=2.2\pm0.4$ degrees. As expected, the best values of $\lambda$ are both more accurate and precise than the ones found by fitting the \acp{RV}. We note that the $v \sin{i}$ value is compatible with that found by \citet{Zhou_2017} within 3$\sigma$. 

\begin{figure*}
    \centering
    \includegraphics[width=\textwidth]{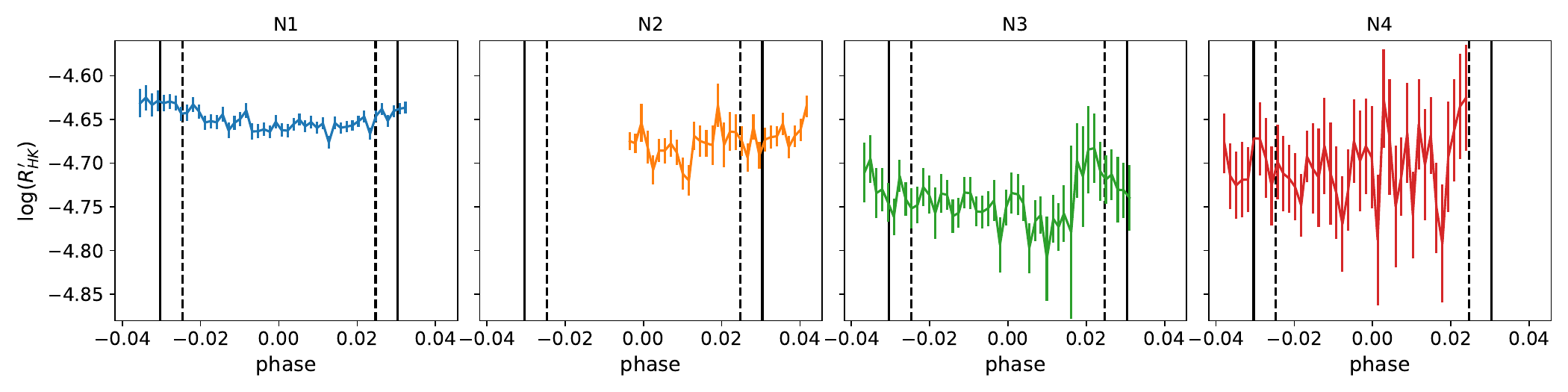}
    \caption{The activity indicator $\log\,R^{\prime}_{\rm HK}$ as a function of the time for each night. The continuous black lines indicate the points of first and fourth contact of the transit, while the dashed black lines represent the points of second and third contact.}
    \label{fig:rhk_index}
\end{figure*}

\section{Stellar activity}\label{sec:activity}
Stellar activity can mimic spurious features in the retrieved transmission spectra (e.g., \citealt{Oshagh_2014, Apai2018}). To detect potential stellar effects in the analysed nights, we first measured the $\log\,R^{\prime}_{\rm HK}$ chromospheric activity index (Fig. \ref{fig:rhk_index} and Tables \ref{tab:rvs-rhk}-\ref{tab:rvs-rhk4}). We extracted it from the HARPS-N spectra through the YABI platform \citep{Hunter2012}, using a \textit{B-V} color index of 0.441 mag. We derived average values of -4.651 $\pm$ 0.001, -4.675 $\pm$ 0.003, -4.743 $\pm$ 0.004, and -4.704 $\pm$ 0.007 for N1, N2, N3 and N4 respectively. The obtained values
are moderately larger than the solar value at the maximum of activity (-4.75 according to \citealt{Dumusque_2011}; or -4.905 according to \citealt{Egeland_2017}). N1 presents slightly higher activity 
compared with the other nights, while N4 is characterized by greater uncertainty, likely due to the highest S/N and airmass (Fig. \ref{fig:airmass_snr}). However, it is important to note that the chromospheric $\log\,R^{\prime}_{\rm HK}$ index is thought to be significantly depressed in stars with transiting giant planets. This is because of the absorption by a circumstellar torus produced by planetary evaporation that is particularly strong in planets with a very low surface gravity, such as \hatssb. Considering the data and the linear models relating the chromospheric index with the inverse of the surface gravity \citep{Lanza_2014, Fossati_2015}, the true value of $\log\,R^{\prime}_{\rm HK}$ of HAT-P-67 could be larger by at least 0.4-0.5, that is, it can approach -4.2.

\begin{figure}
    \centering
    \includegraphics[width=\linewidth]{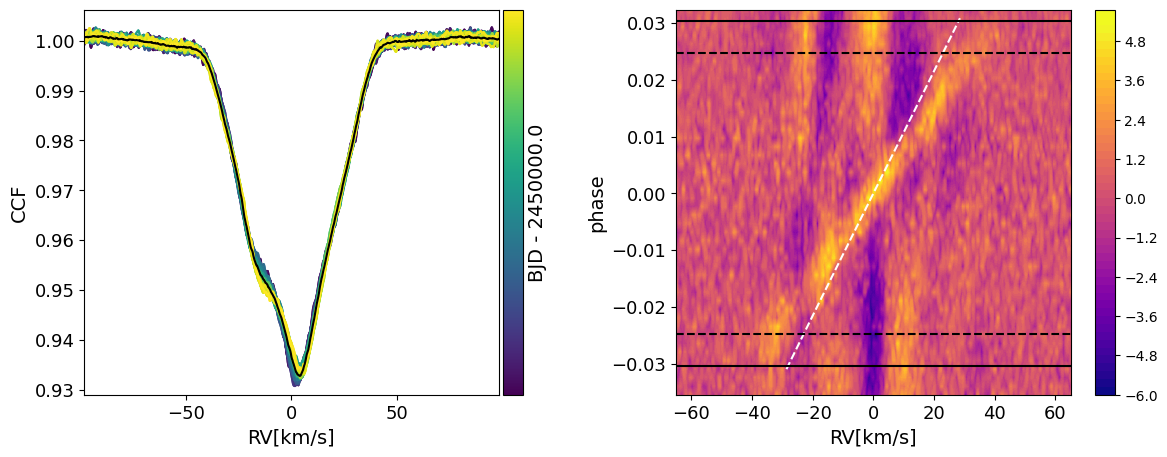}
    \includegraphics[width=\linewidth]{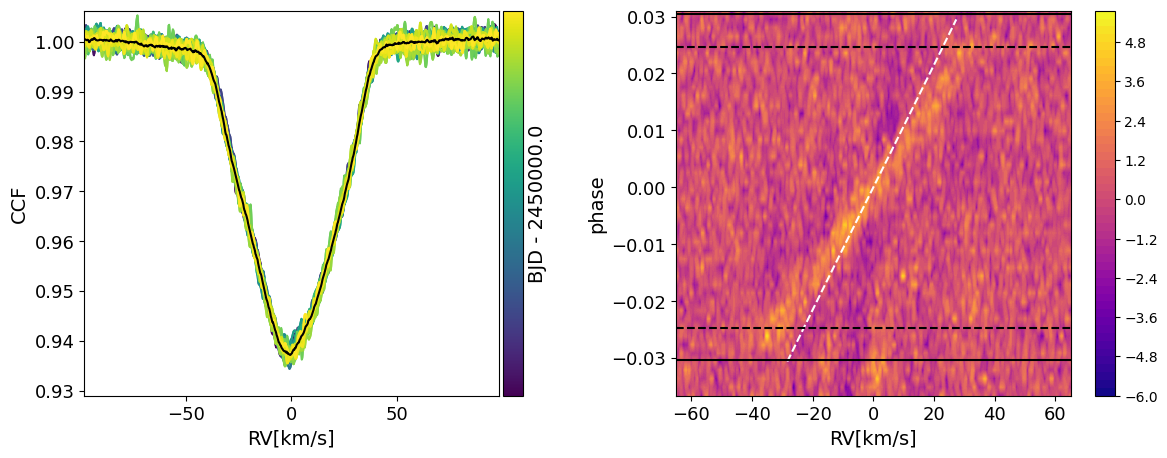}\\
    \includegraphics[width=\linewidth]{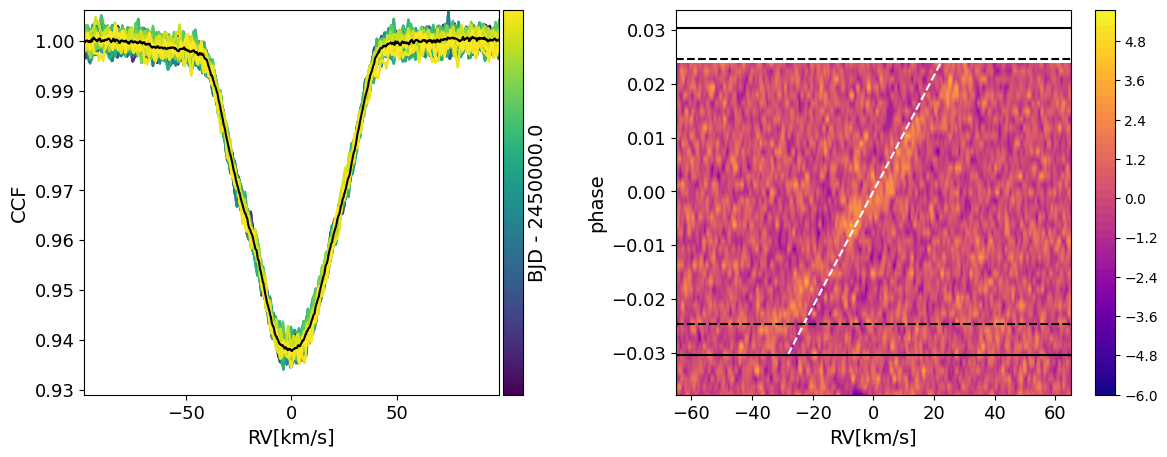}
    \caption{\acp{CCF} for N1 (first row), N3 (second row), N4 (third row) extracted from the DRS using a G2 mask. \textit{Left:} \acp{CCF} linear profiles as a function of time; the black line indicates the average \ac{CCF}. \textit{Right:} two-dimensional maps of the \acp{CCF} as a function of orbital phase and \ac{RV} in the stellar rest frame; the four horizontal black lines show the times of transit contacts. The straight white line shows the expected planetary keplerian velocity \kp\ using a planetary mass value of $\sim$ 0.34 $M\rm_J$ \citep{Zhou_2017}.}
    \label{fig:ccf_profile}
\end{figure}

The occurrence of phenomena associated with stellar activity during N1 is also reflected in the corresponding 
\acp{CCF} linear profile (Fig.~\ref{fig:ccf_profile}, left panels), which is much more distorted and time-varying compared with N3 and N4 (N2 is not considered here since it covers only half transit). 
The distortion of the stellar \ac{CCF} also strongly affects the measured \acp{RV}. Indeed, the systemic velocity obtained fitting the \acp{RV} on N1 presents an offset of $\sim$ 200 m/s compared to N3 and N4 (see Sect. \ref{sec:orbit}). 

Most likely, the origin of this distortion is some kind of stellar variability. To support this hypothesis, for each night we Doppler-shifted the individual \acp{CCF} to the stellar rest frame and computed the average \ac{CCF}. Then we divided each \ac{CCF} by the average one and built the 2-dimensional stack of the residuals, shown in the right panels of Fig.~\ref{fig:ccf_profile}. 
In addition to the evident Doppler shadow due to the planetary transit (the tilted trace), some evident vertical patterns can be seen in N1, even outside the transit. Being at rest in the stellar rest frame, these patterns can not be ascribed to planetary effects. We thus argue that it arises from the evolution of the stellar \ac{CCF} profile during the transit. By applying the \texttt{SpotCCF} tool \citep{SpotCCF}, we could model the deformation of the \ac{CCF} profile: assuming differential stellar rotation, we found compatibility with a dark spot signature, only in the hypothesis that the stellar rotation axis is inclined about 44 degrees from the line of sight. 
Another possible explanation could be the presence of non-radial stellar pulsations (e.g., \citealt{Rieutord_2023}), despite the star having a relatively low temperature ($T_{\rm eff} \sim 6406 $ K). Unfortunately, the time series we have available does not allow us to investigate this further.

\begin{figure}
    \centering
    \includegraphics[width=\linewidth]{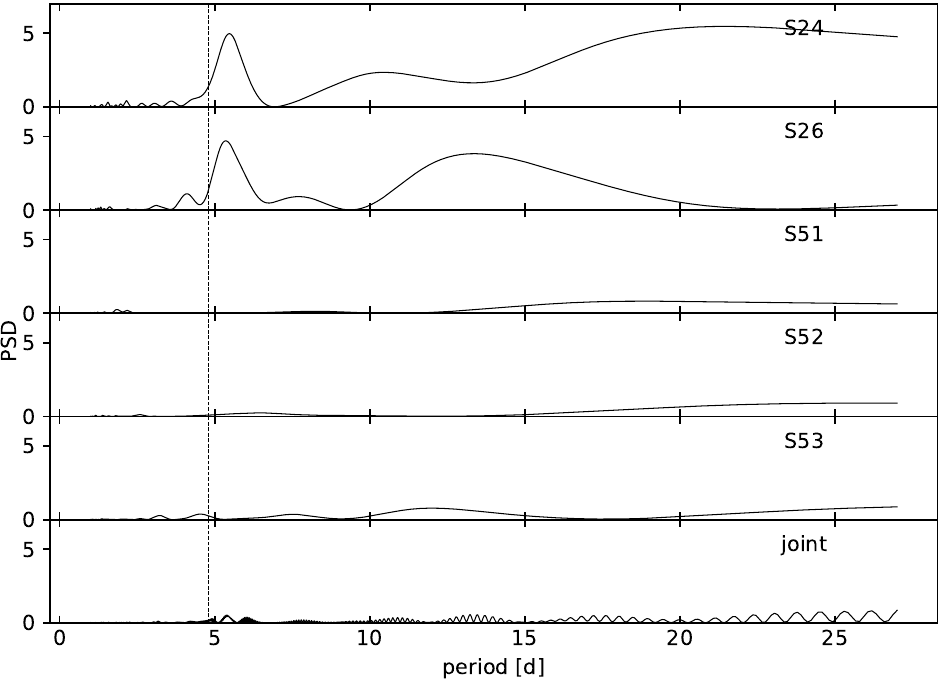}
    \caption{Generalized Lomb-Scargle periodogram of the five \tess\ sectors and the joint \ac{LC} (\textit{bottom box}). The vertical dashed line marks the orbital period of \hatssb\ reported in Table~\ref{tab:orbit}.}
    \label{fig:periodogram}
\end{figure}

We searched for evidence of stellar activity in the \tess \,\acp{LC}. To analyse the photometric variability, we clipped out the in-transit photometry and computed the generalized Lomb-Scargle periodogram \citep{Zechmeister2009, GLS_1981} to detect any periodic signal. We found a clear periodicity at $\sim$5.4 days in S24 and S26 (year 2020) with an amplitude of the order of 1~mmag. In S51, S52 and S53 (year 2022) we did not find any clear indication of periodic signals (Fig.~\ref{fig:periodogram}). As a sanity check, we performed the same analysis using the \tess\ Simple Aperture Photometry (SAP) and obtained the same results.

Using $R\rm_\star$ indicated by \citet{Gully2023} (2.65 $\pm$ 0.12 $R_\odot$), and in the scenario of an equator-on star, 
our estimate of the stellar rotation period leads to an equatorial rotation velocity 
of 25 $\pm$ 7~km$\,$s$^{-1}$.
Due to its large uncertainty, this value is consistent inside 1$\sigma$ with the 
$v \sin{i}$ reported by \citet{Zhou_2017} (30.9 $\pm$ 2 km$\,$s$^{-1}$) and within 2$\sigma$ from our derived $v \sin{i}$. This supports the hypothesis that the periodicity of $\sim$ 5.4 d found in the periodogram (Fig. \ref{fig:periodogram}) is close to the real stellar rotation period. 
If, conversely, we assume an inclination of $\sim$ 44 degrees for the stellar rotation axis, then we derive a rotation period of $\sim$ 3.7 d that does not correspond to any clear peak in the periodogram of the TESS photometry. We thus postulate that the star is most likely seen in an equator-on configuration.
 

\section{Atmospheric characterization}\label{sec:atmosphere}

\subsection{Extraction of the transmission spectra in the VIS}\label{sec:extraction}
The standard data reduction in the VIS range is performed using the HARPS-North dedicated DRS, which produces both 2D and 1D spectra. For each night of observation, we analysed the 2D spectra using the \texttt{SLOPpy} (Spectral Lines Of Planets with python) pipeline \citep{Sicilia}. \texttt{SLOPpy}\footnote{\url{https://github.com/LucaMalavolta/SLOPpy}} is a user-friendly, standard, and reliable tool that is optimized for spectral reduction and the extraction of planetary transmission spectra in the VIS obtained from high-resolution observations. For this purpose, \texttt{SLOPpy} first applies several data reduction steps that are required to correct the input spectra for: \textit{a)} sky emission: the pipeline subtracts the sky spectrum, which is simultaneously retrieved with the science observations, from the stellar spectrum (see subSect. \ref{sec:spectroscopy}); 
\textit{b)} atmospheric dispersion: after dividing each observation with a reference spectrum, the pipeline models this ratio with either a low-order polynomial or a spline and finally divides each observation by this model; \textit{c)} the presence of telluric features: among the different approaches implemented in the pipeline, we decided to apply the one that uses the atmospheric transmission code \texttt{Molecfit}  \citep{Smette_2015, Kausch_2015}.

After applying the aforementioned data reduction steps, each in-transit observation is divided by a master-out spectrum ($M_{\rm out}$, i.e., the integration of the exposures out-of-transit acquired before the ingress and after the planet's egress).
In this way, the pipeline removes the stellar contribution and, in principle, the residuals should contain the exoplanet atmospheric signal.
The $M_{\rm out}$ is built by moving and combining all the out-of-transit spectra to the stellar rest frame. The wavelength shift depends on the barycentric Earth \ac{RV} (BERV) and the \vsys \, of the star. 
While the BERV values are provided by the DRS in the header of the FITS files, we derived \vsys \, (-2.234 $\pm$ 0.027 km$\,$s$^{-1}$) by taking the weighted average of the two values found from the fit of the \acp{RV} of N3 and N4 (we excluded the value found for N1, as it is likely to be contaminated by higher stellar variability, see. Sect. \ref{sec:activity}).

We did not consider the reflex motion of the star since, being a fast rotator, even a wavelength shift of the order of one pixel (corresponding to an RV shift of $\sim 0.8 $ km$\,$s$^{-1}$ in the case of HARPS-N, i.e. the RV variation due to a planet of $\sim 8~M_{\rm J}$) does not change the shape of the spectrum noticeably due to the large broadening of the spectral lines. Even assuming the planet mass upper limit ($M_{\rm p}$ = 0.59 $M_{\rm J}$, \citealt{Zhou_2017}), the maximum stellar RV change is $\sim \pm $ 0.05 km$\,$s$^{-1}$; thus, the out-of-transit spectra can be co-added without taking into account the stellar reflex motion due to the planet ($K_\star$ = 0 km$\,$s$^{-1}$). 

\begin{figure}
    \centering
        \includegraphics[width=\linewidth]{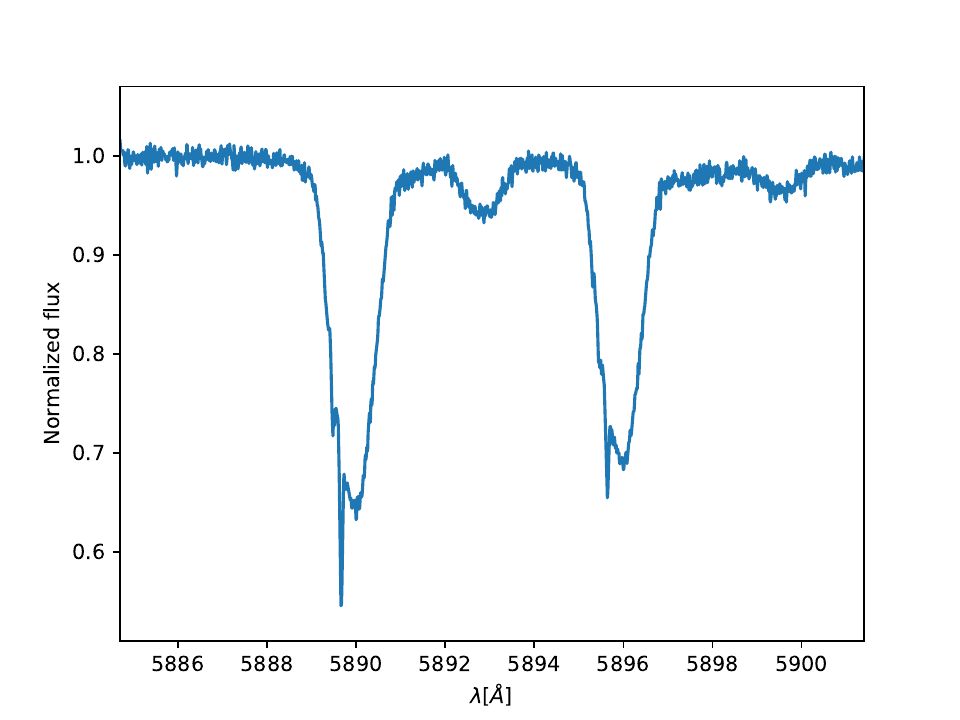}
    \caption{Composite master-out spectrum normalised to unity
around the \ion{Na}{I} doublet lines. Deep and narrow interstellar features peak out from the wider stellar lines.}
    \label{fig:master_out}
\end{figure}

Figure \ref{fig:master_out} shows $M_{\rm out}$ in the region of the sodium doublet, obtained by combining all four nights. 
The presence of interstellar lines is evident, as expected given its large distance from Earth ($\sim$ 320 pc). Any small variation of these lines can mimic a false signature; however, not correcting by the stellar reflex motion, and assuming that interstellar lines remain at the same spectral position and totally stable during the night, they would be automatically removed when dividing each spectrum by $M_{\rm out}$ in the stellar rest frame (e.g., \citealt{CB_2018}).

The two main effects altering the transmission spectra, that are the \ac{CLV} and the \ac{RML} effect, are also taken into account (simultaneously). 
From Spectroscopy Made Easy (SME, \citealt{Piskunov-Valenti}), using a line list from the VALD database \citep{Ryabchikova_2015} and Kurucz ATLAS9 \citep{Kurucz_2005}
 models, we obtained synthetic stellar spectra at different limb angles (ranging from 0 to 1). \texttt{SLOPpy} divides the observed transmission spectrum by a synthetic transmission spectrum computed including the \ac{RML} and \ac{CLV} effects in the stellar models, but without the planetary absorption. This step requires the knowledge of some parameters such as the projected spin-orbit angle ($\lambda$), the differential rotation rate ($\alpha$), and the stellar inclination ($i_\star$). As for the first one, for which \citet{Zhou_2017} measured an upper limit of 12$^{\circ}$, we considered the 
value found through the tomography method (2.2 $\pm$ 0.4 $^\circ$).
For most of the stars, $\alpha$ has not been measured reliably; in this case, the default choice is to exclude the differential rotation in the model and to assume a rigid-body rotation of the star ($\alpha$ = 0). Regarding $i_\star$, on the other hand, 
considering what stated in Sect. \ref{sec:activity}, we expect the star to be close to equator-on ($i_\star \simeq$  90$^\circ$). Anyway, when rigid-body rotation is assumed, the $i_\star$ value is not relevant when modelling the CLV and \ac{RML} effects simultaneously.


\begin{figure*}
    \centering
     \subfloat{%
        \includegraphics[width= 0.5\textwidth]{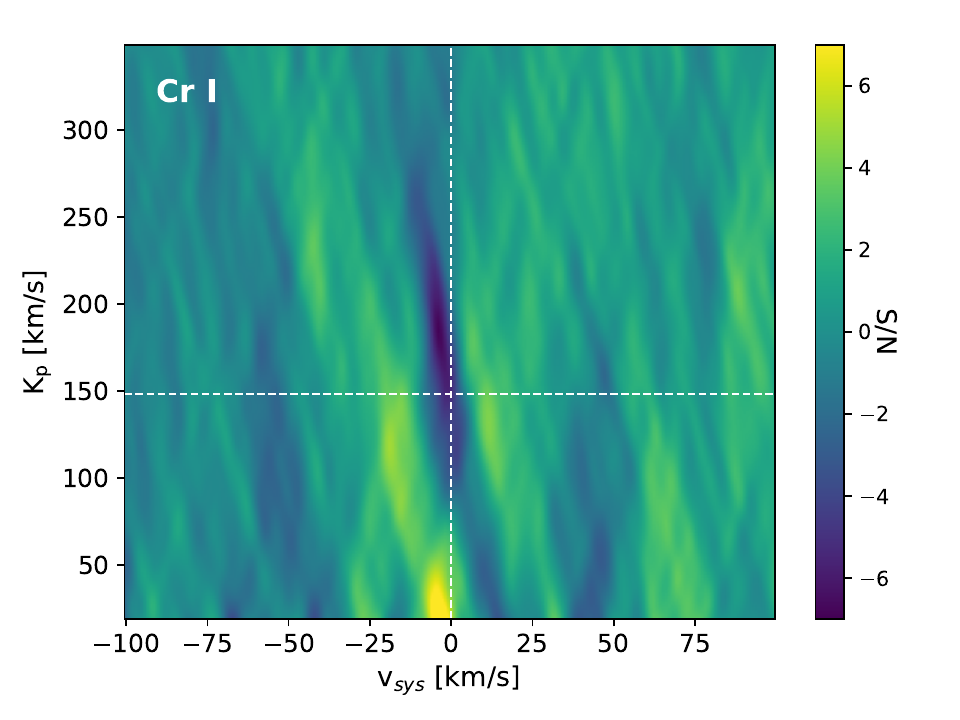}
    }
    \subfloat{%
      \includegraphics[width= 0.5\textwidth]{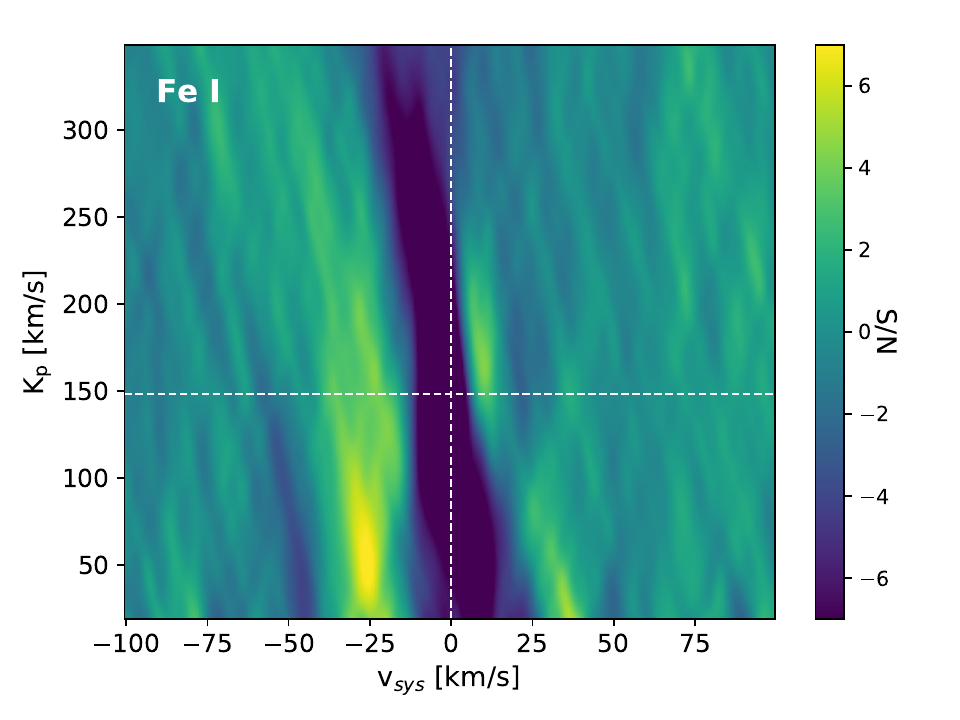}}
\\
    \subfloat{%
    \includegraphics[width= 0.5\linewidth]{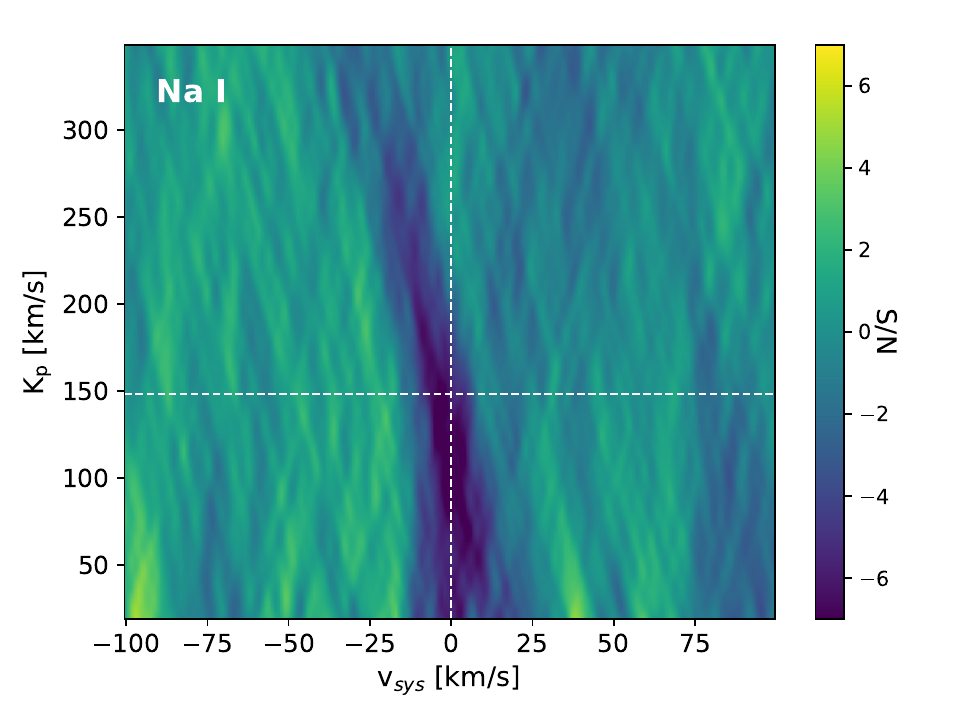}
    }
    \subfloat{%
    \includegraphics[width= 0.5\linewidth]{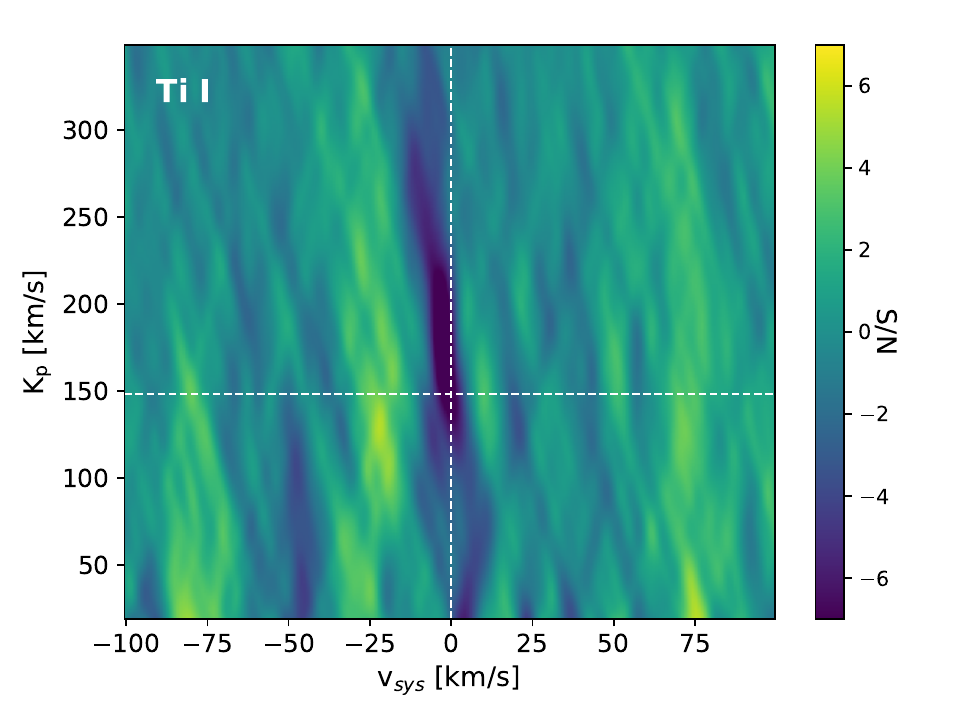}
    }
    \caption{\kp--\vsys\ planes obtained combining all nights for the \ion{Cr}{I} (\textit{top left}), \ion{Fe}{I} (\textit{top right}), \ion{Na}{I} (\textit{bottom left}), and \ion{Ti}{I} (\textit{bottom right}) templates. In each plot, the dashed cross marks the expected \kp\ and \vsys\ of the system.}
    \label{fig:combined}
\end{figure*}

\subsection{Cross-correlation analysis}\label{sec:cross-correlation}
We investigated the atmospheric composition of \hatssb\ using the cross-correlation technique. We used high-resolution model templates at the temperature of 1903 K \citep{Zhou_2017} for the following atoms and molecules: \ion{Ca}{I}, \ion{Cr}{I}, \ion{Fe}{I}, \ion{Fe}{II}, H$_2$O, \ion{K}{I}, \ion{Mg}{I}, \ion{Na}{I}, \ion{Ti}{I}, TiO (using line lists from Exomol and Plez), \ion{V}{I}, VO, \ion{Y}{I}. The synthetic models of studied species were generated using \texttt{petitRADTRANS} \citep{petitRadTrans} assuming equilibrium chemistry, solar abundance from \citet{Asplund_solar_abundance}, and stellar and planetary parameters from Tables \ref{tab:parameters} and \ref{tab:orbit}. Additionally, in order to simulate the continuum opacity produced by H$^-$, we added a cloud layer at $P_0 =1$  mbar. In the final step, the high-resolution synthetic spectra were convolved to the HARPS-N resolution using instrumental (Gaussian kernel) broadening \texttt{instrBroadGaussFast} from \texttt{PyAstronomy}\footnote{\url{https://github.com/sczesla/PyAstronomy}} \citep{PyA}.
 
 
We cross-correlated the planetary transmission spectra extracted in the VIS range with all the templates listed above, and then combined the \acp{CCF} to obtain the \kp--\vsys\ maps. We did not detect any robust signal, except for \ion{Cr}{I}, \ion{Na}{i}, \ion{Fe}{I} and \ion{Ti}{I}. We remark that these species are commonly found in the transmission spectra of \acp{HJ} \citep[see e.g.][]{Hoeijmakers_2018, Ishizuka_2021, Scandariato2023}. The stacked \acp{CCF} show, night by night, an absorption feature stretching between the first and fourth contacts (t$_1$ and t$_4$) and shifting in the velocity space according to the expected planetary keplerian motion (Figs.~\ref{fig:CCF_Cr}-\ref{fig:CCF_Ti}). The corresponding \kp--\vsys\ maps confirm this evidence by showing a clear absorption feature near the expected \kp\ and \vsys\ of the system (Fig. \ref{fig:combined}). Assuming that the median absolute deviation of the \kp--\vsys\ maps is a good estimate of the noise, then, 
the S/N of the absorption features is above 
6, 21, 9 and 11 for \ion{Cr}{i}, \ion{Fe}{i}, \ion{Na}{i} and \ion{Ti}{i} respectively, when combining all nights.

Since the spectral lines of these atomic species are also present in the stellar spectrum, it is unclear whether these detections are due to either stellar residuals in the transmission spectra or planetary absorption lines. To test the origin of the aforementioned detection we checked whether the stellar spectrum has been effectively removed by the reduction pipeline. To do this,
we computed a cross-correlation mask for \hatss\ using \teff = 6500 K, $\log${ g} = 4.0, [Fe/H] = 0 and the theoretical line list provided by the VALD database. Then, we removed all the \ion{Cr}{i}, \ion{Fe}{i}, \ion{Na}{i} and \ion{Ti}{i} spectral lines from the mask, in order to leave only the species that do not lead to any detection. Finally, we cross-correlated the transmission spectra, where in the ideal case there should be no signature of the stellar spectrum, with the modified mask. We found that there is a residual trace in the stacked \acp{CCF} (Fig. \ref{fig:CCF_VALD}) along the expected position of the planetary absorption which closely follows the stellar Doppler shadow (see Fig.~\ref{fig:ccf_profile}). This trace, translated to the \kp--\vsys\ plane, generates an absorption feature close to the expected planetary \kp. The most obvious way to interpret this result is that the stellar spectrum has not been completely removed from the in-transit spectra. To mitigate this problem we tried to re-do the extraction by varying the relevant parameters (\kp, $v\sin{i}$, $\lambda$) within their uncertainties. Unfortunately, the left-over of the stellar spectrum never faded out. 

We thus obtained clear evidence that the removal of the stellar spectrum during the extraction of the transmission spectra is unable to completely remove the stellar signature, casting serious doubts on the reliability of the planetary \ion{Cr}{i}, \ion{Fe}{i}, \ion{Na}{I} and \ion{Ti}{i} detections seen in transmission. This does not exclude the fact that a genuine planetary signal exists. Unfortunately, however, the fact that the planetary trace is expected to follow the Doppler shadow complicates the unequivocal attribution of the signal.

 \begin{figure*}
    \centering
    \includegraphics[width=\textwidth]{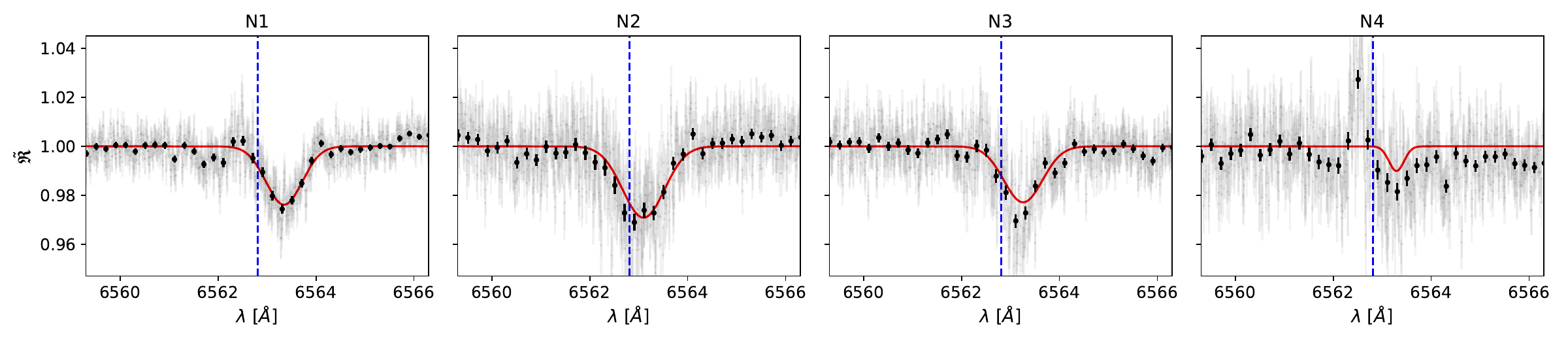}
    \caption{Transmission spectra of \hatssb \, for each night, centered around the H$\rm\alpha$ line in the planetary rest frame (light gray), also binned by 20× (in black circles). The red line is the \ac{MCMC} Gaussian fit performed by \texttt{SLOPpy}, while the vertical blue dashed line indicates the rest frame transition wavelength of the H$\rm\alpha$ line.}\label{fig:ts_Halpha}
\end{figure*}

\subsection{Transmission spectroscopy of the H-alpha line with HARPS-N} \label{sec:halpha}
As already mentioned, recently, the atmosphere of \hatssb \, has been explored by \citet{Bello-Arufe2023}, through the analysis of one full transit retrieved with CARMENES. In addition to the detection of \ion{Na}{I}, which they ascribe to the planetary signal, they also reported a strong absorption near the H-alpha (H$\rm\alpha$), the first spectral line in the Balmer series. We searched for the same signal in our optical transmission spectra ($\tilde{\mathfrak{R}}$) extracted from \texttt{SLOPpy} (see subSect. \ref{sec:extraction}), 
by summing all in-transit observations divided by the $M_{\rm out}$ in the planet's reference system.
The result for each night is shown in Fig. \ref{fig:ts_Halpha}.
The analysis reveals the presence of a strong absorption feature in N1, N2, and N3, while a clear emission feature is visible in N4.

For a measure of the absorption signals, we decided to apply an \ac{MCMC} Gaussian fit in the region of the H$\rm\alpha$ line. A summary of the best-fit parameters obtained for each night is reported in Table \ref{tab:mcmc_Halpha}. Combining all nights, we found an absorption signal with a contrast ($c$) of $\sim$ 2.2\% and a full-width half maximum (FWHM) of $\sim$ 38 km$\,$s$^{-1}$. However, the signal was characterized by a very high red shift ($\sim$ 23 km$\,$s$^{-1}$) with respect to the predicted line position. We point out that we get compatible results even if setting a prior on \kp \, next to the expected value.

Doppler shifts as strong as $\sim$23~km/s have never been claimed in the literature. Moreover, global circulation models of the atmospheres of \acp{HJ} predict the presence of zonal winds flowing from the planetary dayside to the nightside \citep[e.g.,][]{Komacek2016,Parmentier2018,Roman2021}; this would lead to a blue shifted atmospheric signal, in contrast with our finding. We thus postulate that the planetary origin of the H$\rm\alpha$ absorption signal is unlikely. Nonetheless, a more thorough investigation, beyond the scope of this paper, is needed to give an appropriate interpretation of the absorption feature seen during the transits.

Regarding N4, as explained in Appendix \ref{app:Halpha}, we found that the emission feature originates from around mid-transit and seems to be at rest in the stellar reference system.
As shown in Fig. \ref{fig:rhk_index}, N4 presents higher variability in the chromospheric activity index $\log\,R^{\prime}_{\rm HK}$ right at the centre of the transit until the end of the observations. This leads us to state that the observed emission signal is probably due to some stellar activity event. 

\begin{table}
 \caption{\label{tab:mcmc_Halpha} Summary of the best-fit parameters and 1-$\sigma$ error bars obtained with the MCMC fitting procedure for the H$\rm\alpha$ line.}
 \centering
 \begin{spacing}{1.4}
    \begin{tabular}{c|c|c|c}
    \hline
    \hline
    \multirow{2}{*}{Night} & $c$ & FWHM  & $v_{wind}$ \\
    & [\%] & [km s$^{-1}$] & [km s$^{-1}$] \\
    \hline
    N1 & -2.39$_{-0.15}^{+0.15}$& 39.9$_{-4.2}^{+4.8}$ & +24.5$_{-0.7}^{+0.4}$ \\
    N2 & -2.92$_{-0.28}^{+0.26}$ & 46.0$_{-4.0}^{+4.7}$ & +13.1$_{-2.2}^{+2.1}$ \\
    N3 & -2.29$_{-0.25}^{+0.22}$ & 42.5$_{-6.2}^{+6.5}$ & +20.5$_{-2.6}^{+2.6}$\\
    N4 & -1.01$_{-2.23}^{+0.56}$ & 16.5$_{-15.3}^{+10.1}$ & +22.0$_{-9.4}^{+2.3}$\\
    Combined & -2.19$_{-0.10}^{+0.10}$ & 37.7$_{-2.2}^{+2.4}$ & +22.9$_{-1.0}^{+1.0}$ \\
    \hline
    \end{tabular}
\end{spacing}
\end{table} 

\subsection{Transmission spectroscopy of the nIR \Hei triplet with GIANO-B}
\begin{figure}
\centering
 \includegraphics[width=0.7\linewidth]{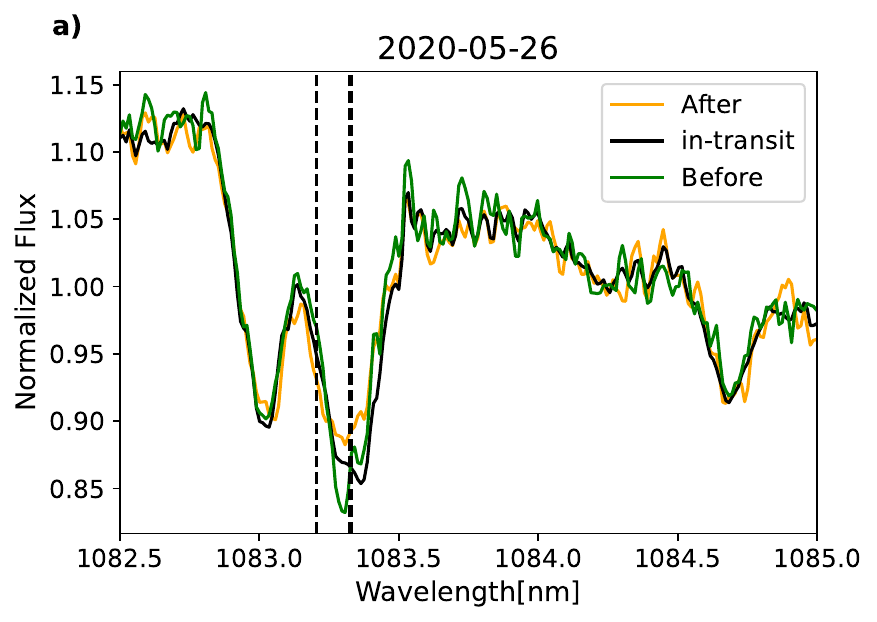}\\ 
 \includegraphics[width=0.7\linewidth]{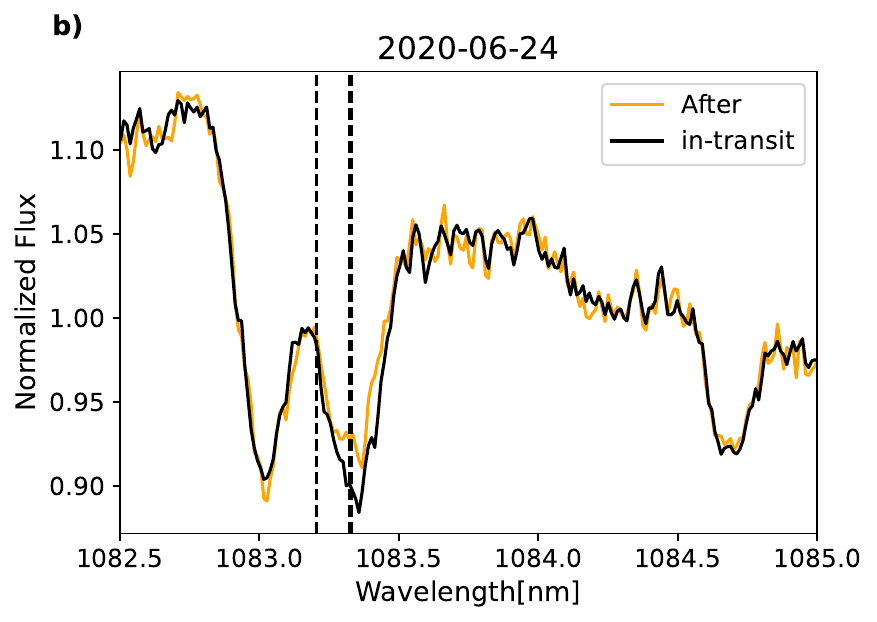}\\
 \includegraphics[width=0.7\linewidth]{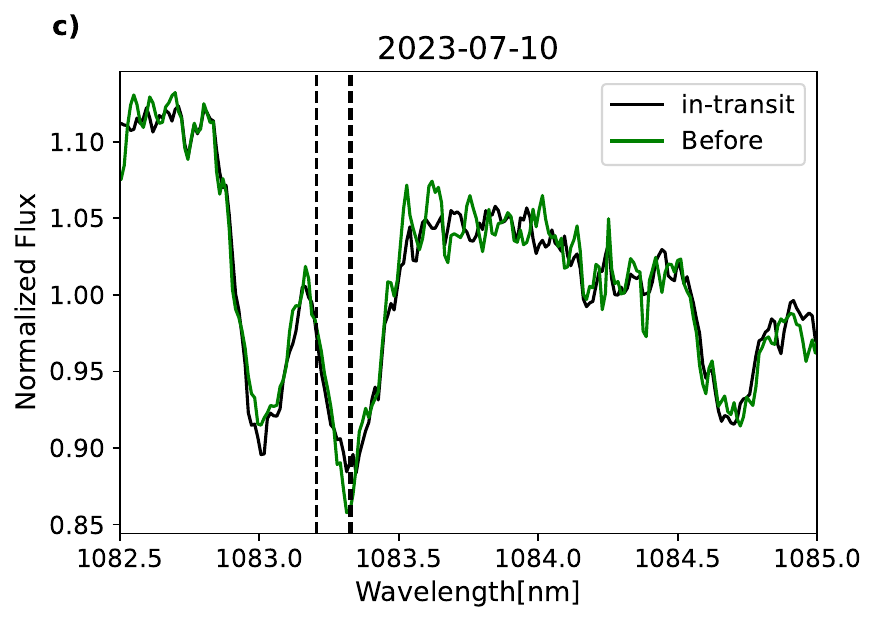}\\
 \includegraphics[width=0.7\linewidth]{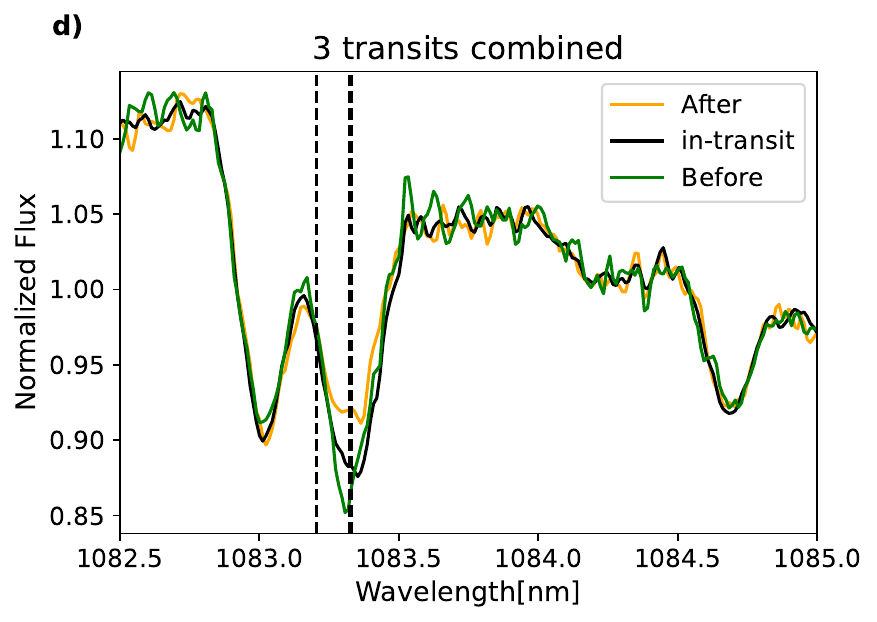}\\
 \includegraphics[width=0.7\linewidth]{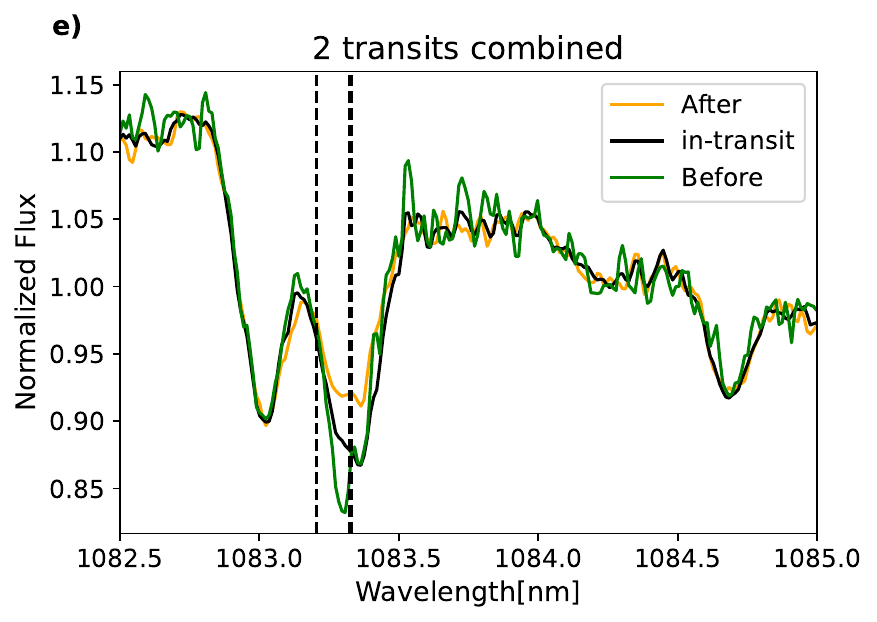}\\
\caption{Master-after, master-before, and master-in spectra in the star frame. a), b), and c) panels represent the GIANO-B nights considered individually. Panel d) shows all three transits together, while in panel e) only the two observations with the highest S/N are considered.  Vertical lines indicate the position of the \ion{He}{I}. In- and before-transit absorption is  visible by eye.}
 \label{real_data}
\end{figure}
Similar to the optical analysis, in order to separate the potential planetary \Hei signal from the stellar contribution, we performed transmission spectroscopy on the GIANO-B spectra using the approach outlined in \citet{Guilluy2023} and \citet{Guilluyinprep}.\\
We employed the GOFIO pipeline \citep{Rainer2018} to extract the spectra from the raw GIANO-B images, and a preliminary wavelength calibration using a U-Ne lamp spectrum as a template in the vacuum wavelength frame. We then refined this initial wavelength solution, by employing the same two-steps approach described in our previous works \citep[e.g.,][]{Guilluy2020, Giacobbe2021}, which consists in aligning all the spectra to the Earth's atmospheric rest frame, assuming it as the frame of the observer (disregarding any $\sim$10 m s$^{-1}$ differences due to winds).
We thus focussed on order \#39, which includes the \Hei triplet. 

We employed 
\texttt{Molecfit} to remove the telluric absorption contamination. Additionally, we masked the emission telluric line at around 1083.43~nm, following the methodology described in \citet{Guilluy2023} and \citet{Guilluyinprep}.

We moved the telluric-corrected spectra to the stellar rest frame using parameters listed in Tables~\ref{tab:orbit} and  \ref{tab:parameters}. 
We normalised each spectrum to the continuum by median division, neglecting the spectral region near the \Hei triplet, and excluding spectra with significantly lower signal-to-noise ratios compared to the other exposures.
Similar to \citet{Gully2023}, we then created a master-before, master-in, master-after spectra, by averaging the before- (i.e., with an orbital phase smaller than $t_1$), after- (greater than $t_4$) and in-transit (between $t_2$ and $t_3$) frames. From the visual comparison of these master spectra, an absorption feature is readily discernible both in the in-transit and before-transit spectra at the position of the \ion{He}{I} triplet. This is particularly evident when the nights are considered together (last two panels of Fig.~\ref{real_data}). This conspicuous excess absorption ($\sim$ 8\%) is visually compatible with what is found in \citet{Gully2023} (see their Fig. 8). 

Since not all the observed nights provided complete transit coverage, as seen in Fig.~\ref{fig:airmass_snr}, we decided to consider all transits collectively. This is particularly crucial for N1, where the only available comparison stellar spectrum was obtained before the transit. However, considering that \citet{Gully2023} identified a helium tail preceding the planet by several hours, we are unable to employ these spectra for comparison due to the potential influence of the planetary tail.
In this section, we present the results we obtained by considering only the two nights with the highest S/N, namely N1 and N3. For completeness, the findings obtained for all the investigated nights are illustrated in the Appendix (see Fig.~\ref{he_result_mask_3} and Table~\ref{tab_result_He_3}).

\begin{figure*}
    \centering
    \includegraphics[width=\textwidth]{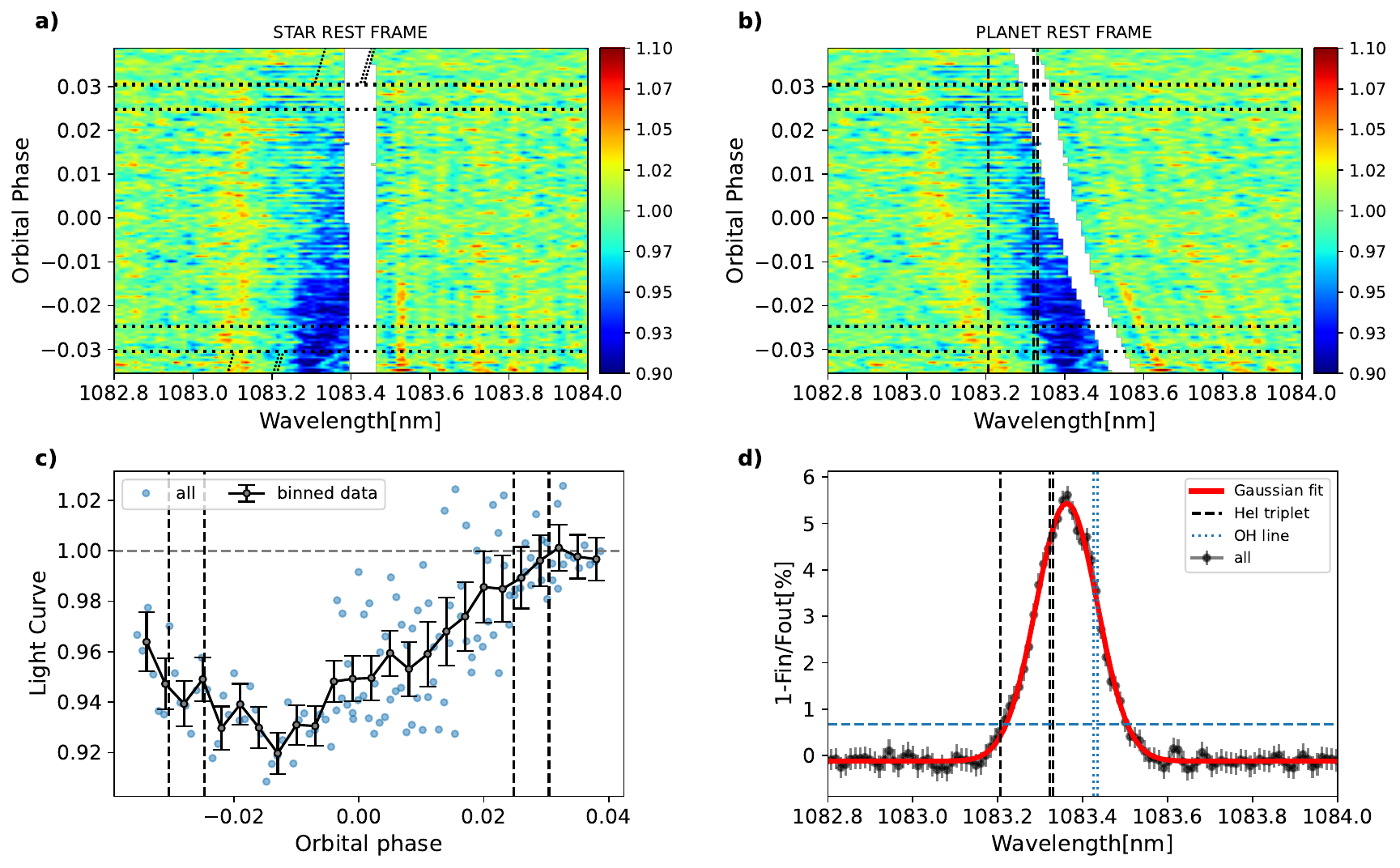}
    \caption{Transmission spectroscopy near the \Hei triplet. a) and b) the 2D transmission spectroscopy maps in the stellar and planet rest frames, respectively. Tilted and vertical lines denote the position of the \Hei triplet, while horizontal dotted lines mark the transit contact points $t_1$, $t_2$, $t_3$, and $t_4$. c) the spectroscopic light curve computed in a 0.075~nm band centered around the peak of absorption. Vertical dashed lines mark the transit contact points. d) the full in-transit averaged transmission spectrum.  }
    \label{he_result_mask}
\end{figure*}

We considered as a comparison spectrum, an average spectrum of all the observations acquired after the transit $S_\mathrm{aft}(\lambda)$. We thus derived the individual transmission spectra, $T\mathrm{(\lambda,i)}$, by dividing each spectrum by $S_\mathrm{aft}(\lambda)$. Finally, we linearly interpolated the transmission spectra in the planet's rest frame. 
The upper panels of Fig.~\ref{he_result_mask} display the 2D transmission spectroscopy map in the star ($a$ panel) and planet ($b$ panel) rest frame, where an absorption signal 
is visible. 
With the goal of monitoring the variation of the \Hei signal during transit, we additionally performed spectrophotometry of the helium triplet within a passband of 0.075~nm centered at the peak of excess absorption in the planet rest frame \citep{Allart2019}. The computed transit light curve is presented in the $c)$ panel of Fig.~\ref{he_result_mask}. We finally obtained the fully in-transit transmission spectrum in the planet's rest frame $T_\mathrm{mean}(\lambda)$ by averaging the transmission spectra with an orbital phase between $t_2$ and $t_3$.

Following \citet{Guilluyinprep}, we estimated the contrast $c$ of the excess absorption at the position of the \Hei triplet, by fitting a Gaussian profile to $T_\mathrm{mean}(\lambda)$ with the DE-MCMC method varying the peak position, the FWHM, the peak value ($c$), and an offset for the continuum. We also accounted for the presence of correlated noise by using \ac{GP} regression within the same DE-MCMC tools, using
a covariance matrix described by a squared exponential kernel \citep{Bonomo2023}. We finally considered possibly uncorrelated noise by introducing a jitter term $\sigma_{\rm j}$. Panel $d$) of Fig.~\ref{he_result_mask} shows the final $T_\mathrm{mean}$ corrected with the GP model and with overplotted the best-fit Gaussian model. We detected a \Hei contrast $c$ of 5.56$^{+ 0.29 }_{ -0.30 }$\% at 19$\sigma$, with a velocity shift of v = 10.0$^{+1.1 }_{-1.2}$~km~s$^{-1}$. The best-fit parameters from the DE-MCMC Gaussian analysis are listed in Table~\ref{tab_result_He}, while the posterior distribution and the application of the GP model are reported in Appendix in Fig.~\ref{corner_He}, and Fig.~\ref{GP}, respectively. 

We then translated $c$ into an effective planetary radius $R_\mathrm{eff}$$\sim$3~$R\rm_p$ \citep[e.g.,][]{Chen_2018}, which extension beyond the planet's Roche Lobe radius (2.7~$R_\mathrm{p}$, \citealt{Gully2023}) indicates that the planet is evaporating. 
We found a restricted Jeans escape parameter $\Lambda$$\sim$18 \citep{Fossati2017}. The value obtained supports the presence of an extended atmosphere with significant mass loss, such as one would expect from seeing metastable helium absorption beyond Roche's lobe.  We finally derived the quantity $\delta_\mathrm{R_p}/H_\mathrm{eq}$ \citep{Nortmann2018}, which represents the number of scale heights probed by the atmosphere in the spectral range under consideration (see Table~\ref{tab_result_He}).


\begin{table*}[h]
\caption{Best fit parameters from the DE-MCMC Gaussian analysis.}
	\centering
	\begin{tabular}{ c c c c c c }
		\hline \hline
		Peak position & $c$ &  $R_\mathrm{eff}$ & FWHM & Significance & $\delta_{R\rm_p}$/H$_{\mathrm{eq}}$  \\
                (nm)      &         (\%) & ($R\rm_p$)             & (nm) & ($\sigma$)   &           \\   
                \hline
               1083.3624$^{+ 0.0041 }_{ -0.0043 }$ & 5.56$^{+ 0.29 }_{ -0.30 }$ & 3.04 $\pm$ 0.07 & 0.1701$^{+ 0.011 }_{ -0.009 }$ & 19.0 & 48.5 $\pm$ 35.8\\
		\hline
        \end{tabular}
	\tablefoot{From left to right: the peak position of the \ion{He}{I}, the absorption (expressed both as contrast $c$ and $R_{\rm eff}$), and FWHM obtained from the DE-MCMC analysis, the significance of the detection, and the ratio between the equivalent height of the \Hei atmosphere and the atmospheric scale height. We determined the values and the 1$\sigma$ uncertainties of the derived parameters from the medians and the 16\%-84\% quantiles of their posterior distributions.}
	\label{tab_result_He}
\end{table*}

We need to be cautious with our \Hei detection, and emphasize that our measured signal could be contaminated by potential pseudo-signals of stellar activity, caused by the planet's passage during transit over a non-uniform stellar disk \citep[e.g.,][]{Salz_2018, Guilluy2020, Guilluyinprep}. The variability observed in the H$\rm\alpha$ line (subSect. \ref{sec:halpha}), likely attributed to stellar activity, is a warning signal. However, in the \Hei analysis, we considered all the nights together, primarily due to the absence of post-transit observations in both N1 and N4, which are crucial if a pre-transit tail exists. For this reason, we were unable to analyse possible \Hei night-to-night variability due to stellar activity.

\section{Summary and conclusions}\label{sec:summary}
\hatssb \, is one of the lowest-density exoplanets known to date ($\rho \sim$  0.05 g cm$^{-3}$), making it a prime target for transmission spectroscopy. In the framework of the GAPS programme, we observed four transits of \hatssb \, using the GIARPS mode of the TNG, thus aiming at the simultaneous VIS and nIR study of the exoplanet atmosphere. 

Firstly, we derived a new orbital solution for the \hatss\ system using archival TESS photometry and found it to be compatible within 2$\sigma$ with \citet{Gully2023}, who analysed the same TESS sectors. We combined our orbital solution with the \ac{RV} times series to analyse the \ac{RML} effect \citep{Rossiter, McLaughlin}. From the fit, we estimated 
the systemic velocity of the system and the projected spin-orbit angle. In order to improve the precision on $\lambda$, we applied the Doppler tomographic analysis, finding a value of 2.2$\pm$0.4$^\circ$ when combining all the four transits.
Our result is compatible with \citet{Zhou_2017}, who measured an upper limit of 12$^{\circ}$ through the 
same technique.
The derived value indicates an aligned planetary orbit and suggests that the planet has likely migrated to its current orbit via tidal interactions with a protoplanetary gas disc \citep{Lin_1996, Ward_1997}. It is unlikely that the outer stellar companion is directly involved in the migration of the planet through e.g. Kozai mechanism, because of its very large separation (3400 au); nevertheless, some combinations of the original planetary orbit (e.g., $\ge$ 1 au) and binary orbit may allow a phase of high eccentricity and very close periastron followed by tidal circularization on the current orbit within the age of the system. The tidal timescale for the damping of the obliquity, computed for a planetary mass of 0.32 $M\rm_J$ \citep{Gully2023}, is of about 0.6 Gyr, while for the mass upper limit of 0.59 $M\rm_J$, it is of 0.3 Gyr. We have adopted a strong tidal interaction assuming a modified tidal quality factor of the star ($Q^{\prime}_{\star}$) of 2.5 $\times$ 10$^4$ as in the case of Kepler-1658, another system with a similar subgiant star \citep{Vissapragada_2022}. Therefore, an alternative interpretation could be that the planet mass is close to 0.59 $M\rm_J$, and the low obliquity has been reached only relatively recently, when the star was in the final phase of its main-sequence evolution, and experienced a strong tidal interaction with \hatssb. 
This is in agreement also with a stellar rotation period close to the orbital period because tides would act both to damp the obliquity and to synchronize the rotation of the star. 

Our first night of observation shows signatures of higher stellar 
variability than on the other nights analysed. As a matter of fact, compared with the remaining part of our survey, we found an \ac{RV} offset of $\sim$ 200 m s$^{-1}$, a higher $\log\,R^{\prime}_{\rm HK}$ activity index and a distorted \ac{CCF} profile. 
Furthermore, although the TESS photometry does not cover our first night of observation, we nevertheless extracted a clear periodicity consistent with stellar rotation from the available TESS \acp{LC}, spanning a period fairly close to that observed. By modelling the \ac{CCF} profiles, we found that the stellar variability observed does not appear to be due to stellar spots, which would have generated a different periodicity than what was instead extracted from TESS photometry. It is more likely that the source of this variability is related to the presence of non-radial pulsations on the stellar surface.

Using the cross-correlation technique, we searched for the presence of atomic or molecular species in the optical transmission spectrum of \hatssb. Despite the expected high absorption signal, we did not get any robust detection, with the exception of the \ion{Cr}{I}, \ion{Fe}{I}, \ion{Na}{I} and \ion{Ti}{I}, finding 
a formal statistical significance of
$\sim 6\sigma$, 21$\sigma$, 9$\sigma$ and 11$\sigma$ respectively in the combined signals. However, when we tested the robustness of the detections (with that of sodium compatible with what is reported by \citealt{Bello-Arufe2023}), we could not confirm their planetary origin. By carrying out a series of tests, we realised that we were not able to completely remove the stellar contribution from the in-transit spectra, leading to an unreliable transmission signal. After checking that the same analysis technique works well on other targets with clear detections (e.g. KELT-20 b), our assumption is that the models used for the data reduction, especially the removal of the contribution of the \ac{RML} and \ac{CLV} effects, poorly match the real spectrum of \hatss. The use of 3D models could improve the characterization of the exoplanetary atmosphere. Evidence for this was recently reported by \citet{Canocchi_2023}, who applied 3D non-LTE synthetic spectra to estimate the stellar RM+CLV effects in transmission spectra of solar-like planet hosts, showing that 1D models seem to overestimate the CLV signature. Even earlier, \citet{Chiavassa-Brogi} showed that removing the stellar spectrum using 3D radiative hydrodynamical simulations leads to a significant improvement in planet detectability, both in solar-type and K-dwarf stars.

Due to the strong irradiation from the host star combined with internal heating within the planet, and its very low escape velocity ($v_{\rm esc} \sim$ 25 km$\,$s$^{-1}$, \citealt{Zhou_2017}), \hatssb's atmosphere has been undergoing an intense radius inflation; this should have led to the formation of an extended hydrogen atmosphere. Gas giants with masses below Jupiter, and temperatures above 1800 K, like \hatssb \, ($T_{\rm eq} \sim$ 1900 K), are so inflated and puffed out that they are all on unstable evolutionary paths which eventually lead to Roche-Lobe overflow and the evaporation and loss of the planet's atmosphere \citep{Batygin_2011}.
In the VIS wavelength range, the H$\rm\alpha$ line is a powerful probe of the escaping atmosphere \citep{Yan-Henning, Borsa_2021, Czesla_2022}. On \hatssb, our results for the first three nights analysed are compatible with what found by \citet{Bello-Arufe2023}, that is a highly red shifted ($\sim$23 km/s) strong absorption signal ($\sim$ 2.5\%); however, this finding is unexpected by the current global circulation models of \acp{HJ}. We thus argue that such an absorption signal can be hardly ascribed to planetary origin.
Only for the last night of observations, we find a clear emission 
signal in the transmission spectrum; the correlation with the chromospheric index $\log\,R^{\prime}_{\rm HK}$ leads us to think that the origin of the emission feature is most likely stellar activity.

\citet{Gully2023} found no detectable variability in other hydrogen lines, but confirmed the presence of a large and variable \ion{He}{I} tail preceding the planet. Thanks to the GIARPS observing mode of the TNG, we were able to extract the transmission spectrum in the region of the nIR \ion{He}{I} triplet.
Using the average of all the spectra acquired after the transit as the master-out spectrum, 
we found a clear absorption signal. 
Applying a Gaussian fit to the full in-transit mean transmission spectrum, we estimated a contrast of the excess absorption of 5.56$^{+ 0.29 }_{ -0.30 }$\% (19.0$\sigma$). This value corresponds to an effective planetary radius of $\sim$ 3 $R\rm_p$ ($\sim$ 6 $R\rm_J$), indicating that the planet's atmosphere is evaporating. Assuming a scale height ($H$) of $\sim$ 3500 km, and considering 1$\sigma$ uncertainties, we found that the \ion{He}{I} atmosphere probes a number of scale heights varying between $\sim$ 13 and $\sim$ 84 $H$.
 
In conclusion, 
despite \hatssb \, is one of the most favorable target for transmission spectroscopy studies, there are factors, such as stellar variability and observational constraints (i.e., the long transit duration), 
that make analysis challenging and, therefore, less accurate. Furthermore, the fact that the orbital trace of the planet is almost identical to that of the Doppler shadow complicates the unequivocal attribution of the signal found to either planet or star.

\begin{acknowledgements}
    The authors acknowledge financial contribution from PRIN INAF 2019 and from the European Union - Next Generation EU RRF M4C2 1.1 PRIN MUR 2022 project 2022CERJ49 (ESPLORA). We thank the anonymous referee for their thoughtful comments which helped to improve the quality of this work.
    This paper includes data collected by the TESS mission, which are publicly available from the Mikulsky Archive for Space Telescopes (MAST). Funding for the TESS mission is provided by the NASA's Science Mission directorate.
    D.S. thanks L. Pino for discussions and insights. L.\,M. acknowledges support from the MIUR-PRIN project No. 2022J4H55R.
\end{acknowledgements}

\bibliographystyle{aa}
\bibliography{sample} 

\begin{appendix}
\section{Additional Figures and Tables} \label{add}

\begin{figure*}
    \centering
     \subfloat{%
        \includegraphics[width= 0.5\textwidth]{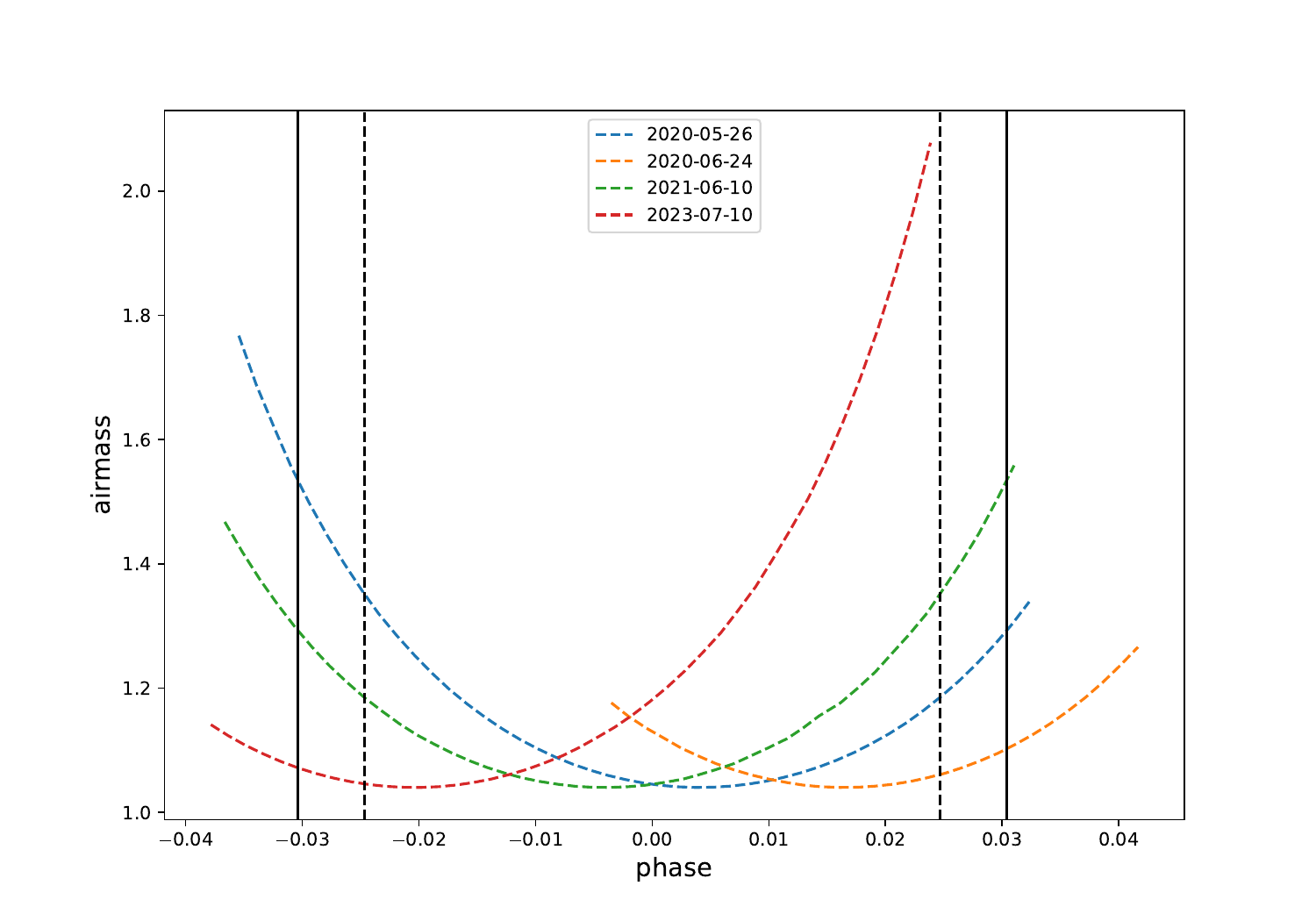}
    }
    \subfloat{%
      \includegraphics[width= 0.5\textwidth]{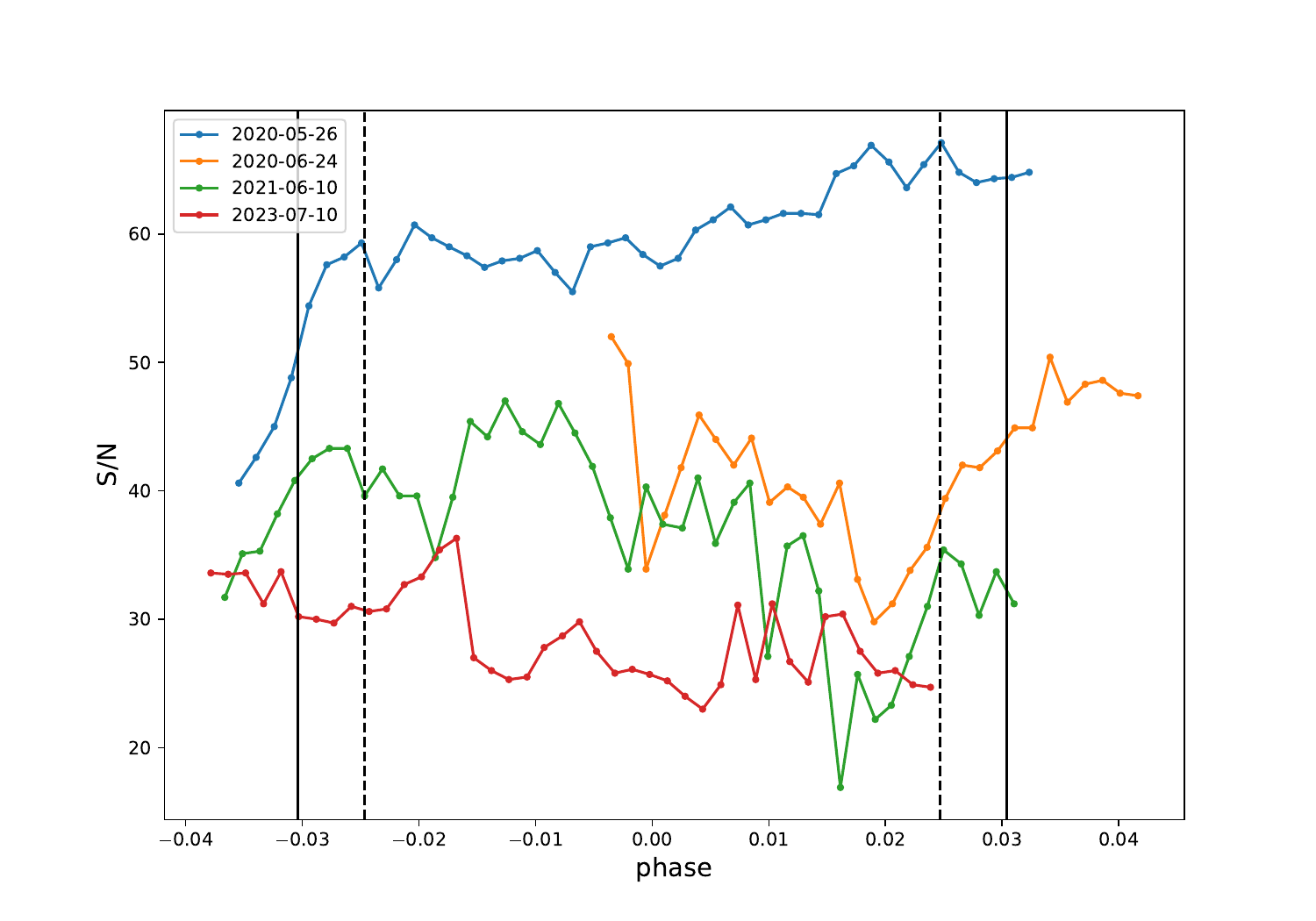}}
    \caption{Variation of airmass (\textit{left}) and signal-to-noise ratio (S/N) (\textit{right}) for each night of \hatssb. The S/N is extracted from the FITS header of the HARPS-N spectra on the 53rd order that contains the sodium feature. The continuous and the dashed vertical black lines represent the four points of contact of the transit. }
    \label{fig:airmass_snr}
\end{figure*}

\begin{table*}
    \centering
\caption{Time series of \hatssb \, from HARPS-N data of night 2020-05-26: BJD, \acp{RV}, and the $\log\,R^{\prime}_{\rm HK}$ values with their related uncertainties.}\label{tab:rvs-rhk}
    \centering
    \begin{tabular}{c c c c c}
    \hline 
    \hline
        BJD & \ac{RV} [km$\,$s$^{-1}$] & $\sigma\rm_{RV}$ [km$\,$s$^{-1}$]& $\log\,R^{\prime}_{\rm HK}$ &$\sigma\rm_{\log\,R^{\prime}_{HK}}$\\
        \hline
        
2458996.390744  & -2.186  & 0.025& -4.6316 & 0.0163	\\
2458996.397862  & -2.084  & 0.025& -4.6250 & 0.0145 \\	
2458996.405362  & -2.029  & 0.024& -4.6335 & 0.0135	\\
2458996.412387  & -2.013  & 0.023& -4.6287 & 0.0116	\\
2458996.419644  & -2.007  & 0.022& -4.6314 & 0.0096	\\
2458996.426994  & -2.007  & 0.022& -4.6295 & 0.0087	\\
2458996.434193  & -1.942  & 0.021& -4.6316 & 0.0085	\\ 
2458996.441369  & -1.921  & 0.022& -4.6435 & 0.0083	\\
2458996.448417  & -1.891  & 0.022& -4.6426 & 0.0092	\\
2458996.455802  & -1.826  & 0.021& -4.6328 & 0.0084	\\
2458996.463186  & -1.877  & 0.021 & -4.6411 & 0.0078\\	
2458996.470350  & -1.767  & 0.021 & -4.6535	& 0.0081\\	
2458996.477480  & -1.832  & 0.021 & -4.6520	& 0.0082\\
2458996.484806  & -1.824  & 0.021 & -4.6533	& 0.0082\\
2458996.492086  & -1.833  & 0.021 & -4.6442 & 0.0085\\
2458996.499309  & -1.815  & 0.022 & -4.6626	& 0.0087\\
2458996.506531  & -1.765  & 0.021 & -4.6540	& 0.0083\\
2458996.513823  & -1.820  & 0.021 & -4.6493 & 0.0081\\
2458996.521149  & -1.817  & 0.021 & -4.6398	& 0.0082\\	
2458996.528348  & -1.886  & 0.021 & -4.6638	& 0.0091\\
2458996.535721  & -1.906  & 0.020 & -4.6635 &0.0080	\\
2458996.542908  & -1.875  & 0.021 & -4.6613	& 0.0077\\
2458996.550211  & -1.945  & 0.020 & -4.6642	& 0.0076\\
2458996.557329  & -1.992  & 0.021 & -4.6528	& 0.0075\\	
2458996.564471  & -1.991  & 0.021 & -4.6622	& 0.0078\\
2458996.571878  & -2.074  & 0.020 & -4.6629 & 0.0076\\
2458996.579100  & -2.108  & 0.020 & -4.6555 & 0.0070\\	
2458996.586334  & -2.097  & 0.019 & -4.6500	& 0.0067\\	
2458996.593522  & -2.100  & 0.020 & -4.6579	& 0.0066\\
2458996.608047  & -2.136  & 0.020 & -4.6531	& 0.0069\\
2458996.600767  & -2.181  & 0.019 & -4.6597 & 0.0069\\ 
2458996.615258  & -2.209  & 0.019 & -4.6546	& 0.0068\\
2458996.622549  & -2.161  & 0.020 & -4.6763	& 0.0072\\
2458996.629853  & -2.110  & 0.020 & -4.6540	& 0.0070\\
2458996.637063  & -2.098  & 0.019 & -4.6596	& 0.0065\\
2458996.644309  & -2.131  & 0.020 & -4.6583	& 0.0065\\ 
2458996.651461  & -2.050  & 0.019 & -4.6569 & 0.0062\\
2458996.658799  & -2.127  & 0.020 & -4.6521	& 0.0065\\	
2458996.666079  & -2.051  & 0.020 & -4.6468	& 0.0069\\
2458996.673186  & -2.061  & 0.020 & -4.6665	& 0.0069\\
2458996.680397  & -1.963  & 0.019 & -4.6463	& 0.0064\\
2458996.687700   &-2.002  & 0.020 & -4.6380	& 0.0067\\
2458996.694899   &-1.964  & 0.020 & -4.6517	& 0.0067\\	
2458996.702179  & -1.915  & 0.020 & -4.6405	& 0.0066\\	
2458996.709447  & -1.946  & 0.020 & -4.6377	& 0.0066\\
2458996.716647  & -1.945  & 0.020 & -4.6367	& 0.0067\\
 \hline
    \hline
    \end{tabular}
\end{table*}

\begin{table*}
    \centering
\caption{As Table \ref{tab:rvs-rhk} but for night 2020-06-24.}\label{tab:rvs-rhk2}
    \centering
    \begin{tabular}{c c c c c}
    \hline 
    \hline
        BJD & \ac{RV} [km$\,$s$^{-1}$] & $\sigma\rm_{RV}$ [km$\,$s$^{-1}$]& $\log\,R^{\prime}_{\rm HK}$ &$\sigma\rm_{\log\,R^{\prime}_{HK}}$\\
        \hline
2459025.404989  & -1.841  & 0.024 & -4.6748	& 0.0104\\
2459025.411922  & -1.870  & 0.023 & -4.6773	& 0.0113\\
2459025.419317  & -2.016  & 0.028 & -4.6539	& 0.0218\\
2459025.426991  & -1.984  & 0.026 & -4.6818	& 0.0186\\
2459025.433773  & -1.912  & 0.025 & -4.7076	& 0.0165\\	
2459025.441180  & -2.022  & 0.024 & -4.6857 & 0.0132\\
2459025.448078  & -2.024  & 0.025 & -4.6857	& 0.0144\\
2459025.455462  & -2.111  & 0.026 & -4.6775	& 0.0155\\
2459025.462800  & -2.164  & 0.024 & -4.6877	& 0.0143\\
2459025.470254  & -2.147  & 0.026 & -4.7123	& 0.0183\\
2459025.477661  & -2.151  & 0.025 & -4.7198	& 0.0175\\
2459025.484166  & -2.096  & 0.026 & -4.6689 & 0.0162\\
2459025.491237  & -2.076  & 0.027 & -4.6746	& 0.0181\\
2459025.499096  & -2.108  & 0.025 & -4.6773	& 0.0154\\
2459025.506515  & -2.124  & 0.027 & -4.6794	& 0.0225\\
2459025.513343  & -2.113  & 0.028 & -4.6337	& 0.0248\\
2459025.520901  & -2.000  & 0.028 & -4.6791	& 0.0254\\
2459025.528193  & -2.051  & 0.027 & -4.6639	& 0.0224\\
2459025.535206  & -2.098  & 0.027 & -4.6646	& 0.0205\\	
2459025.542695  & -2.082  & 0.025 & -4.6747	& 0.0171\\	
2459025.549766  & -2.040  & 0.025 & -4.6936	& 0.0160\\	
2459025.556896  & -1.976  & 0.026 & -4.6585	& 0.0148\\	
2459025.564245  & -1.907  & 0.024 & -4.6911	& 0.0150\\
2459025.571340  & -2.028  & 0.024 & -4.6730	& 0.0134\\
2459025.578620  & -2.009  & 0.024 & -4.6702	& 0.0133\\	
2459025.585888  & -2.013  & 0.024 & -4.6693	& 0.0114\\	
2459025.593053  & -1.941  & 0.024 & -4.6550 & 0.0128\\	
2459025.600367  & -1.938  & 0.024 & -4.6817	& 0.0129\\	
2459025.607566  & -2.089  & 0.024 & -4.6691	& 0.0124\\
2459025.614777  & -1.980  & 0.024 & -4.6619	& 0.0128\\	
2459025.622172  & -2.006  & 0.023 & -4.6351	& 0.0122\\
 \hline
    \hline
    \end{tabular}
\end{table*}

\begin{table*}
    \centering
\caption{As Table \ref{tab:rvs-rhk} but for night 2021-06-10.}\label{tab:rvs-rhk3}
    \centering
    \begin{tabular}{c c c c c}
    \hline 
    \hline
        BJD & \ac{RV} [km$\,$s$^{-1}$] & $\sigma\rm_{RV}$ [km$\,$s$^{-1}$]& $\log\,R^{\prime}_{\rm HK}$ &$\sigma\rm_{\log\,R^{\prime}_{HK}}$\\
        \hline
2459376.383654  & -2.102  & 0.028 & -4.7104	& 0.0341\\
2459376.390923  & -2.298  & 0.026 & -4.6950	& 0.0273\\
2459376.398122  & -2.299  & 0.026 & -4.7342	& 0.0289\\	
2459376.405355  & -2.214  & 0.026 & -4.7300	& 0.0248\\
2459376.412520  & -2.205  & 0.025 & -4.7450 & 0.0223\\	
2459376.419672  & -2.287  & 0.024 & -4.7614	& 0.0211\\
2459376.426790  & -2.229  & 0.025 & -4.7147	& 0.0183\\	
2459376.434163  & -2.206  & 0.026 & -4.7410	& 0.0190\\
2459376.441316  & -2.122  & 0.025 & -4.7518	& 0.0233\\
2459376.448688  & -2.098  & 0.025 & -4.7480	& 0.0211\\
2459376.455702  & -2.076  & 0.026 & -4.7273	& 0.0219\\
2459376.462855  & -2.104  & 0.027 & -4.7359	& 0.0220\\
2459376.470447  & -2.083  & 0.027 & -4.7571	& 0.0293\\	
2459376.477681  & -1.919  & 0.025 & -4.7348	& 0.0223\\
2459376.484880  & -2.099  & 0.025 & -4.7361	& 0.0173\\
2459376.491987  & -2.094  & 0.026 & -4.7607	& 0.0193\\ 
2459376.499243  & -2.040  & 0.025 & -4.7569 & 0.0169\\	
2459376.506338  & -2.043  & 0.024& -4.7336	& 0.0177\\	
2459376.513757  & -2.092  & 0.025 & -4.7343	& 0.0184\\
2459376.521234  & -2.099  & 0.024 & -4.7551	& 0.0170\\	
2459376.528039  & -2.117  & 0.025 & -4.7559	& 0.0187\\
2459376.535238  & -2.125  & 0.025 & -4.7518	& 0.0202\\	
2459376.542623  & -2.246  & 0.025 & -4.7427	& 0.0232\\
2459376.549972  & -2.226  & 0.026 & -4.7933	& 0.0317\\
2459376.557368  & -2.351  & 0.025 & -4.7509	& 0.0204\\
2459376.564231  & -2.277  & 0.025 & -4.7341	& 0.0232\\
2459376.572310  & -2.319  & 0.026 & -4.7358	& 0.0238\\	
2459376.578757  & -2.357  & 0.026 & -4.7459	& 0.0203\\	
2459376.585956  & -2.351  & 0.026 & -4.7971	& 0.0286\\
2459376.593629  & -2.272  & 0.026 & -4.7658	& 0.0234\\	
2459376.600134  & -2.499  & 0.025 & -4.7594	& 0.0214\\
2459376.607564  & -2.362  & 0.029 & -4.8090	& 0.0481\\	
2459376.615527  & -2.390  & 0.027 & -4.7636	& 0.0273\\
2459376.622101  & -2.368  & 0.027 & -4.7728	& 0.0272\\
2459376.628571  & -2.420  & 0.030 & -4.7577	& 0.0326\\
2459376.637494  & -2.413  & 0.030 & -4.7790 & 0.0994\\
2459376.644589  & -2.295  & 0.028 & -4.6970	& 0.0425\\	
2459376.651846  & -2.407  & 0.029 & -4.7152 & 0.0575\\
2459376.658490  & -2.311  & 0.031 & -4.6845	& 0.0501\\
2459376.665862  & -2.383  & 0.027 & -4.6829	& 0.0395\\
2459376.673374  & -2.387  & 0.028 & -4.7096	& 0.0338\\	
2459376.680006  & -2.301  & 0.025 & -4.7188	& 0.0275\\	
2459376.687297  & -2.344  & 0.027 &	-4.7130	& 0.0289\\ 
2459376.694658  & -2.190  & 0.028 & -4.7304	& 0.0379\\	
2459376.701765  & -2.248  & 0.027 & -4.7309	& 0.0317\\	
2459376.709103  & -2.267  & 0.026 & -4.7395	& 0.0376\\
\hline
    \hline
    \end{tabular}
\end{table*}

\begin{table*}
    \centering
\caption{As Table \ref{tab:rvs-rhk} but for night 2023-07-10.}\label{tab:rvs-rhk4}
    \centering
    \begin{tabular}{c c c c c}
    \hline 
    \hline
        BJD & \ac{RV} [km$\,$s$^{-1}$] & $\sigma\rm_{RV}$ [km$\,$s$^{-1}$]& $\log\,R^{\prime}_{\rm HK}$ &$\sigma\rm_{\log\,R^{\prime}_{HK}}$\\
        \hline
2460136.375287  & -2.343 &  0.027 & -4.6775	& 0.0341\\	
2460136.382428  & -2.279 &  0.028 & -4.7147	& 0.0375\\
2460136.389615  & -2.209 &  0.027 & -4.7254	& 0.0386\\
2460136.397011  & -2.246 &  0.027 & -4.7191	& 0.0432\\ 
2460136.404233  & -2.219 &  0.027 & -4.7187	& 0.0374\\
2460136.411455  & -2.199 &  0.027 & -4.6715	& 0.0411\\	
2460136.418769  & -2.185 &  0.029 & -4.6722 & 0.0418\\	
2460136.426026  & -2.234 &  0.027 & -4.6940	& 0.0453\\
2460136.433213  & -2.186 &  0.027 & -4.7258	& 0.0451\\	
2460136.440540  & -2.055  & 0.028 & -4.6986	& 0.0427\\
2460136.447715  & -2.097  & 0.028 & -4.7120	& 0.0428\\
2460136.455042  & -2.190  & 0.027 & -4.7177	& 0.0390\\
2460136.462148  & -2.020  & 0.028 & -4.7273 & 0.0384\\	
2460136.469567  & -2.138  & 0.026 & -4.7484	& 0.0358\\
2460136.476592  & -2.108  & 0.025 & -4.6923	& 0.0291\\
2460136.483687  & -2.116  & 0.028 & -4.7060	& 0.0520\\	
2460136.490990  & -1.927  & 0.029 & -4.7164	& 0.0562\\	
2460136.498177  & -2.025  & 0.030 & -4.6803	& 0.0550\\
2460136.505642  & -1.997  & 0.027 & -4.7121	& 0.0583\\	
2460136.512725  & -2.068  & 0.027 & -4.7339	& 0.0528\\	
2460136.520260  & -2.123  & 0.028 & -4.7696 & 0.0544\\	
2460136.527297  & -2.206  & 0.029 & -4.7307	& 0.0465\\	
2460136.534299  & -2.211  & 0.028 & -4.6749	& 0.0475\\
2460136.541868  & -2.330  & 0.030 & -4.6974	& 0.0579\\
2460136.549044  & -2.331  & 0.029 & -4.6806	& 0.0547\\	
2460136.556196  & -2.212  & 0.028 & -4.6933	& 0.0587\\	
2460136.563511  & -2.346  & 0.029 & -4.7877	& 0.0748\\
2460136.570814  & -2.318  & 0.029 & -4.6262	& 0.0564\\	
2460136.578059  & -2.434  & 0.030 & -4.6700	& 0.0663\\	
2460136.585571  & -2.445  & 0.029 & -4.7495	& 0.0699\\
2460136.592561  & -2.431  & 0.027 & -4.7155	& 0.0447\\
2460136.599934  & -2.457  & 0.029 & -4.6655	& 0.0575\\	
2460136.606751  & -2.408  & 0.028 & -4.7594 & 0.0497\\
2460136.613996  & -2.515  & 0.030 & -4.6555	& 0.0514\\	
2460136.621611  & -2.550  & 0.029 & -4.7017	& 0.0637\\	
2460136.628706  & -2.472  & 0.027 & -4.6684	& 0.0437\\
2460136.635824  & -2.397  & 0.029 & -4.7512	& 0.0515\\
2460136.643023  & -2.393  & 0.029 & -4.7921	& 0.0675\\	
2460136.650291  & -2.374  & 0.029 & -4.6927	& 0.0629\\	
2460136.657467  & -2.575  & 0.028 & -4.6626	& 0.0586\\	
2460136.664735  & -2.415  & 0.029 & -4.6351	& 0.0604\\	
2460136.672027  & -2.221  & 0.029 & -4.6254	& 0.0607\\
    \hline
    \hline
    \end{tabular}
\end{table*}

\begin{figure*}
    \centering
    \includegraphics[width=0.7\linewidth]{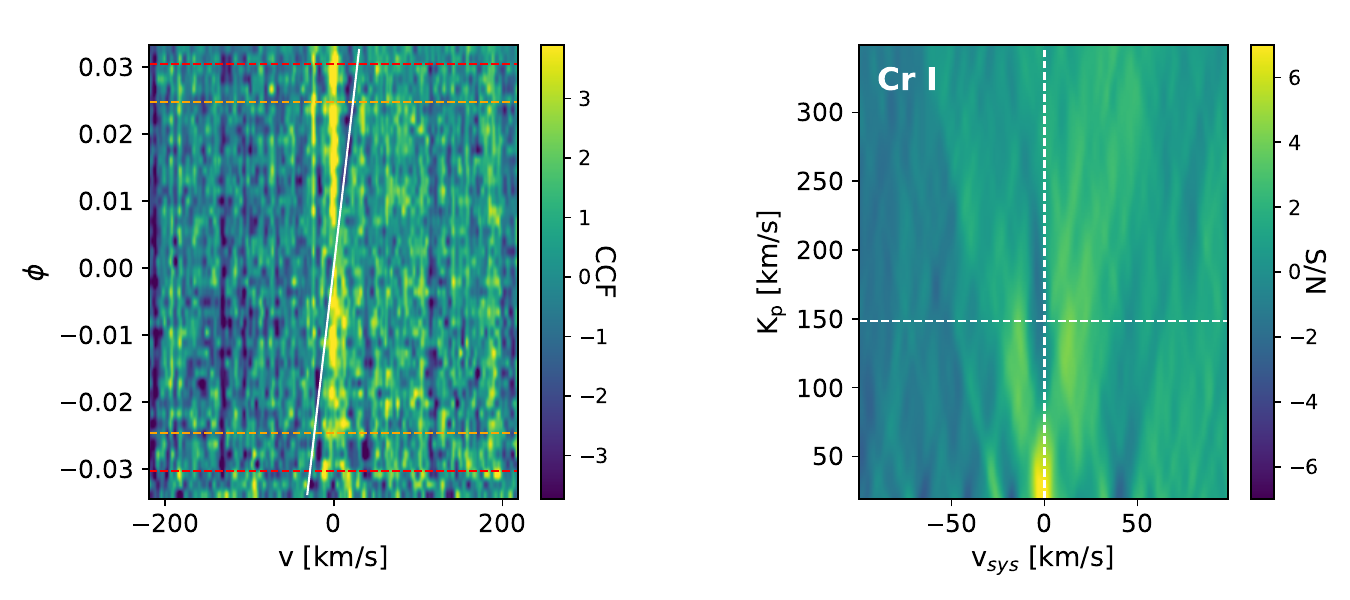}
    \includegraphics[width=0.7\linewidth]{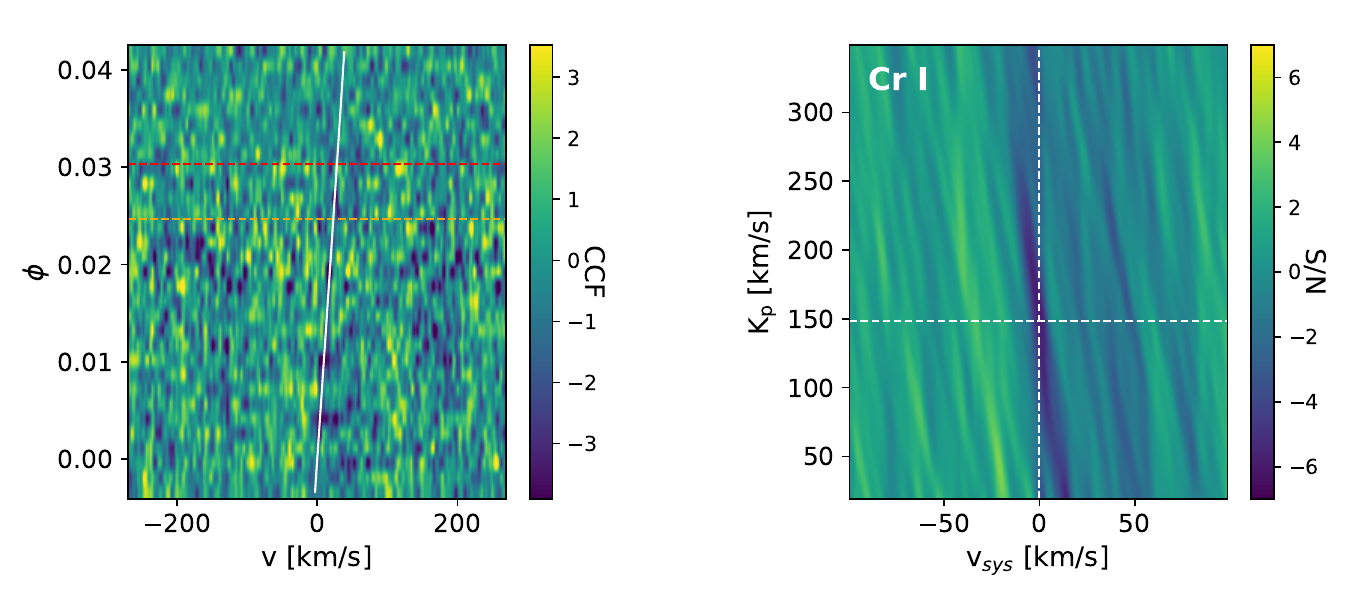}
    \includegraphics[width=0.7\linewidth]{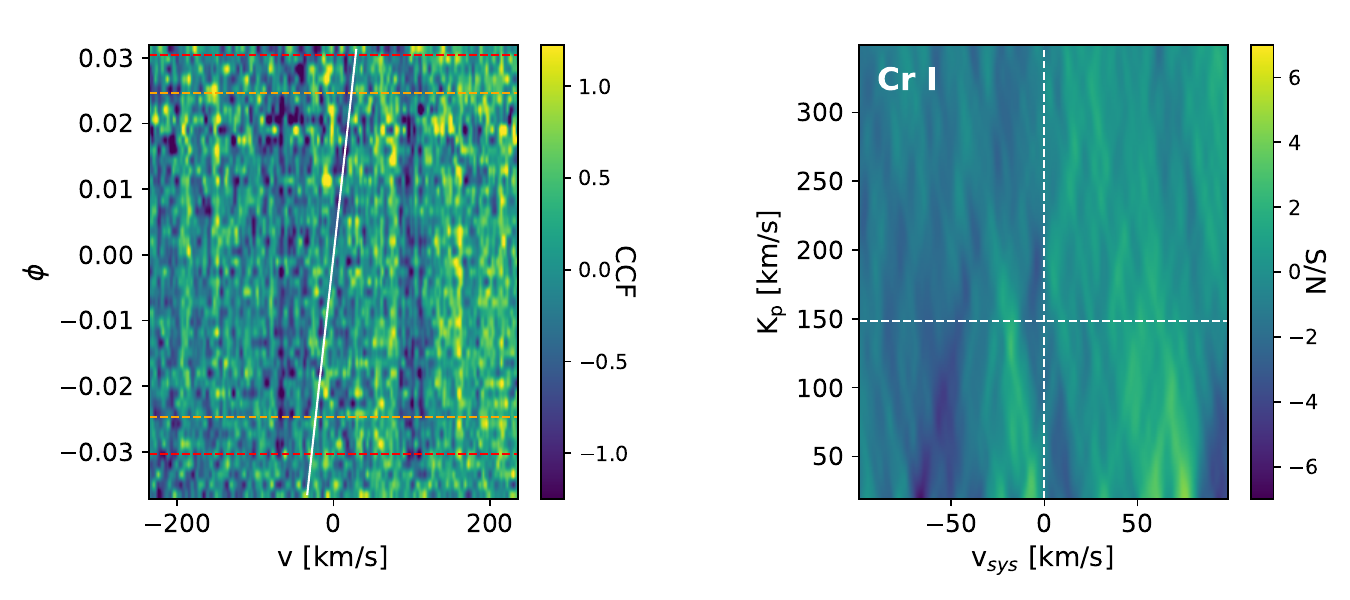}
    \includegraphics[width=0.7\linewidth]{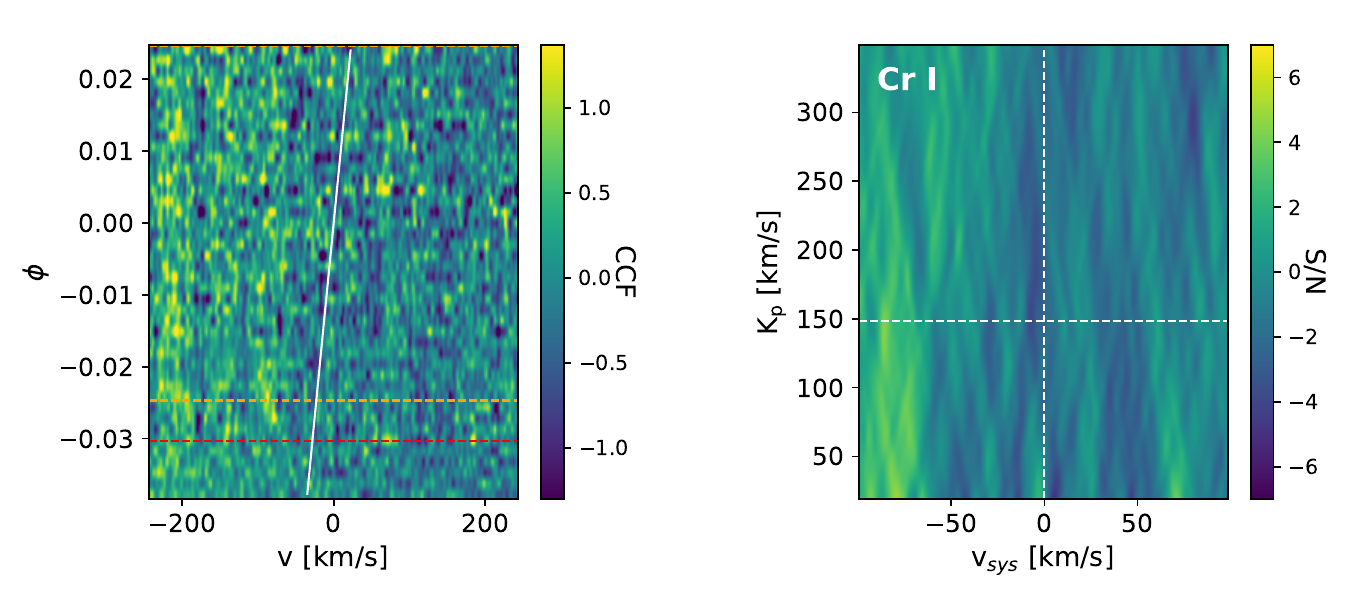}
    \caption{\textit{Left column - }Phase-stack of the cross-correlation between the transmission spectra and the Chrome mask. The nights N1, N2, N3 and N4 are shown from top to bottom respectively. In each plot, the orange dashed lines mark the orbital phases corresponding to the second and third contacts. Similarly, the red dashed lines mark the first and fourth contacts. The white solid line traces the expected planetary signal. \textit{Right column - }\kp--\vsys\ planes corresponding to the \acp{CCF} shown in the left column. In each plot, the dashed cross marks the expected \kp\ and \vsys\ of the system.}\label{fig:CCF_Cr}
\end{figure*}

\begin{figure*}
    \centering
    \includegraphics[width=0.7\linewidth]{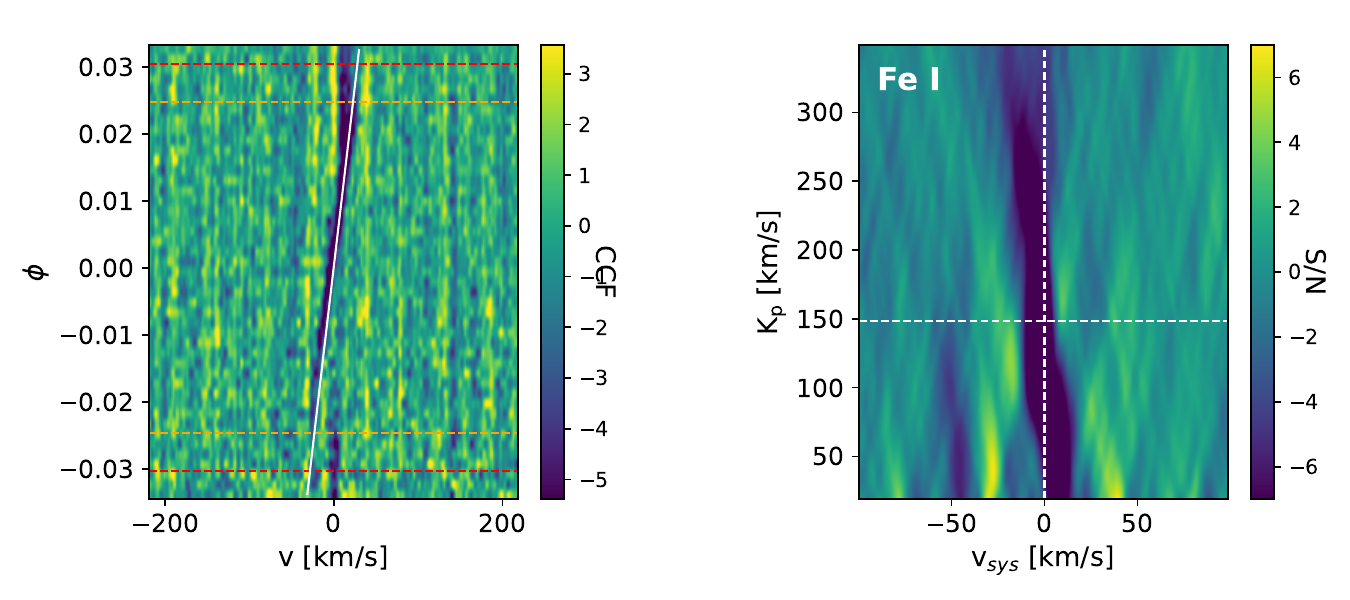}
    \includegraphics[width=0.7\linewidth]{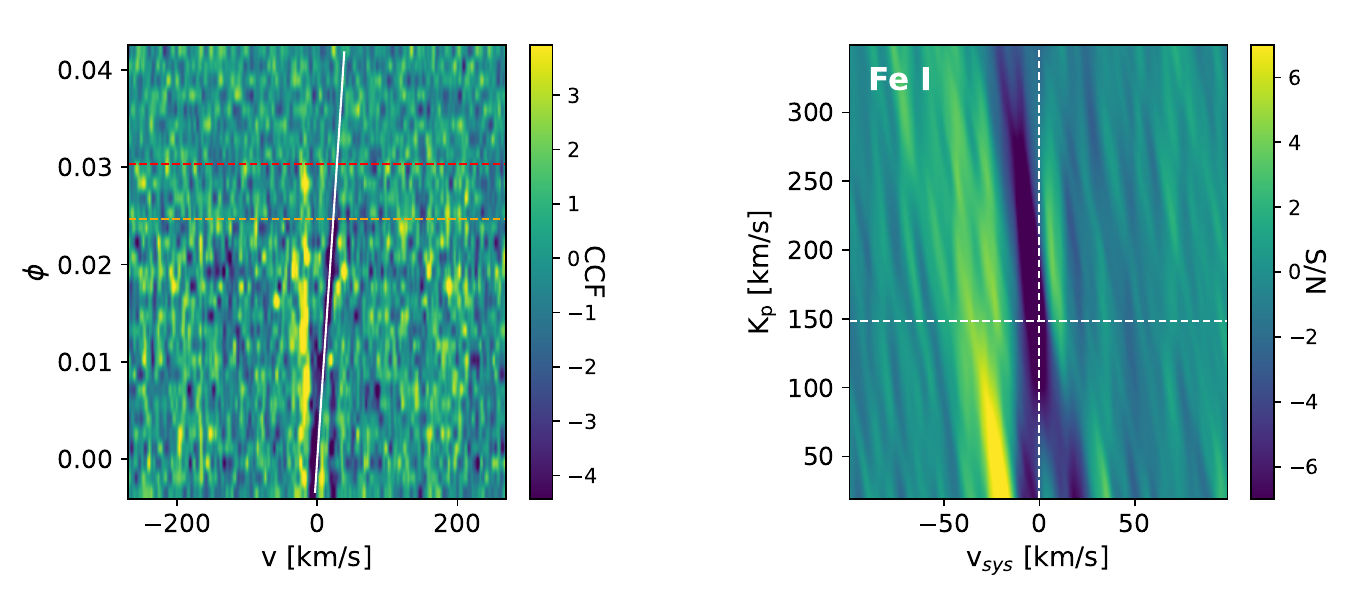}
    \includegraphics[width=0.7\linewidth]{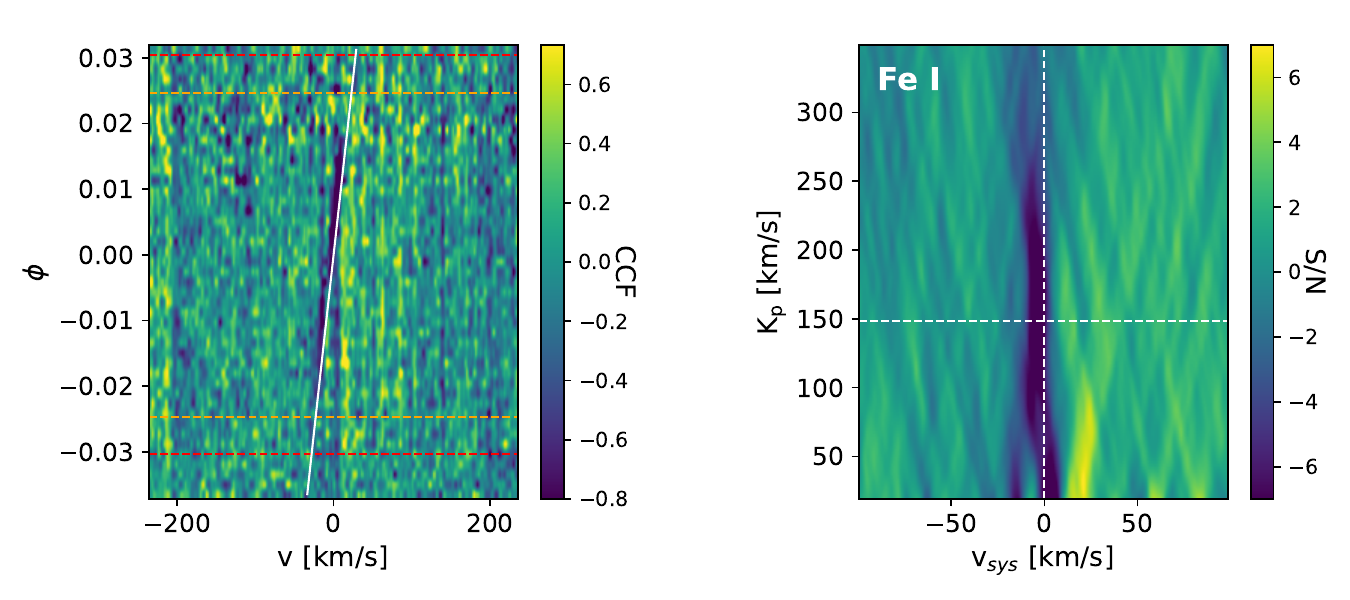}
    \includegraphics[width=0.7\linewidth]{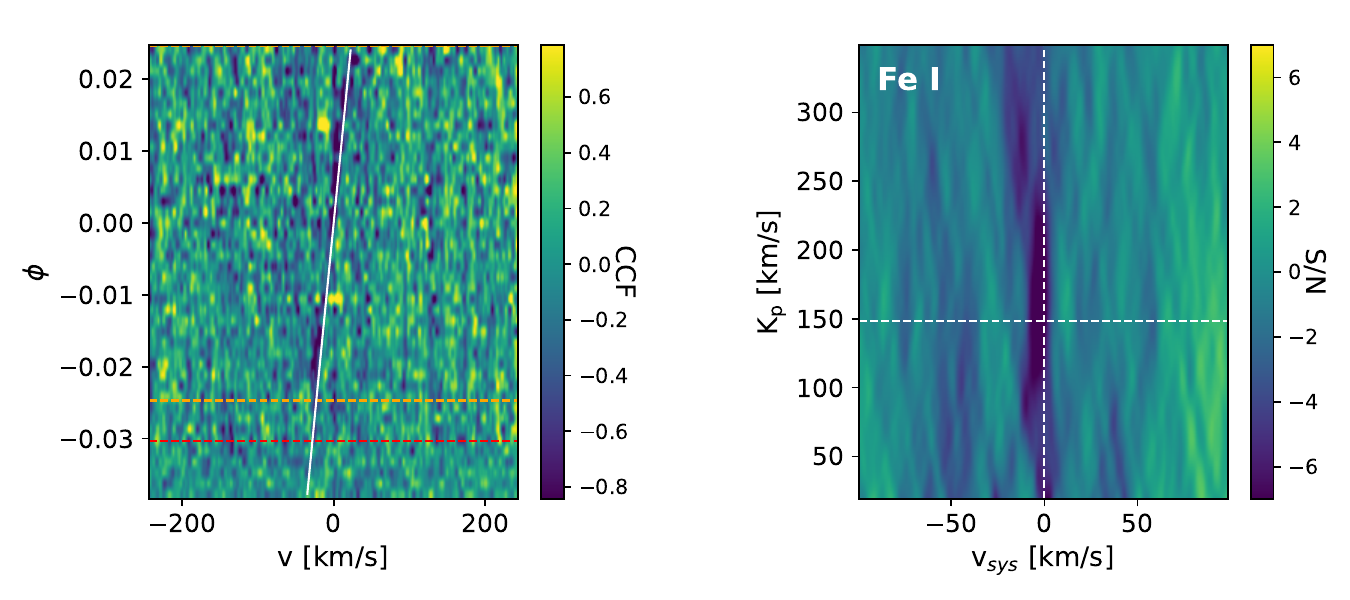}
    \caption{As Figure \ref{fig:CCF_Cr} but for the \ion{Fe}{I} mask.}\label{fig:CCF_fe}
\end{figure*}

\begin{figure*}
    \centering
    \includegraphics[width=0.7\linewidth]{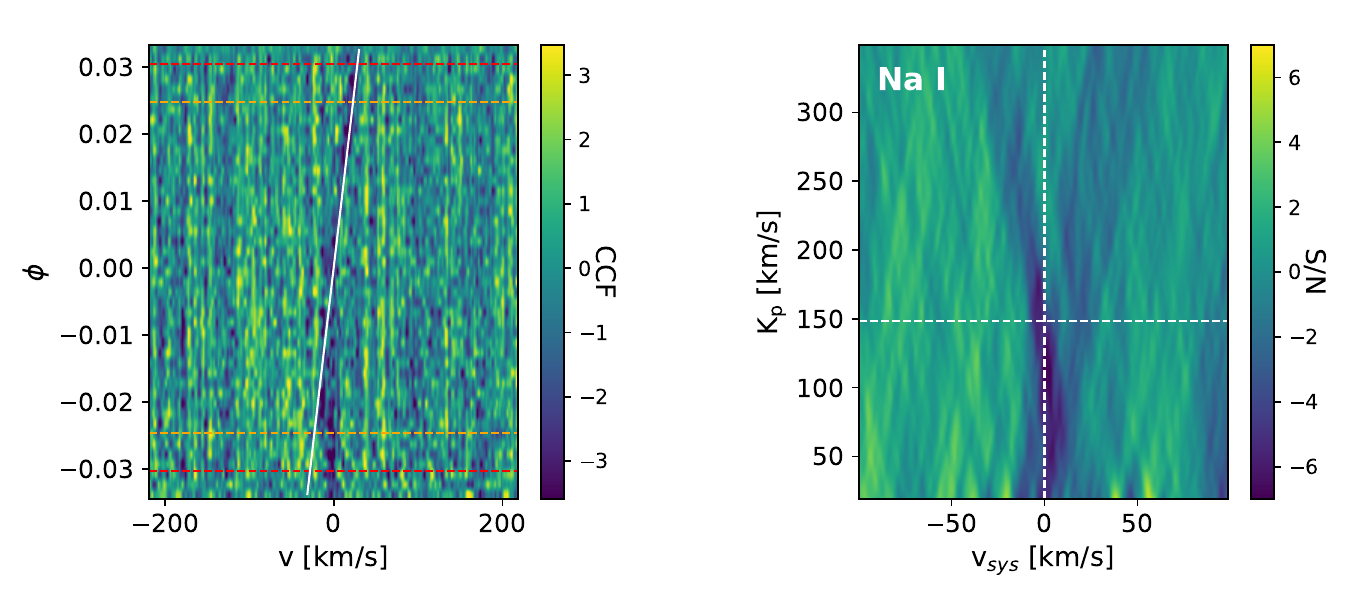}
    \includegraphics[width=0.7\linewidth]{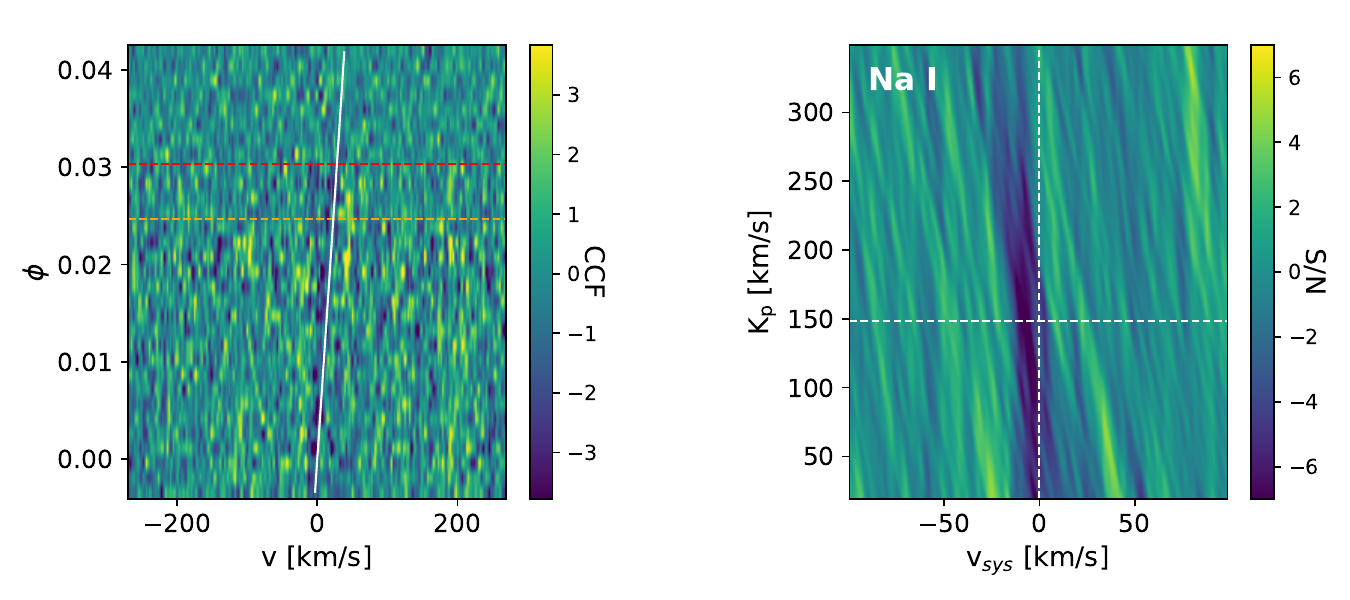}
    \includegraphics[width=0.7\linewidth]{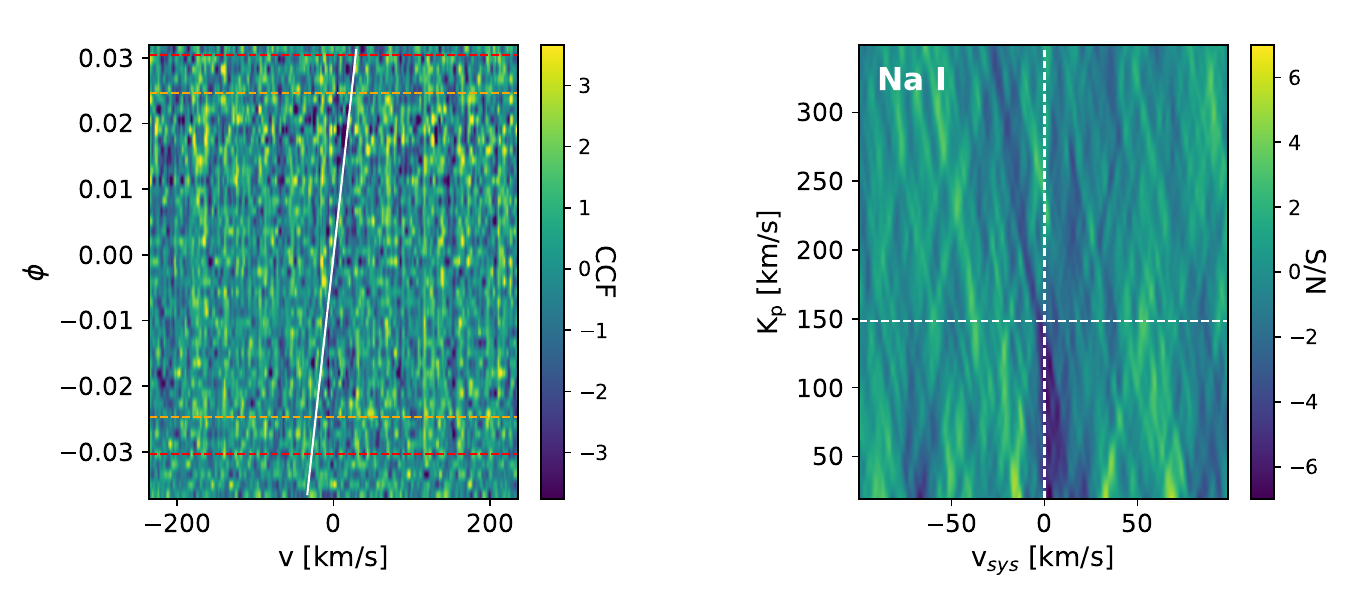}
    \includegraphics[width=0.7\linewidth]{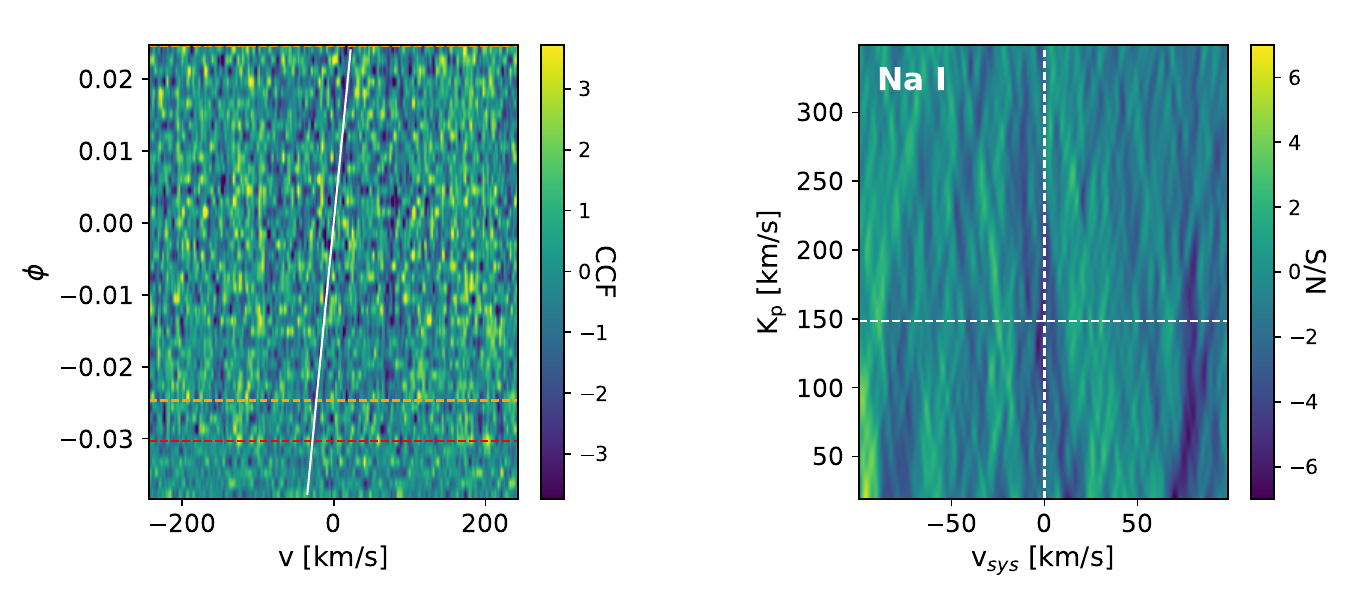}
    \caption{As Figure \ref{fig:CCF_Cr} but for the \ion{Na}{I} mask.}\label{fig:CCF_Na}
\end{figure*}

\begin{figure*}
    \centering
    \includegraphics[width=0.7\linewidth]{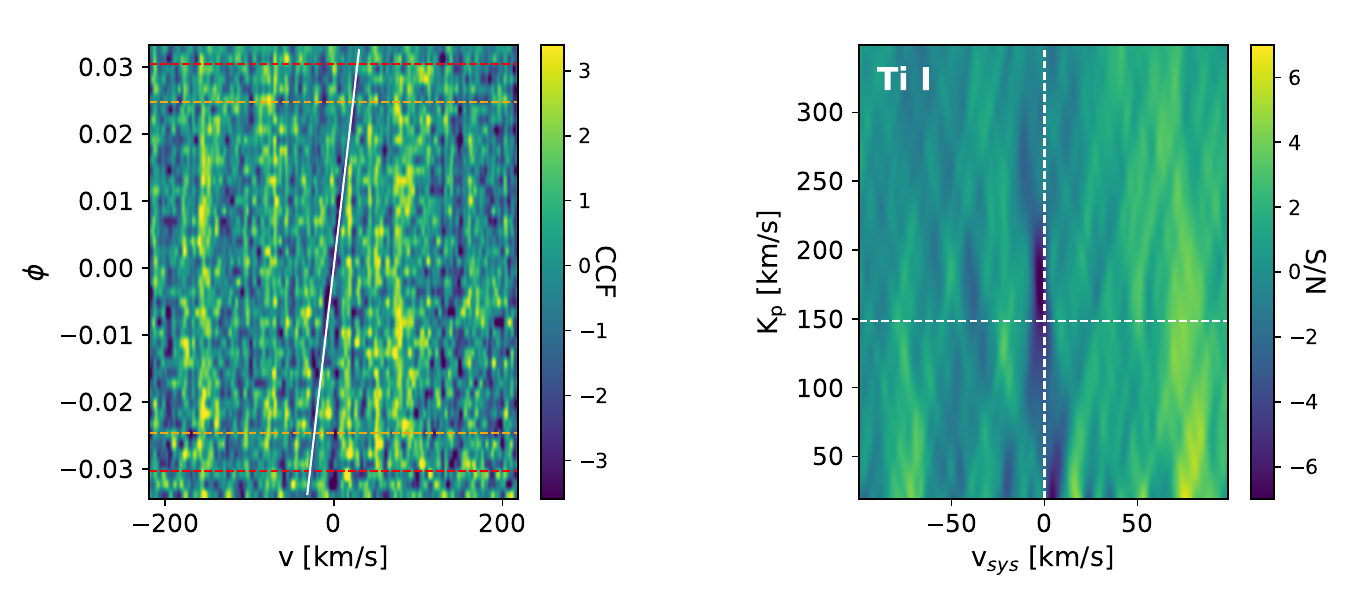}
    \includegraphics[width=0.7\linewidth]{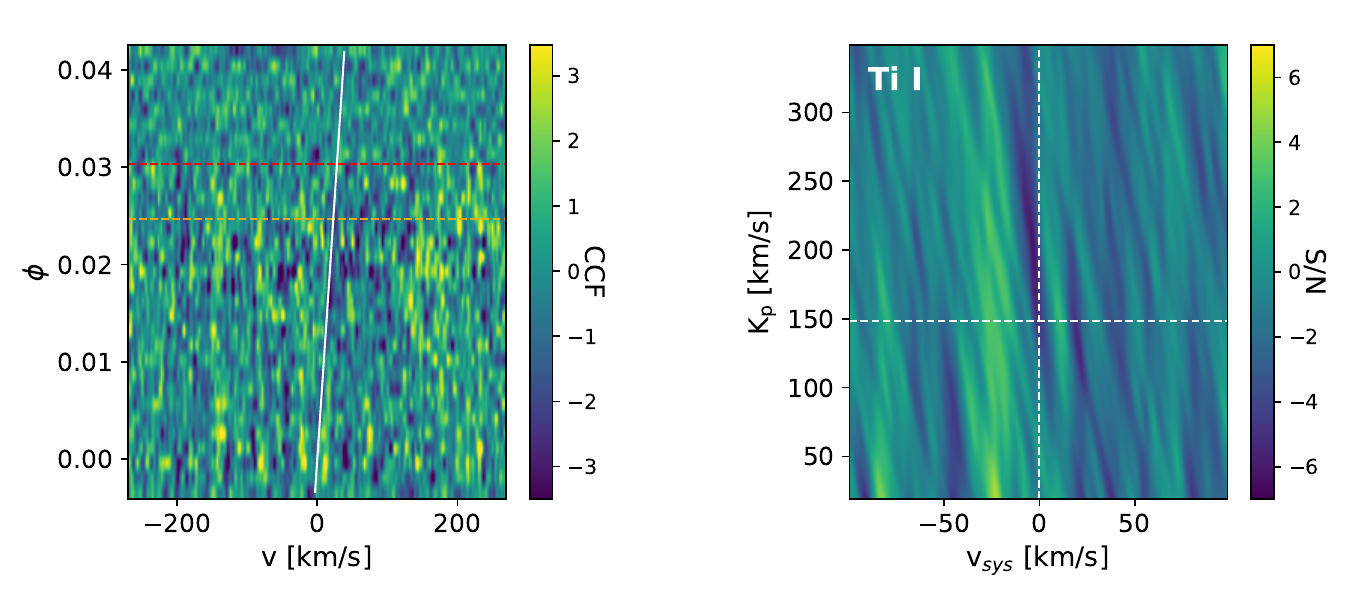}
    \includegraphics[width=0.7\linewidth]{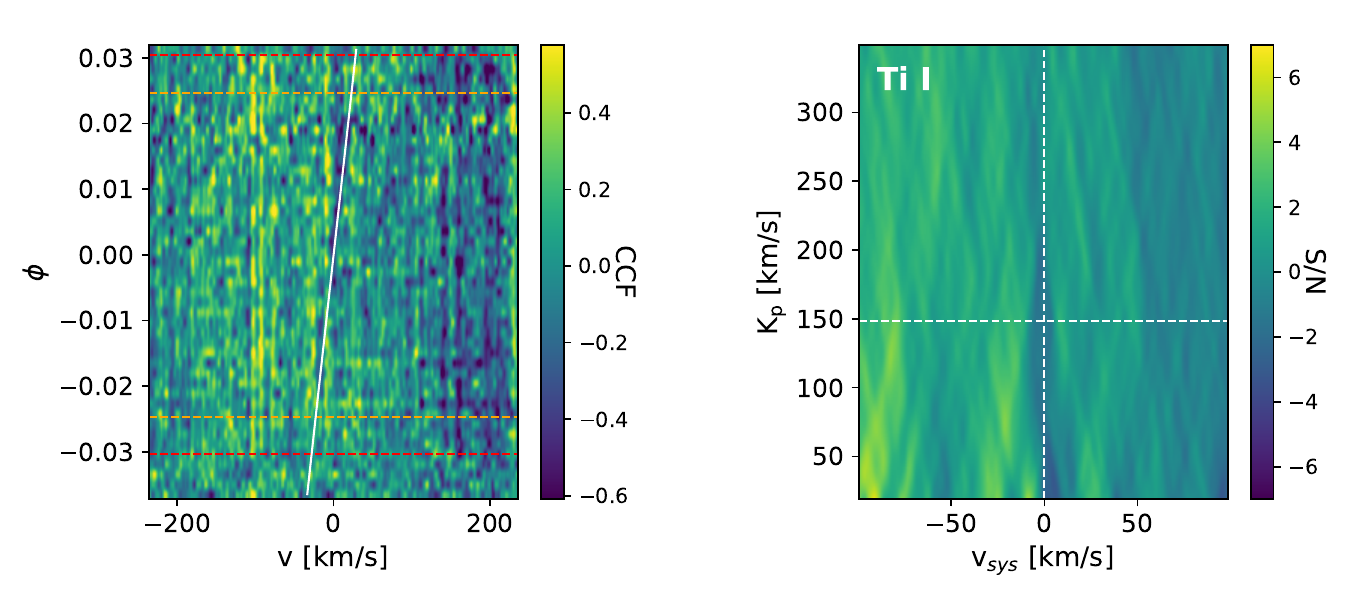}
    \includegraphics[width=0.7\linewidth]{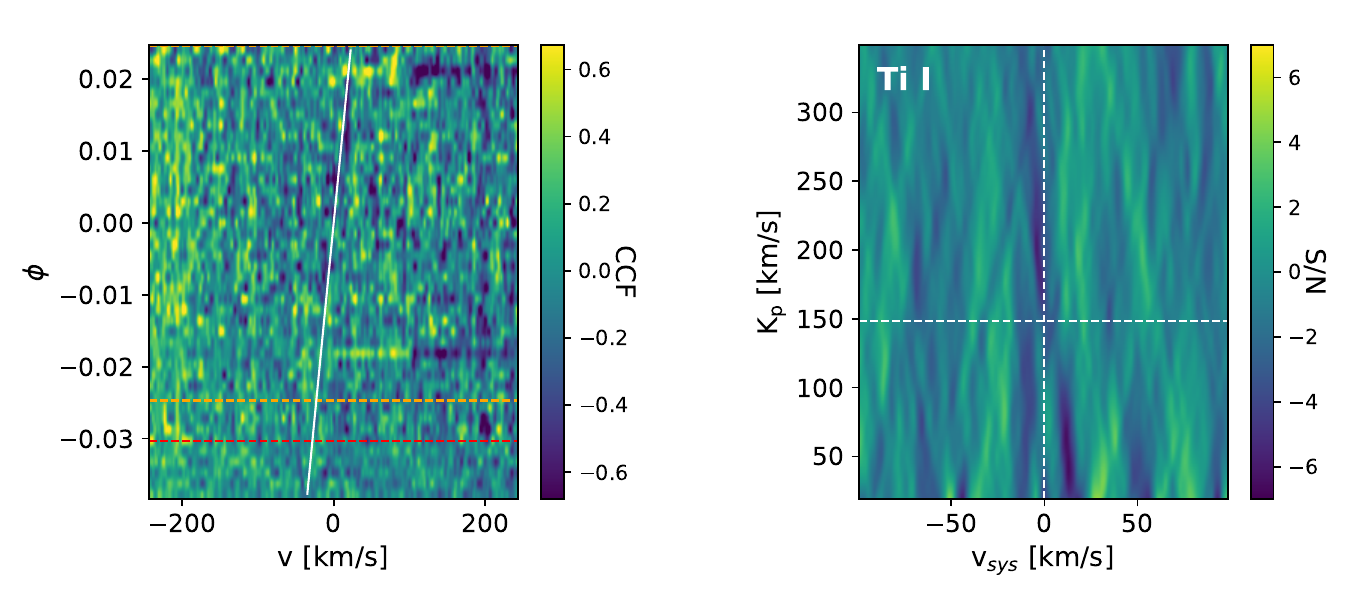}
    \caption{As Figure \ref{fig:CCF_Cr} but for the \ion{Ti}{I} mask.}\label{fig:CCF_Ti}
\end{figure*}

\begin{figure*}
    \centering
    \includegraphics[width=0.7\linewidth]{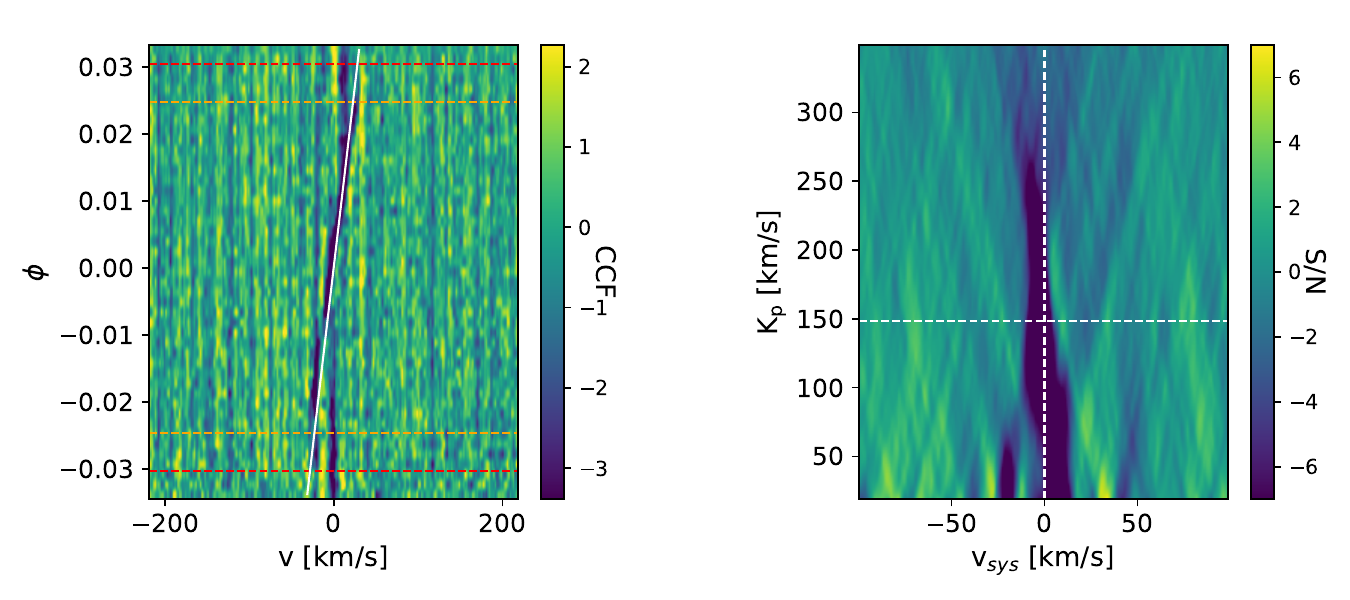}
    \includegraphics[width=0.7\linewidth]{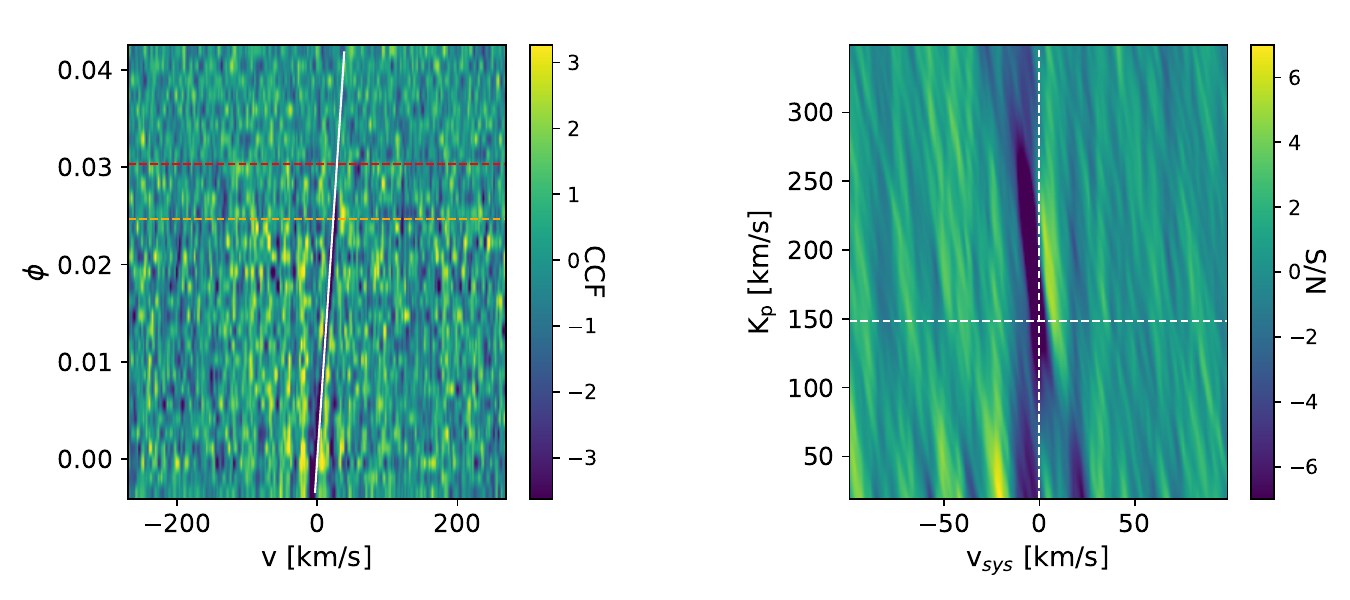}
    \includegraphics[width=0.7\linewidth]{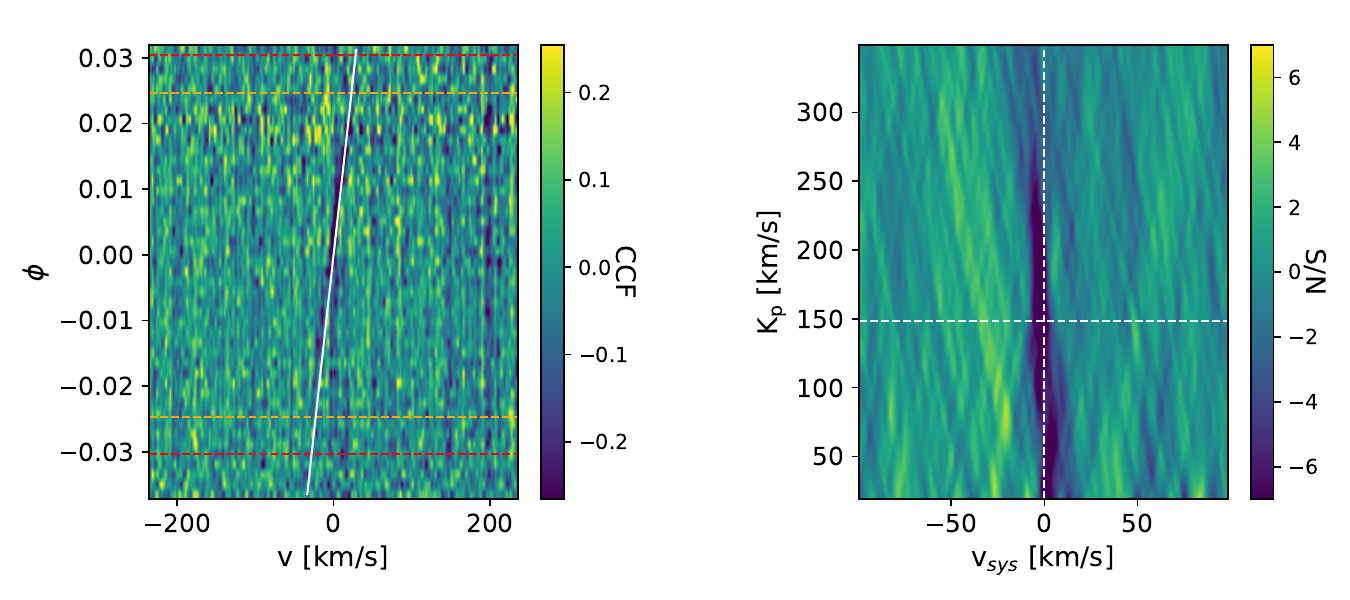}
    \includegraphics[width=0.7\linewidth]{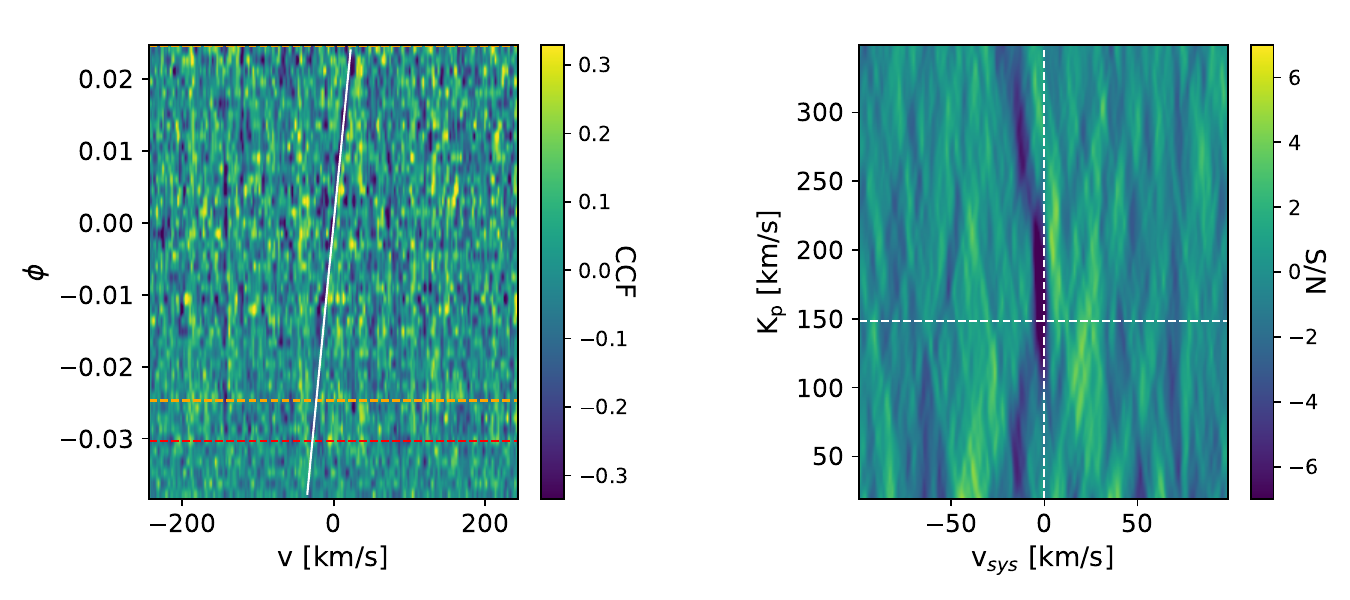}
    \caption{As Figure \ref{fig:CCF_Cr} but for the VALD mask devoid of the \ion{Cr}{i}, \ion{Fe}{i}, \ion{Na}{i}, and \ion{Ti}{i} spectral lines.}\label{fig:CCF_VALD}
\end{figure*}

\begin{figure*}
    \centering
    \includegraphics[width=\textwidth]{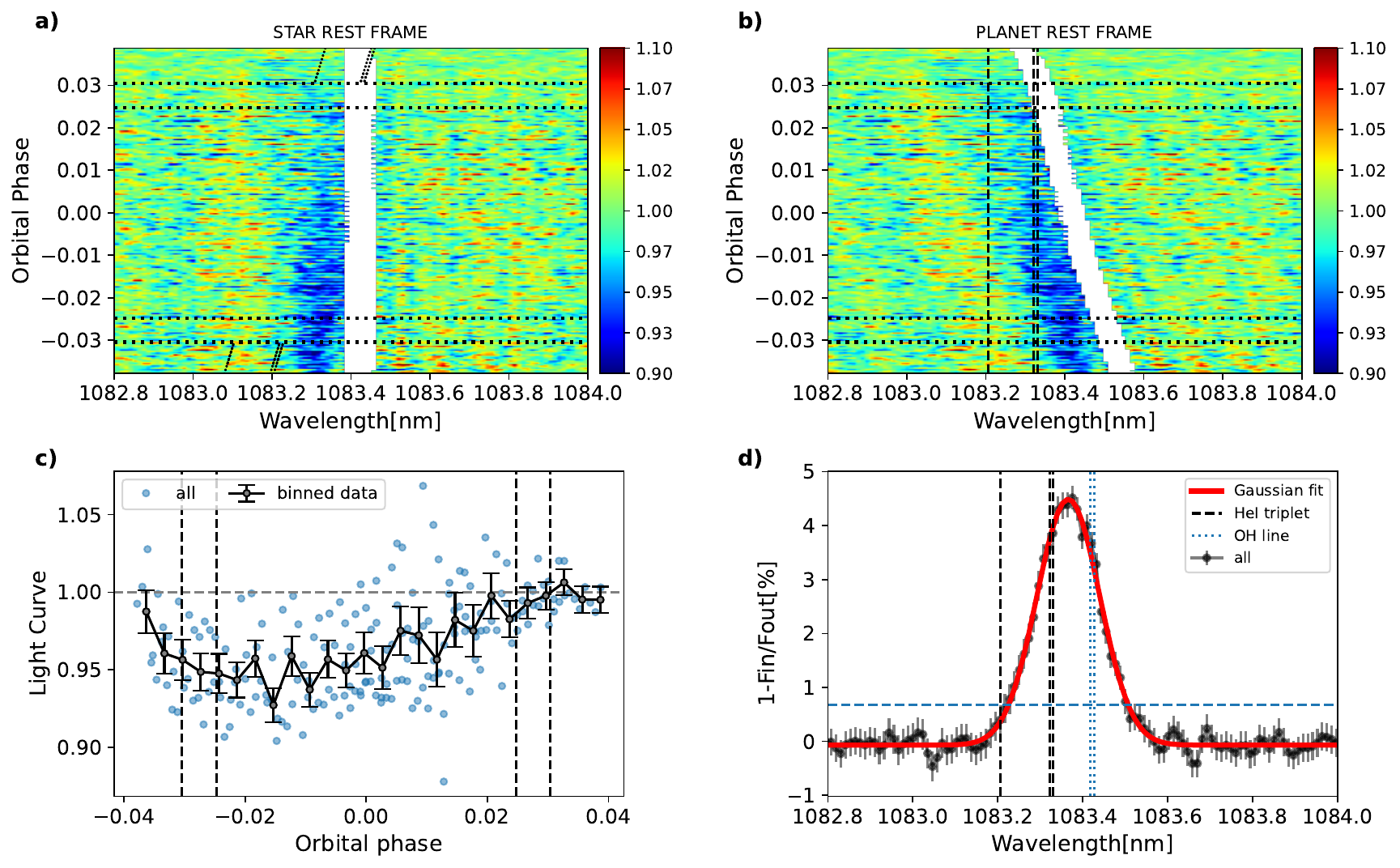}
    \caption{Same as Fig.~\ref{he_result_mask}, but considering all the available GIANO-B nights. }
    \label{he_result_mask_3}
\end{figure*}

\begin{table*}[h]
\caption{Best fit parameters.}
	\centering
	\begin{tabular}{ c c c c c c}
		\hline \hline
		Peak position & Contrast $c$ &  $R_\mathrm{eff}$ & FWHM & Significance  & $\delta_{R\rm_p}$/H$_{\mathrm{eq}}$  \\
                (nm)      &         (\%) & ($R\rm_p$)  & (nm) & ($\sigma$)      &      \\   
                \hline

            1083.3698$^{+ 0.0045 }_{ -0.0055 }$ & 4.56$^{+ 0.32 }_{ -0.31 }$ & 2.79 $\pm$ 0.08 & 0.1730$^{+ 0.014 }_{ -0.012 }$ & 14.5 & 42.4 $\pm$ 31.3 \\ 
         \hline
        \end{tabular}
	\tablefoot{Same as Table~\ref{tab_result_He}, but considering all the available GIANO-B nights. From left to right: the peak position of the \ion{He}{I}, the absorption (expressed both as contrast $c$ and $R_\mathrm{eff}$), and FWHM obtained from the DE-MCMC analysis, the significance of the detection, and the ratio between the equivalent height of the \Hei atmosphere and the atmospheric scale height. We determined the values and the 1$\sigma$ uncertainties of the derived parameters from the medians and the 16\%-84\% quantiles of their posterior distributions.}
	\label{tab_result_He_3}
\end{table*}

\begin{figure*}
    \centering
    \includegraphics[width=\textwidth]{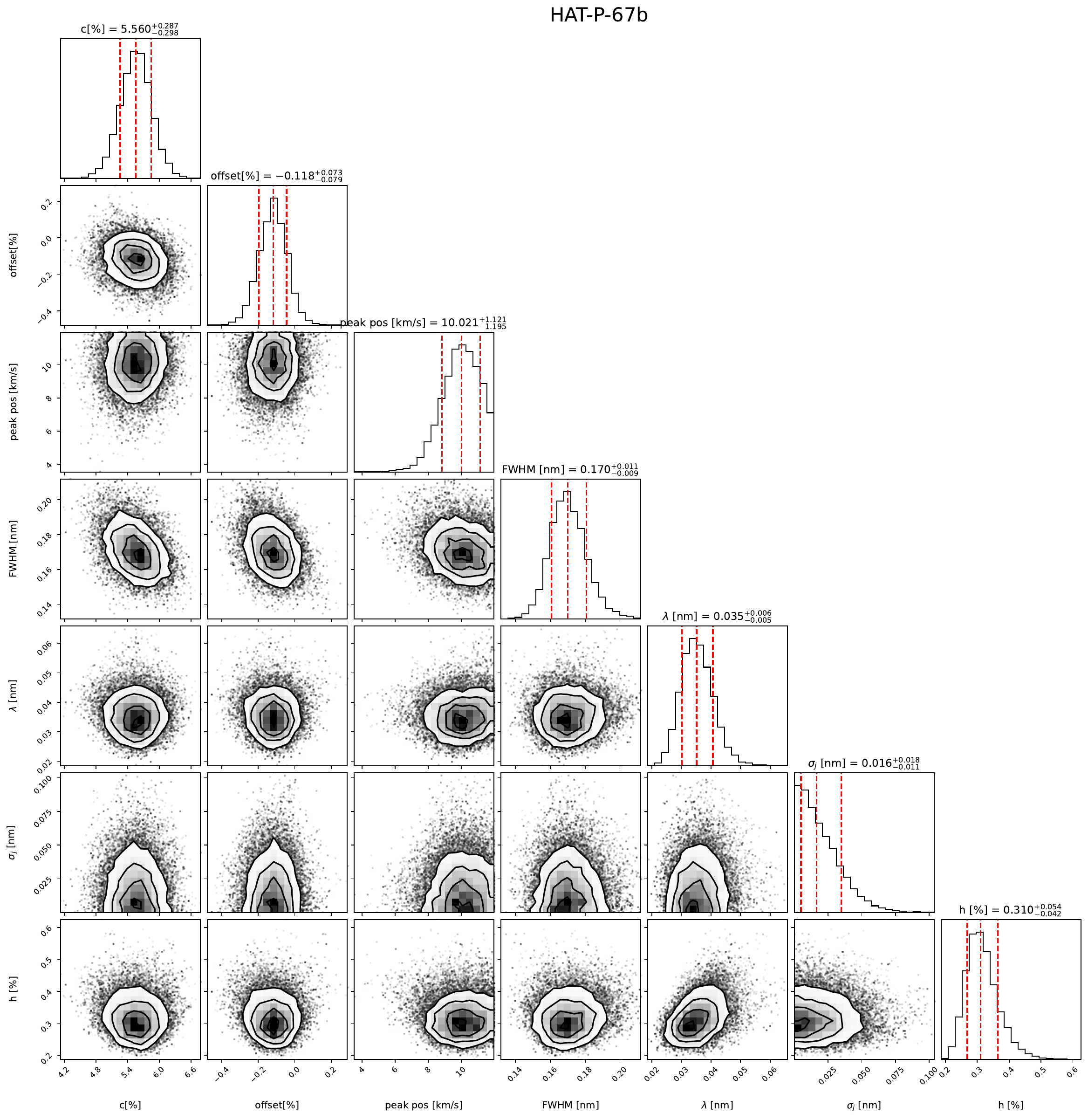}
    \caption{Posterior distribution of the investigated parameters in the DE-MCMC analysis of the \Hei triplet. The excess of absorption $c$ [\%], offset [\%], peak position, and FWHM correspond to the parameters we used in the Gaussian fit, while the jitter term $\sigma_\mathrm{j}$, the semi-amplitude of the correlated noise $h$, the correlation length $\lambda$ were used to parametrize the SE kernel within the GP. }
    \label{corner_He}
\end{figure*}
\begin{figure*}
    \centering
    \includegraphics[width=\textwidth]{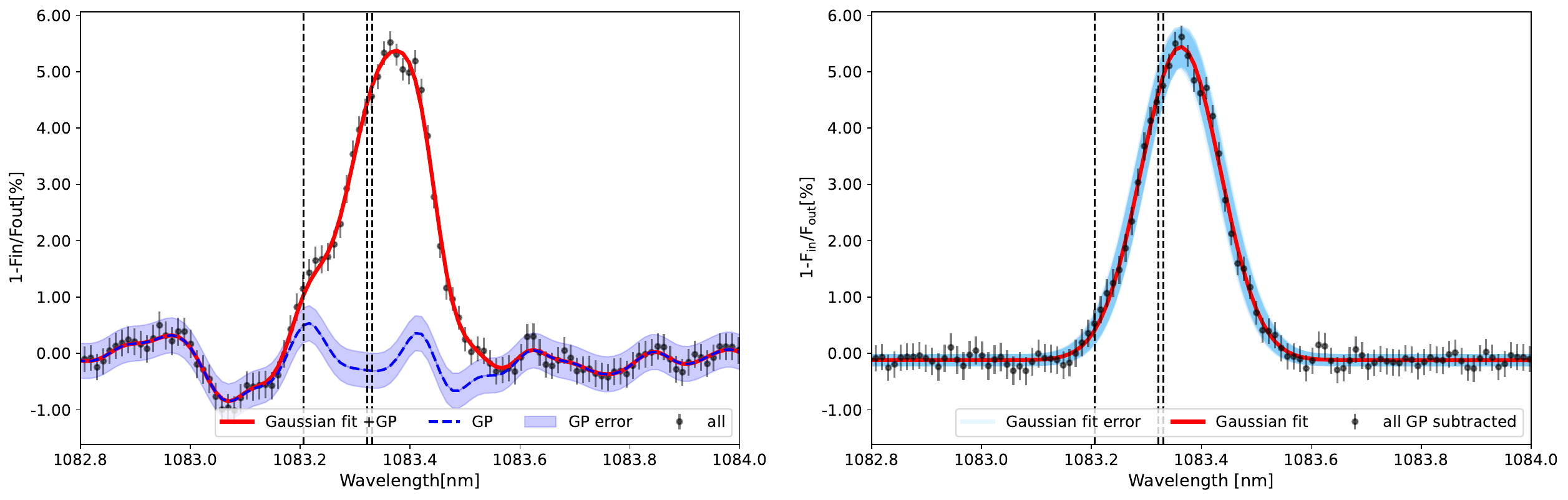}
    \caption{ GP correction. Right panel: transmission spectrum centered on the \ion{He}{I} triplet (in the planet rest frame) with overplotted the GP regression model, along with the 1$\sigma$ uncertainty intervals (in blue) and the Gaussian+GP model (in red). Left Panel: Final transmission spectrum after removing the GP model. The error intervals
for the Gaussian fit were computed by displaying 1000 Gaussian fits within the 1$\sigma$ uncertainties of the derived parameters, spanning the 16\%-84\%
quantiles. Vertical black dotted lines indicate the position of the \ion{He}{I} triplet.}
    \label{GP}
\end{figure*}

\section{Cross-correlation analysis of the H$\rm\alpha$ line} \label{app:Halpha}
\begin{figure*}
    \centering
    \includegraphics[width=0.7\linewidth]{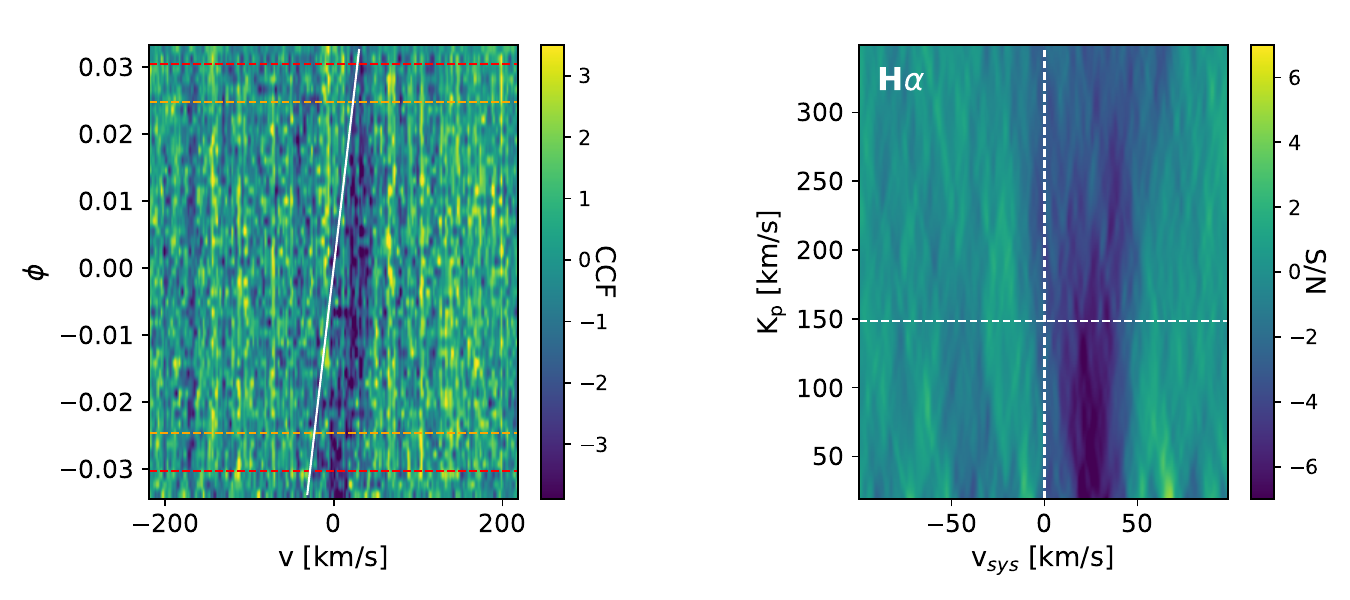}
    \includegraphics[width=0.7\linewidth]{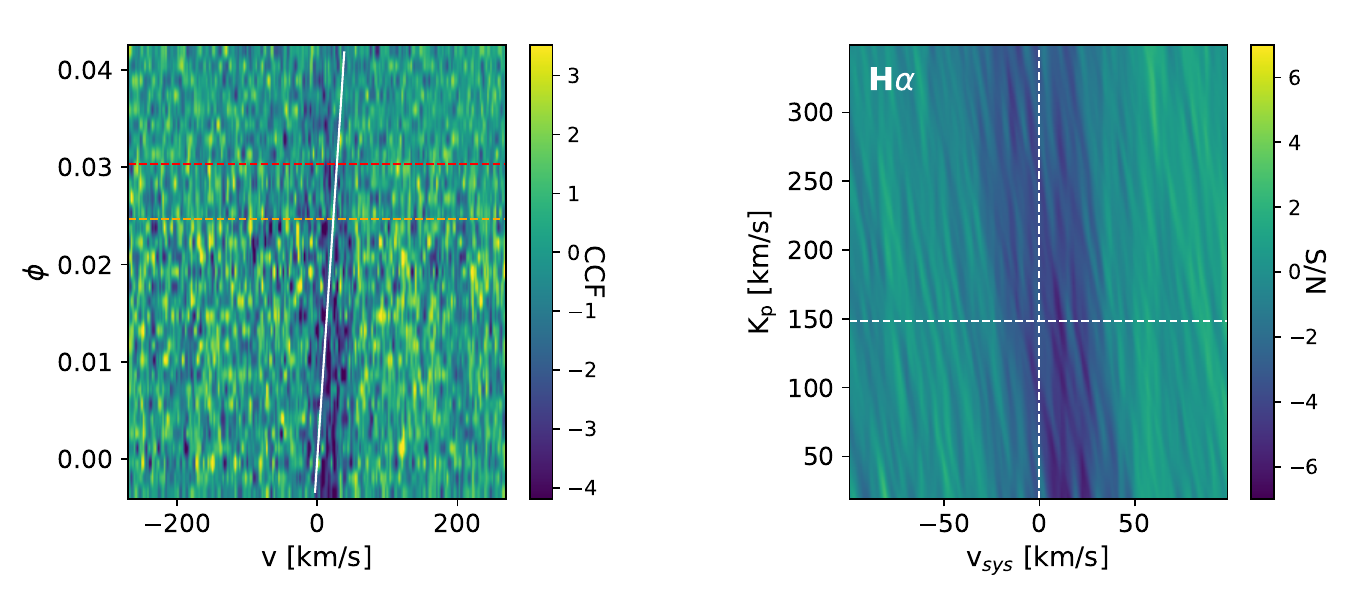}
    \includegraphics[width=0.7\linewidth]{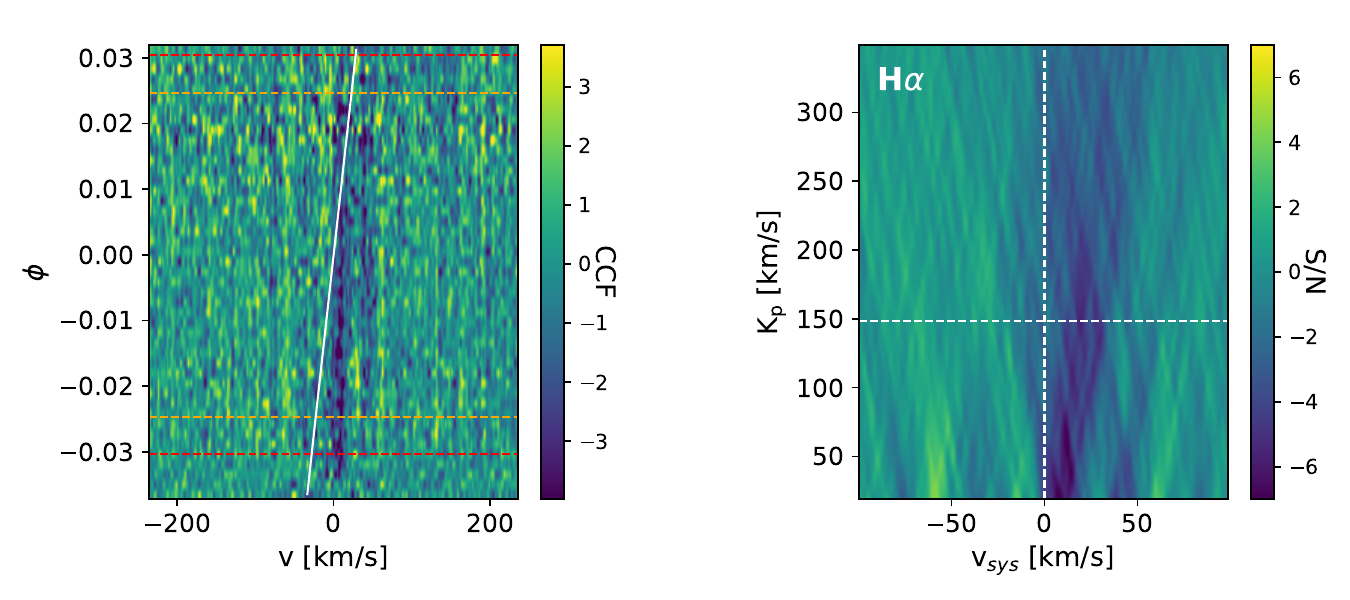}
    \includegraphics[width=0.7\linewidth]{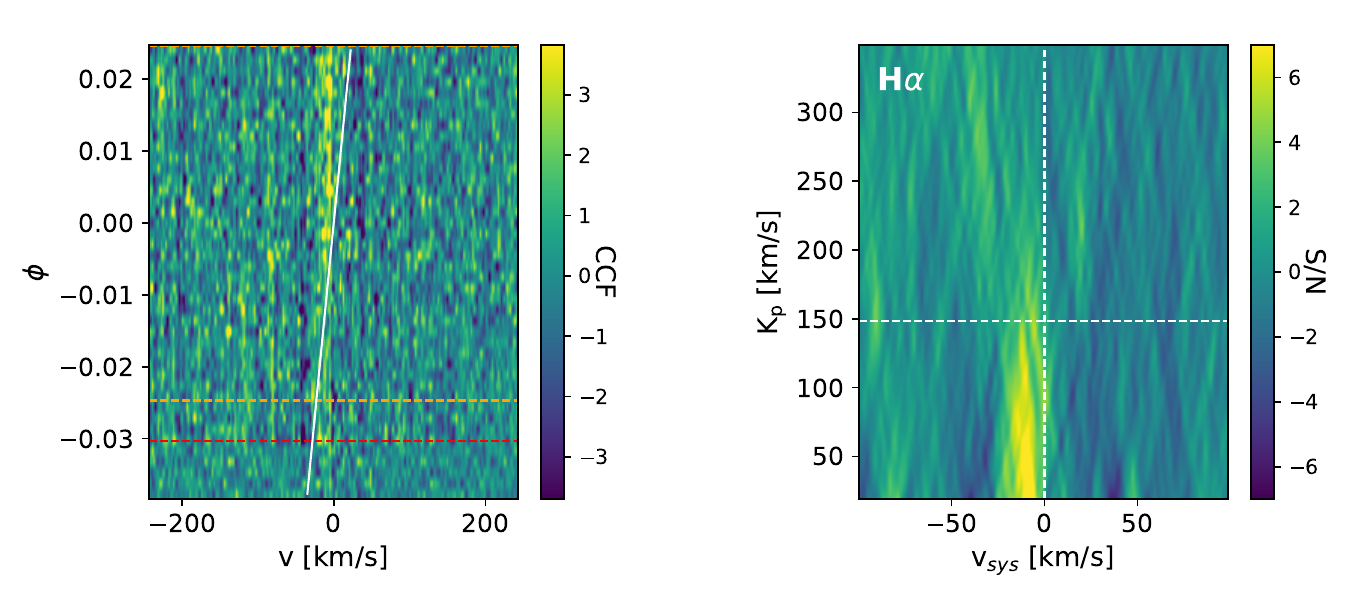}
    \caption{As Fig. \ref{fig:CCF_Cr} but for a binary mask containing only the H$\rm\alpha$ line. }\label{fig:CCF_Halpha}
\end{figure*}

The transmission spectrum of the H$\rm\alpha$ line reveals the presence of a clear emission feature during N4 (see Fig. \ref{fig:ts_Halpha}).
Trying to understand the origin of this feature, we replicated the same \ac{CCF} analysis discussed in Sect. \ref{sec:cross-correlation} by creating a binary mask containing only the H$\rm\alpha$ line. The result for each night is shown in Fig. \ref{fig:CCF_Halpha}.
As expected, the cross-correlation analysis reveals the presence of a strong absorption feature in N1, N2, and N3, well-aligned in the stellar rest frame, with a S/N above 5 in all three nights. Besides, the peaks of these signals are not in correspondence with the expected \kp \, of the system. 
From the cross-correlation analysis of N4, we can see the emission feature starting from around mid-transit. Moreover, even on this night, the signal does not follow the track expected by the planet, but is well aligned in the stellar reference system. 

\end{appendix}

\end{document}